\documentclass[aps,plb,letterpaper,twocolumn,reprint,preprintnumbers,floatfix,nofootinbib,showpacs]{revtex4-1}
\usepackage{graphicx}
\usepackage{dcolumn}
\usepackage{epsfig,cancel}
\usepackage{epstopdf}
\usepackage{amsmath}
\usepackage{amssymb}
\usepackage{color}
\usepackage{hyperref}
\usepackage{url}
\usepackage{breakurl}
\usepackage[normalem]{ulem}
\usepackage{subcaption}
\usepackage{float}

\newcommand{\be}{\begin{equation}}
\newcommand{\ee}{\end{equation}}
\newcommand{\bea}{\begin{eqnarray}}
\newcommand{\eea}{\end{eqnarray}}
\newcommand{\nn}{\nonumber}

\begin{document}

\title{Phenomenology of single-inclusive jet production with jet radius and threshold resummation}

\begin{flushright}
DESY 18-007 \\
\end{flushright}

\author{Xiaohui Liu}
\email{xiliu@bnu.edu.cn}
\affiliation{Center of Advanced Quantum Studies, Department of Physics, Beijing Normal University, Beijing 100875, China}

\author{Sven-Olaf Moch}
\email{sven-olaf.moch@desy.de}
\affiliation{II. Institut f\"ur Theoretische Physik, Universit\"at Hamburg, Luruper Chaussee 149, D-22761 Hamburg, Germany} 

\author{Felix Ringer}
\email{fmringer@lbl.gov}
\affiliation{Nuclear Science Division, Lawrence Berkeley National Laboratory, Berkeley, California 94720, USA}

\begin{abstract}
We perform a detailed study of inclusive jet production cross sections at the LHC 
and compare the QCD theory predictions based on the recently developed formalism 
for threshold and jet radius joint resummation at next-to-leading logarithmic accuracy 
to inclusive jet data collected by the CMS collaboration at $\sqrt{S} = 7$ and $13$~TeV. 
We compute the cross sections at next-to-leading order in QCD with and without the joint resummation 
for different choices of jet radii $R$ and observe that the joint resummation
leads to crucial improvements in the description of the data. 
Comprehensive studies with different parton distribution functions demonstrate 
the necessity of considering the joint resummation in fits 
of those functions based on the LHC jet data. 
\end{abstract}

\maketitle

\section{Introduction}

Long term persistence in achieving higher order calculations in 
perturbative Quantum Chromodynamics (pQCD) paves the way to the precision frontier at
the Large Hadron Collider (LHC). 
With many Standard Model processes now being measured with an impressive
accuracy at the LHC, 
theoretical predictions beyond next-to-leading order (NLO) in pQCD, 
nowadays considered the standard in phenomenological analyses, 
are often mandatory.
During the past three years, there have been a burst of publications on 
complete next-to-next-to-leading order (NNLO) calculations 
for various hadro-production processes involving jets~\cite{Boughezal:2015dva,Boughezal:2015dra,Boughezal:2015aha,
  Chen:2014gva,Ridder:2015dxa,Boughezal:2015ded,Currie:2016bfm,Currie:2017eqf,Campbell:2016lzl}.
The list of those processes includes the hadro-production of gauge bosons $V$+jet ($V = W^\pm, Z, \gamma$) 
as well as single-inclusive jets and dijets, but it is limited to $2 \to 2$ reactions at Born level due to the enormous
computational complexity at NNLO.
In particular the calculations for $V$+jet production have already been shown 
to greatly improve the description of the available LHC data~\cite{Boughezal:2016yfp,Lindert:2017olm,Boughezal:2017nla}. 

For the hadro-production of jets at the LHC the experimental collaborations have provided
very precise data for the single-inclusive jet production cross sections
$pp\to\text{jet}+X$ at all collider energies and differential in the jet
transverse momentum $p_T$ and the rapidity $\eta$. 
Specifically, ALICE~\cite{Abelev:2013fn}, ATLAS~\cite{Aad:2013lpa} and  CMS~\cite{Khachatryan:2016jfl}
have collected data at $\sqrt{S} = 2.76$~TeV and ATLAS and CMS at $\sqrt{S} = 7$~TeV~\cite{Aad:2014vwa,Chatrchyan:2014gia}, 
$8$~TeV~\cite{Aaboud:2017dvo,Khachatryan:2016mlc} and $13$~TeV~\cite{Aaboud:2017jcu,ATLAS-CONF-2017-048,Khachatryan:2016wdh}.
These data allow for important consistency tests of pQCD as well as a precise 
extraction of the value of the strong coupling constant $\alpha_s(M_Z)$~\cite{Britzger:2017maj} 
and they provide very valuable constraints on parton distribution functions (PDFs) 
which govern the parton luminosity of the colliding initial protons~\cite{Nocera:2017zge,Harland-Lang:2017ytb}. 

In order to fully utilize the available data, a precise understanding of the
corresponding theoretical calculations within pQCD is very important. 
The current accuracy for fixed order pQCD predictions is NNLO where the $\alpha_s^2$ coefficient 
is known in the leading-color approximation~\cite{Currie:2016bfm}, i.e. 
for large values of $N_c$ for a general SU$(N_c)$ gauge group.
Any additional corrections are parametrically suppressed as $1/N_c^2$, so that
the results of~\cite{Currie:2016bfm} are supposed to approximate the full NNLO
calculation very well. 
Preliminary comparisons of those NNLO results with some of the LHC data, however, 
have not been entirely satisfactory. 
Refs.~\cite{Currie:2017ctp,ATLAS-CONF-2017-048} have shown that the agreement between theory and data heavily depends on the
choices for the renormalization and factorization scales $\mu_R$ and $\mu_F$.
Moreover, for some natural scale choices, such as identifying $\mu_R$ and $\mu_F$
with the transverse momentum $p_T^{\text{max}}$ of the leading jet in the event, i.e. $\mu_R=\mu_F=p_T^{\text{max}}$,  
the theory description of the data at NNLO deteriorates compared to NLO.
This situation implies the existence of potentially large higher order
corrections beyond fixed NNLO. 

Improvements beyond fixed order in pQCD are possible by supplementing the
fixed order calculations with resummation results where dominant classes of
logarithmic corrections are summed up to all orders in the strong coupling
constant. 
Recently, a joint resummation framework was developed~\cite{Liu:2017pbb} 
that allows to resum both threshold and jet radius logarithms simultaneously.  
Threshold logarithms appear in the partonic cross
section at $n$-th order as $\alpha_s^n(\ln^k(z)/z)_+$ where $z=s_4/s$
and $k\le 2n-1$. Here, $s_4$ is the invariant mass of the partonic system recoiling against the
observed jet and $s$ is the partonic center-of-mass energy~\cite{Kidonakis:1998bk}.
Since these logarithms are integrated over the specified parton kinematics together with the
steeply falling parton luminosity, threshold logarithms can dominate the
entire cross section in a wide kinematic range. 
Instead, the jet radius $R$ is an external quantity and the dependence of the
cross section is single-logarithmic $\alpha_s^n\ln^k(R)$ with $k\le n$ instead of double-logarithmic~\cite{Dasgupta:2014yra,Kang:2016mcy,Dai:2016hzf}. 
The framework developed in~\cite{Liu:2017pbb} addresses both these logarithmic
corrections on the same footing and it was shown that numerically the
threshold and the jet radius logarithmic terms do account for the dominant
bulk of the NLO corrections. 
The explicit resummation of these logarithms to next-to-leading logarithmic (NLL) accuracy  
was also realized within the joint resummation framework derived in~\cite{Liu:2017pbb} 
and the subsequent matching to fixed order NLO results leads to theory predictions at the combined 
NLO $+$ NLL accuracy.
The approach of~\cite{Liu:2017pbb}, however, is not limited to this logarithmic accuracy 
and the framework is ready for a systematic extension 
to the next-to-next-to leading logarithmic (NNLL) accuracy which may then be matched
to the available fixed order NNLO results to achieve a combined accuracy of 
NNLO $+$ NNLL. We leave the extension to NNLL for future work and instead focus here on
the phenomenological results at NLO $+$ NLL accuracy. 

In general, one expects competing effects from threshold and small-$R$
resummation. As it was observed in~\cite{Kidonakis:2000gi,deFlorian:2007fv,Kumar:2013hia,deFlorian:2013qia}
threshold resummation leads to an enhancement
whereas small-$R$ resummation alone leads to a decrease of the cross section~\cite{Dasgupta:2016bnd,Kang:2016mcy},
see also~\cite{Kang:2018qra} for studies on jet angularities.
Depending on the non-trivial interplay within the joint resummation framework, one or
the other effect will dominate. For certain kinematics and values of $R$, the
two effects may even largely cancel out. In order to obtain a good
understanding of the convergence of the perturbative series expansion it is
important to disentangle these two effects. A closely related issue is the 
dependence of the fixed order and the resummed calculations 
on the renormalization and factorization scales $\mu_R$ and $\mu_F$, 
collectively denoted by $\mu$ in the following.
As it was pointed out in~\cite{Currie:2017ctp,ATLAS-CONF-2017-048}, the fixed order results
change significantly depending on whether the hard scale is chosen as
$\mu=p_T$ of the individual jet or as the transverse momentum
$p_T^{\text{max}}$ of the leading jet in the event. At the same time, the
residual scale dependence is very small and even vanishes for some kinematic
configurations. In~\cite{Dasgupta:2014yra,Dasgupta:2016bnd,Kang:2016mcy}, it
was argued that this is generally an artifact of results at fixed order in
perturbation theory. Here, we address this issue within the joint resummation
formalism.

In this work, we provide a detailed comparison with LHC data and find that the
inclusion of the resummation generally yields a much better description of those data. 
In addition, our studies highlight possible improvements that can be
obtained by using a resummed calculation in fits of PDFs. 
The constraints from inclusive jet data on PDF fits are most significant for the gluon PDF
$g(x)$ in the large-$x$ region. 
In this endpoint region the cross sections from which PDFs are extracted can
be subject to large logarithmic corrections that need to be taken into account
to all orders.  
Improvements in the precision of the extracted PDFs eventually have direct impact
on all PDF sensitive analyses at the LHC and 
recent progress on PDFs in the large-$x$ region has been made in~\cite{Accardi:2014qda,Bonvini:2015ira}. 

The remainder of this work is organized as follows. In section~\ref{sec:th},
we briefly review the theoretical framework of~\cite{Liu:2017pbb}. 
In section~\ref{sec:pheno}, we present detailed phenomenological studies of the
resummation effect and the scale dependence of the resummed cross section.  
We study cross section ratios for different jet radii to discriminate the
predictive power of the NLL $+$ NLO and the NLO results.
Finally, we present a comprehensive comparison to the inclusive jet data from
the LHC together with the impact of different PDF sets. 
We conclude in section~\ref{sec:conc} with a summary and an outlook.

\section{Theoretical framework \label{sec:th}}
 
First we review the theoretical formalism which allows us to achieve
the threshold and small-$R$ joint resummation used in this work. The
resummation is based on the factorization theorem~\cite{Liu:2017pbb} developed
within the framework of the Soft Collinear Effective
Theory~\cite{Bauer:2000ew,Bauer:2000yr,Bauer:2001ct,Bauer:2001yt}, in which
the single-inclusive jet cross section with jet transverse momentum $p_T$, jet
rapidity $y$ and a small anti-$k_T$~\cite{Cacciari:2008gp} jet radius $R$ near
the partonic threshold can be written as  
\bea
\label{eq:crs}
\frac{ p_T^2 \mathrm{d}^2 \sigma    }{   \mathrm{d}p_T^2 \mathrm{d} y } 
& =  & \sum_{i_1i_2} 
\int_0^{V(1-W)}  \!\! \mathrm{d} z 
\int_{\frac{VW}{1-z}}^{1- \frac{1-V}{1-z}} \!\! \mathrm{d}v \,
x_1^2\,  f_{i_1}(x_1) \, x_2^2 \,  f_{i_2}(x_2)     \nn  \\ 
 & & \times \frac{\mathrm{d}^2 \hat{\sigma}_{i_1i_2}}{ \mathrm{d} v \, \mathrm{d} z } (v,z,p_T,R)   \,,  
\eea
where the partonic cross sections $\hat{\sigma}_{i_1i_2}$ are further factorized as
\bea\label{eq:fac}
&& \frac{\mathrm{d}^2 \hat{\sigma}_{i_1i_2}}{ \mathrm{d} v \, \mathrm{d} z }   = 
 s \int \mathrm{d} s_X \, \mathrm{d}s_c \mathrm{d}s_G \, \delta(z s -s_X -s_G - s_c)   \nn \\
& &  \hspace{-1.5mm} 
\times {\rm Tr} \left[   { \bf H}_{i_1i_2}(v,p_T\,, \mu_h\,, \mu) \,  {\bf S}_G (s_G\,,\mu_{sG}\,, \mu) \right] 
J _X(s_X\,, \mu_{X} \,, \mu ) 
 \nn  \\
&&  \hspace{-1.5mm} \times  \sum_m 
{\rm Tr}\left[  J_{m}(p_T R \,,\mu_J\,, \mu) \otimes_{\Omega} S_{c,m}(s_c R\,, \mu_{sc} \,,  \mu )  \right]  \,. 
\eea
In Eq.~(\ref{eq:crs}) the PDFs are denoted by $f_i$ which are evaluated at the
momentum fractions $x_1 = VW/v/(1-z)$ and $x_2 = (1-V)/(1-v)/(1-z)$,  
where $V = 1- p_T e^{-y}/\sqrt{S}$, $VW = p_T e^{y}/\sqrt{S} $ and $\sqrt{S}$ is the hadronic center-of-mass energy. 
The sum $i$ runs over all partonic channels initiating the subprocesses and $m$ runs over the collinear splitting history. 
The associated angular integrals are denoted by `$\otimes_{\Omega}$\!'~\cite{Becher:2015hka} to resum non-global
logarithms~\cite{Dasgupta:2001sh,Caron-Huot:2015bja,Becher:2015hka,Larkoski:2015zka,Larkoski:2016zzc,Neill:2016stq}. 
Besides the jet $p_T$, the partonic cross sections depend on the partonic kinematic
variables $s = x_1 x_2 S$, $z$ and $v = u/(u+t)$ with $t = (p_1 - p_3)^2$ and
$u = (p_2 - p_3)^2$. Here, $p_{1,2}$ are the momenta of the incoming partons
and $p_3$ is the momentum of the parton which initiates the signal-jet.

The $2 \to 2$ hard scattering functions in Eq.~(\ref{eq:fac}) are denoted by
${\bf H}_{i_1i_2}$ which are available to two loops~\cite{Broggio:2014hoa}. 
The inclusive jet function $J_X(s_X)$ is also known to order
$\alpha_s^2$~\cite{Becher:2006qw,Becher:2010pd} and the NLO jet function can
be extracted from~\cite{Ellis:2010rwa,Liu:2012sz}. 
The global soft function ${\bf S}_G$ and the soft collinear~\cite{Becher:2015hka,Chien:2015cka} function
$S_{c}$ have been derived to NLO in~\cite{Dai:2017dpc,Liu:2017pbb}. 
The global soft function and the soft collinear function can be readily
calculated to two loops following~\cite{Becher:2012za} and~\cite{Kelley:2011ng,Boughezal:2015eha}. 
All the functions are evolved from their natural scales $\mu_i$ to the common
scale $\mu$ according to their renormalization group equations in order to
obtain the NLL resummation used in this work.

The factorization formalism in Eq.~(\ref{eq:fac}) holds in the threshold
regime in which $z \to 1$ and $R \ll 1$. 
To extend the region of validity, we combine the NLL resummed results with the
NLO predictions using an additive matching procedure and define
 \bea
 \label{eq:sigma-NLLres}
 \sigma_{\rm NLO + NLL} = \sigma_{\rm NLO} - \sigma_{\rm NLO_{sing}} + \sigma_{\rm NLL} \,.
 \eea
Here, the logarithmically enhanced contributions at NLO are obtained within the resummation framework and denoted by $\sigma_{\rm NLO_{sing}}$. They are subtracted from the full NLO calculation 
and replaced by the NLL resummed results $\sigma_{\rm NLL}$. 
For the phenomenological studies presented in the next section, we use as a
default scale choice the leading jet transverse momentum 
$\mu_R = \mu_F = p_T^{\rm max}$ for the fixed NLO calculations~\cite{Gao:2012he}. 
We vary the scales around the central scale up and down by a factor of two
and take the maximal deviation as our NLO scale uncertainties. 
For the resummed results, we make the central scale choices 
$\mu = \mu_h = p_T^{\rm max}$, $\mu_J = p_T^{\rm max} R$ for the hard and the signal-jet functions,
respectively, and we set $\mu_X =  p_T^{\rm max}  (1 - 2p_T^{\rm max}/\sqrt{S})$, see also~\cite{Becher:2006nr,Becher:2009th}. 
The other scales are determined in the seesaw way: 
$\mu_{sG} = \mu_X^2 / \mu_h $ and $\mu_{sc} = \mu_J \times  \mu_{sG} / \mu_h$ for the global soft and the soft collinear functions, respectively. 
Our uncertainty estimates are obtained by varying $\mu$, $\mu_h$, $\mu_J$ and
$\mu_X$ independently by a factor of two around their central values while
keeping the seesaw relations for $\mu_{sG}$ and $\mu_{sc}$ in terms of
$\mu_X$. The final scale uncertainty is obtained by taking the envelope of the
scale variations. 

\renewcommand{\thetable}{\arabic{table}}
\setcounter{table}{0}

\begin{table}[th!]
\renewcommand{\arraystretch}{1.3}
\begin{center}
\begin{tabular}{|c|c|c|c|c|}
    \hline
    \multicolumn{5}{l}{$\mathbf{ 0 \le |y| < 0.5, \sqrt{S}=7~\mbox{\bf TeV}}$} \\ 
    \hline
    \hline
\multicolumn{1}{|c|}{$p_T$[GeV]} &
\multicolumn{1}{c|}{$200$} &
\multicolumn{1}{c|}{$300$} &
\multicolumn{1}{c|}{$500$} &
\multicolumn{1}{c|}{$1000$} \\
\hline
$K_{0.2}$
& $0.76$
& $0.78$
& $0.81$
& $0.83$
\\
$K_{0.5}$
& $0.90$
& $0.90$
& $0.92$
& $0.94$
\\
$K_{0.7}$
& $0.95$
& $0.96$
& $0.96$
& $0.98$
\\
$K_{0.9}$
& $0.98$
& $0.99$
& $1.00$
& $1.02$
\\
\hline
\end{tabular}
\vspace*{-3mm}
\begin{tabular}{c}
  \phantom{xxxxxxxxxxxxxxxxxxxxxxxxxxxxxxxxxxxxxxxxxxxxxxxxxxxxxxxxxxx}
\end{tabular}
\begin{tabular}{|c|c|c|c|c|c|}
  \hline
    \multicolumn{5}{l}{$\mathbf{ 0 \le |y| < 0.5, \sqrt{S}=13~\mbox{\bf TeV}}$} \\ 
    \hline
    \hline
\multicolumn{1}{|c|}{$p_T$[GeV]} &
\multicolumn{1}{c|}{$200$} &
\multicolumn{1}{c|}{$300$} &
\multicolumn{1}{c|}{$500$} &
\multicolumn{1}{c|}{$1000$} &
\multicolumn{1}{c|}{$1500$} \\
\hline
$K_{0.2}$
& $0.75$
& $0.77$
& $0.81$
& $0.83$
& $0.85$
\\
$K_{0.4}$
& $0.85$
& $0.87$
& $0.88$
& $0.91$
& $0.91$
\\
$K_{0.7}$
& $0.94$
& $0.95$
& $0.96$
& $0.97$
& $0.97$
\\
$K_{0.9}$
& $0.98$
& $0.98$
& $0.99$
& $1.01$
& $1.01$
\\
\hline
\end{tabular}
\caption{\label{tab:KR}
   Cross section ratios $K_R$ of Eq.~(\ref{eq:KR}) for different jet radii 
   at $\sqrt{S}=7$~TeV (top) and 13~TeV (bottom) for selected values of the 
   the signal jet $p_T$ using the MMHT PDF set~\cite{Harland-Lang:2014zoa} at NLO.}
\end{center}
\end{table}



%
\vspace*{-3mm}

\begin{table}[h!]
\renewcommand{\arraystretch}{1.3}
\begin{center}
\begin{tabular}{|c|c|c|c|c|}
  \hline
  \multicolumn{5}{l}{$\mathbf{ 0 \le |y| < 0.5, \sqrt{S}=7~\mbox{\bf TeV}}$} \\ 
  \hline
  \hline
\multicolumn{1}{|c|}{$p_T$[GeV]} &
\multicolumn{1}{c|}{$200$} &
\multicolumn{1}{c|}{$300$} &
\multicolumn{1}{c|}{$500$} &
\multicolumn{1}{c|}{$1000$} \\
\hline
$D^{{\rm NLO} + {\rm NLL}}_{0.2}$
& $0.62$
& $0.65$
& $0.69$
& $0.73$
\\
$D^{{\rm NLO} + {\rm NLL}}_{0.5}$
& $1.00$
& $1.00$
& $1.00$
& $1.00$
\\
$D^{{\rm NLO} + {\rm NLL}}_{0.7}$
& $1.16$
& $1.16$
& $1.14$
& $1.12$
\\
$D^{{\rm NLO} + {\rm NLL}}_{0.9}$
& $1.27$
& $1.26$
& $1.22$
& $1.21$
\\
\hline
$D^{\rm NLO}_{0.2}$
& $0.73$
& $0.76$
& $0.78$
& $0.82$
\\
$D^{\rm NLO}_{0.5}$
& $1.00$
& $1.00$
& $1.00$
& $1.00$
\\
$D^{\rm NLO}_{0.7}$
& $1.10$
& $1.09$
& $1.08$
& $1.07$
\\
$D^{\rm NLO}_{0.9}$
& $1.14$
& $1.14$
& $1.13$
& $1.11$
\\
\hline
\end{tabular}
\vspace*{-3mm}
\begin{tabular}{c}
  \phantom{xxxxxxxxxxxxxxxxxxxxxxxxxxxxxxxxxxxxxxxxxxxxxxxxxxxxxxxxxxx}
\end{tabular}
\begin{tabular}{|c|c|c|c|c|c|}
  \hline
  \multicolumn{5}{l}{$\mathbf{ 0 \le |y| < 0.5, \sqrt{S}=13~\mbox{\bf TeV}}$} \\ 
  \hline
  \hline
\multicolumn{1}{|c|}{$p_T$[GeV]} &
\multicolumn{1}{c|}{$200$} &
\multicolumn{1}{c|}{$300$} &
\multicolumn{1}{c|}{$500$} &
\multicolumn{1}{c|}{$1000$} &
\multicolumn{1}{c|}{$1500$} \\
\hline
$D^{{\rm NLO} + {\rm NLL}}_{0.2}$
& $0.70$
& $0.74$
& $0.76$
& $0.79$
& $0.80$
\\
$D^{{\rm NLO} + {\rm NLL}}_{0.4}$
& $1.00$
& $1.00$
& $1.00$
& $1.00$
& $1.00$
\\
$D^{{\rm NLO} + {\rm NLL}}_{0.7}$
& $1.30$
& $1.27$
& $1.24$
& $1.21$
& $1.19$
\\
$D^{{\rm NLO} + {\rm NLL}}_{0.9}$
& $1.46$
& $1.40$
& $1.37$
& $1.31$
& $1.29$
\\
\hline
$D^{\rm NLO}_{0.2}$
& $0.80$
& $0.81$
& $0.84$
& $0.86$
& $0.86$
\\
$D^{\rm NLO}_{0.4}$
& $1.00$
& $1.00$
& $1.00$
& $1.00$
& $1.00$
\\
$D^{\rm NLO}_{0.7}$
& $1.17$
& $1.15$
& $1.15$
& $1.12$
& $1.11$
\\
$D^{\rm NLO}_{0.9}$
& $1.26$
& $1.23$
& $1.22$
& $1.19$
& $1.16$
\\
\hline
\end{tabular}
\caption{\label{tab:DR}
  Ratios of Eq.~(\ref{eq:DR}) for the cross sections at  
  NLO and NLO $+$ NLL accuracy denoted by $D^{\rm NLO}_R$ and $D^{{\rm NLO} + {\rm NLL}}_R$, respectively,
  at $\sqrt{S}=7$~TeV (top) and 13~TeV (bottom) for selected values of the 
  the signal jet $p_T$ using the MMHT PDF set~\cite{Harland-Lang:2014zoa} at NLO.
  The results at $\sqrt{S}=7$~TeV include NP correction factors which are taken from~\cite{Chatrchyan:2014gia}.}
\end{center}  
\end{table}

\section{Phenomenology \label{sec:pheno}}

We start by studying the overall numerical impact of the joint threshold and
small-$R$ resummation. We then continue by analyzing the scale dependence of
the resummed cross section and provide a detailed comparison to LHC data. 
Finally, we study in detail the impact of different PDF sets. 
The two single-inclusive jet data sets 
%
\newpage
\clearpage
\noindent
%
from CMS that we are comparing to
throughout this section were taken at $\sqrt{S}=7$~TeV~\cite{Chatrchyan:2014gia} and
at 13~TeV~\cite{Khachatryan:2016wdh}. 
For the $\sqrt{S}=7$~TeV data set, the jets were
reconstructed using two different values of the jet radius,
$R=0.5$ and $R=0.7$ covering a rapidity range of $|y|<3$. 
Instead, for the $\sqrt{S}=13$~TeV data set, 
the jet radius parameters were chosen as $R=0.4$ and $R=0.7$ covering $|y|<4.7$. 
For both data sets, the jets were reconstructed using the anti-$k_T$ algorithm~\cite{Cacciari:2008gp} and the
transverse momentum of the identified jets ranges up to $p_T=2$~TeV. 

 \begin{figure*}
  \begin{subfigure}{.5\textwidth}
  \includegraphics[width=0.95\linewidth]{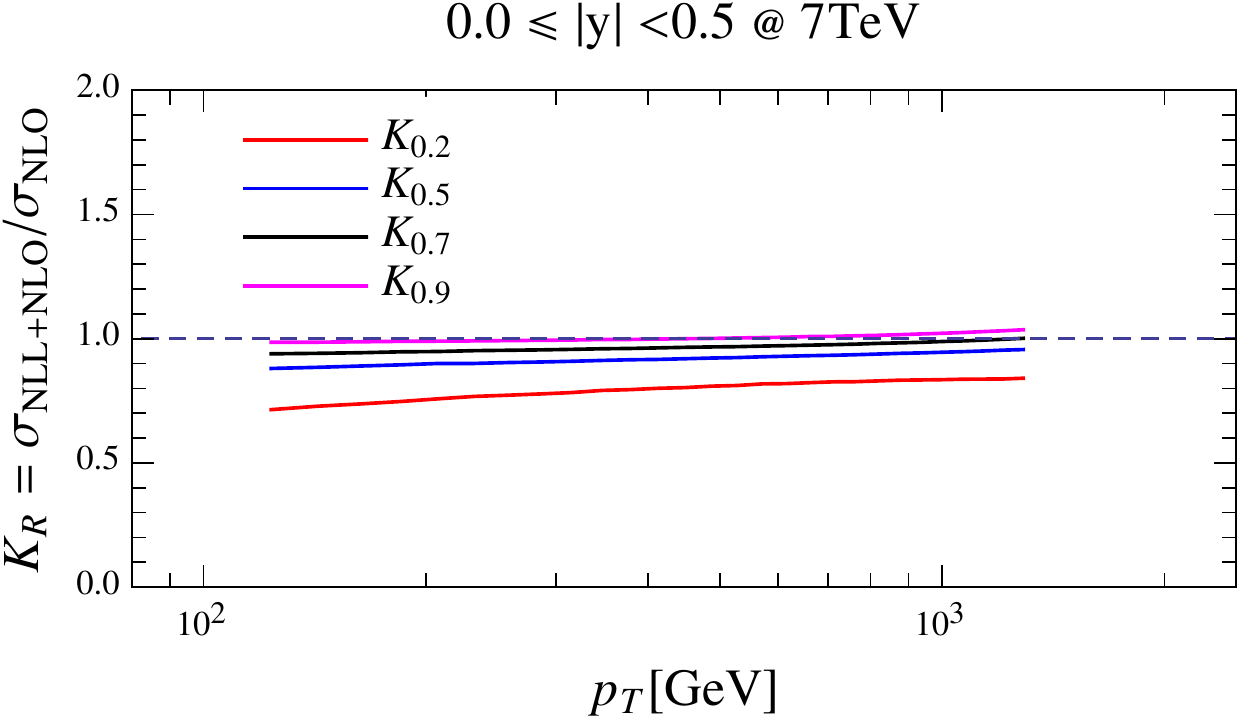}
   \end{subfigure}%
 \begin{subfigure}{.5\textwidth}
   \includegraphics[width=0.95\linewidth]{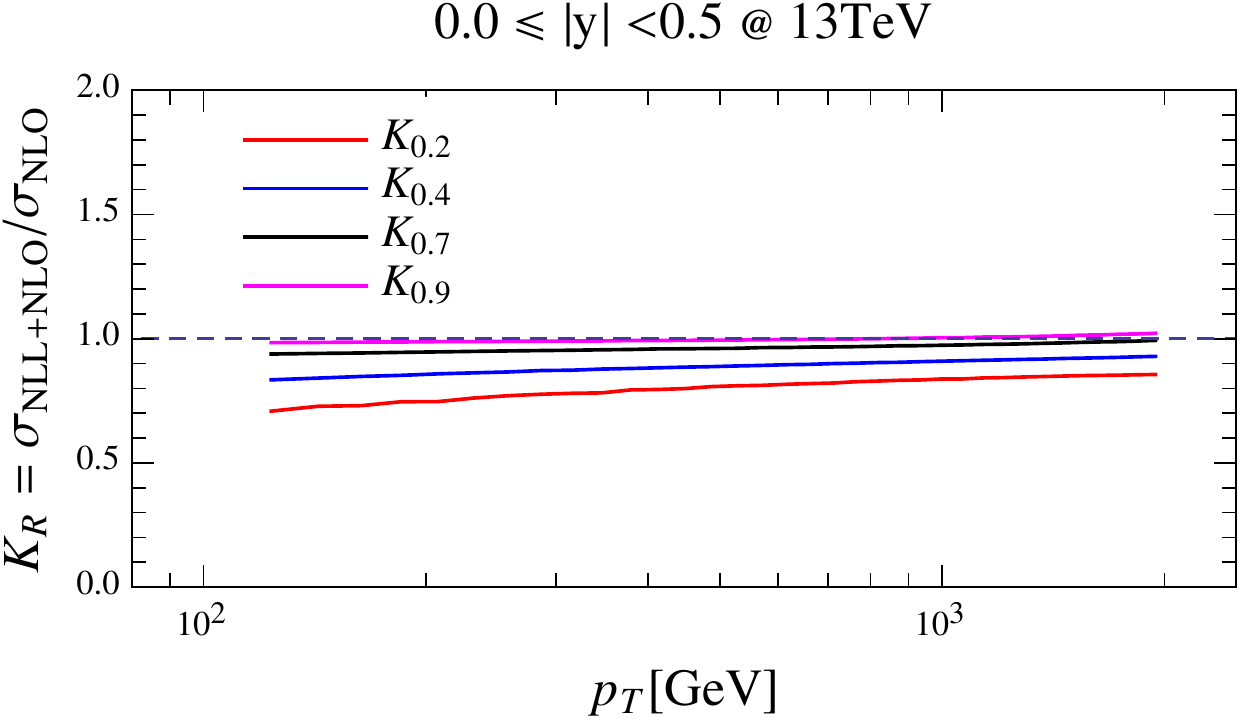}
  \end{subfigure}
 \caption{
   Cross section ratios $K_R$ of Eq.~(\ref{eq:KR}) with different jet radii 
   at $\sqrt{S}=7$~TeV (left) and 13~TeV (right) using the MMHT PDF set~\cite{Harland-Lang:2014zoa} at NLO.}    
 \label{fig:Rscan}
 \end{figure*}
 \begin{figure*}
  \begin{subfigure}{.5\textwidth}
  \includegraphics[width=0.95\linewidth]{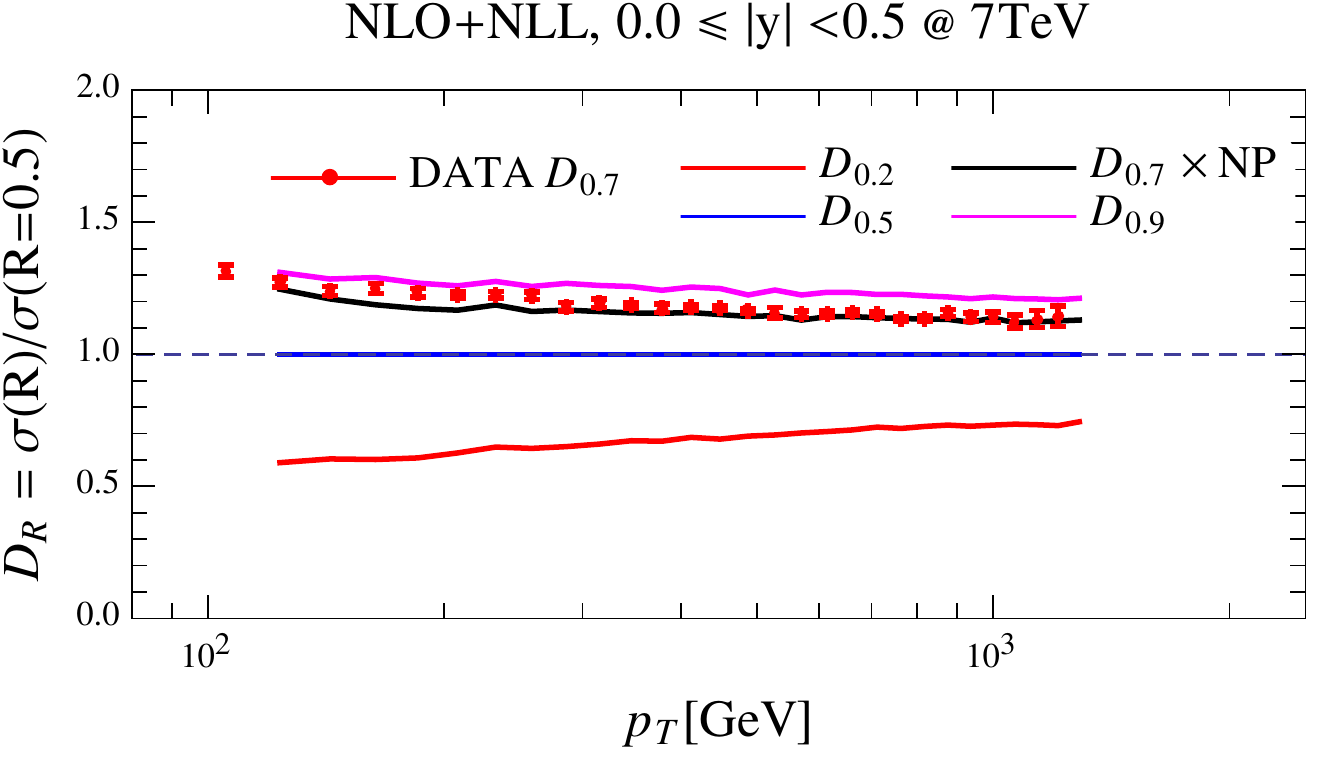}
   \end{subfigure}%
 \begin{subfigure}{.5\textwidth}
   \includegraphics[width=0.95\linewidth]{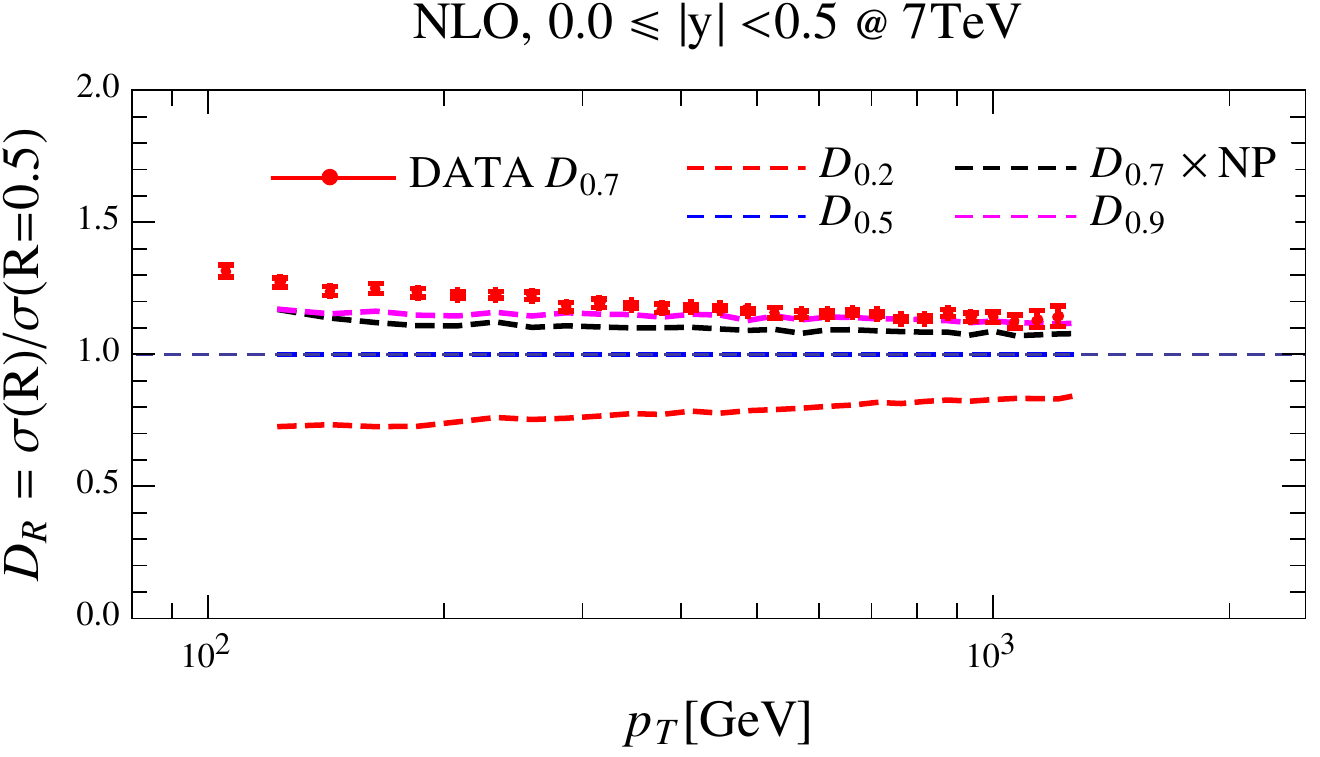}
  \end{subfigure}
 \caption{
   Ratios $D_R$ of Eq.~(\ref{eq:DR}) with $R_{\rm fixed}=0.5$ 
   at $\sqrt{S}=7$~TeV for the cross sections at 
   NLO $+$ NLL (left) and NLO (right) accuracy
   using the MMHT PDF set~\cite{Harland-Lang:2014zoa} at NLO with 
   NP correction factors which are taken from~\cite{Chatrchyan:2014gia}.
   The red dots indicate the single-inclusive jet data for $D_R$ from CMS 
   collected at $\sqrt{S}=7$~TeV with $R=0.7$~\cite{Chatrchyan:2014gia}.}
 \label{fig:Rscan-ratio-7tev}
 \end{figure*}
 \begin{figure*}
  \begin{subfigure}{.5\textwidth}
  \includegraphics[width=0.95\linewidth]{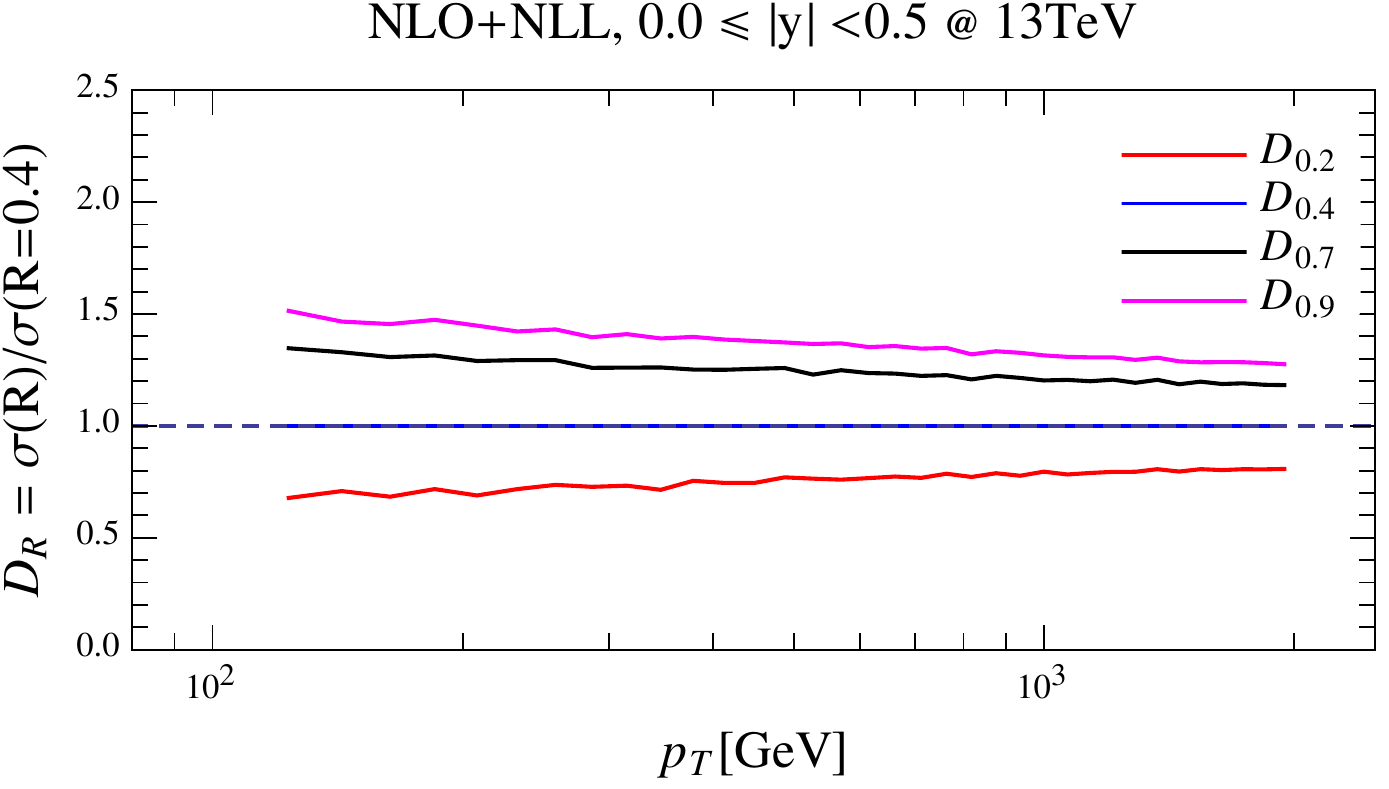}
   \end{subfigure}%
 \begin{subfigure}{.5\textwidth}
   \includegraphics[width=0.95\linewidth]{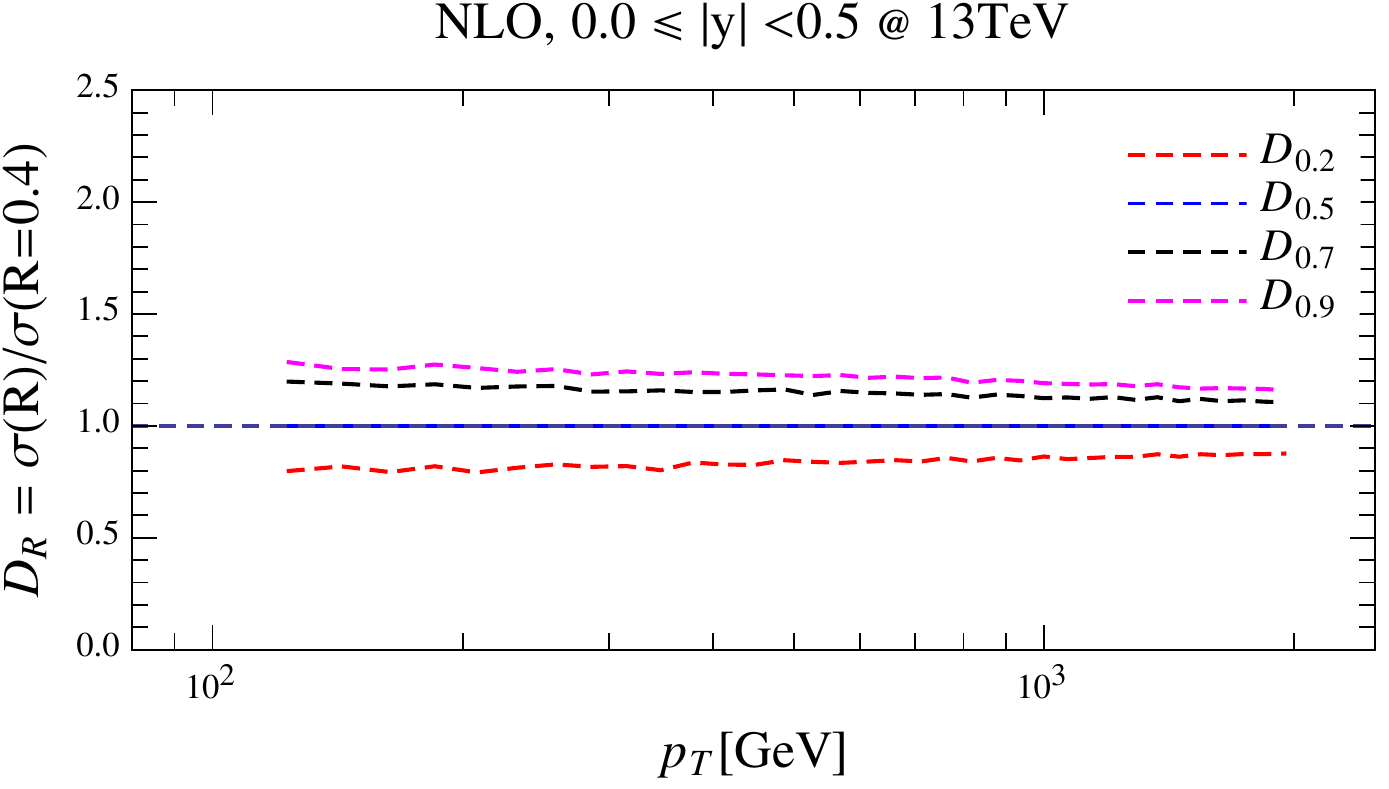}
  \end{subfigure}
 \caption{
   Ratios $D_R$ of Eq.~(\ref{eq:DR}) with $R_{\rm fixed}=0.4$ 
   at $\sqrt{S}=13$~TeV for the cross sections at 
   NLO $+$ NLL (left) and NLO (right) accuracy
   using the MMHT PDF set~\cite{Harland-Lang:2014zoa} at NLO.}
 \label{fig:Rscan-ratio-13tev}
 \end{figure*}
 \begin{figure*}
 \begin{subfigure}{.5\textwidth}
  \includegraphics[width=0.95\linewidth]{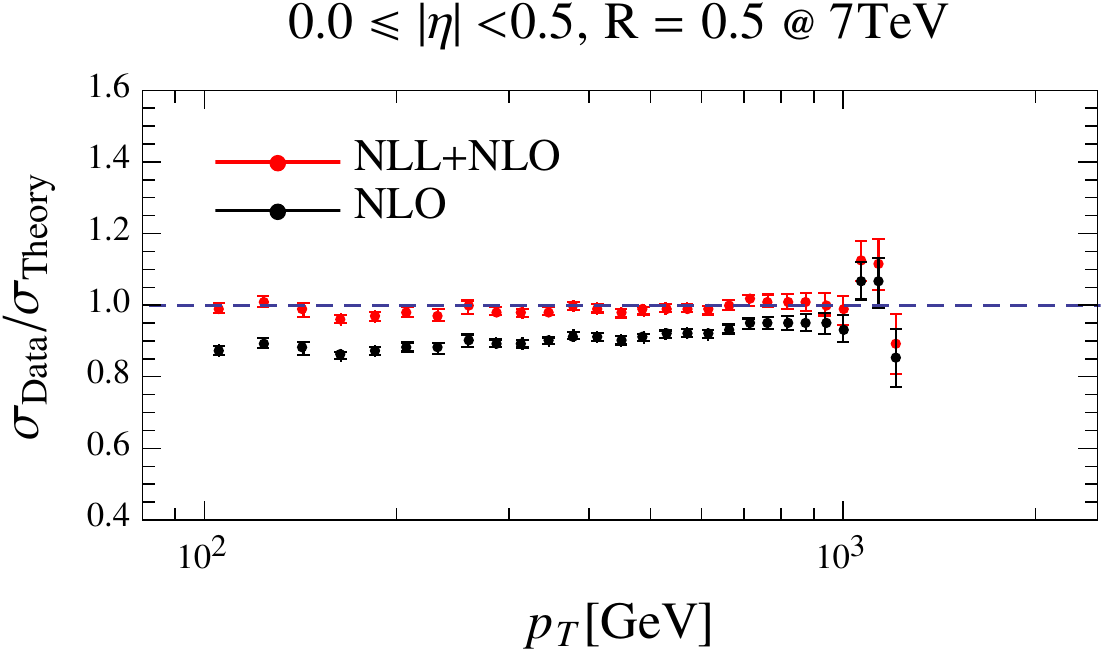}
 \end{subfigure}%
 \begin{subfigure}{.5\textwidth}
  \includegraphics[width=0.95\linewidth]{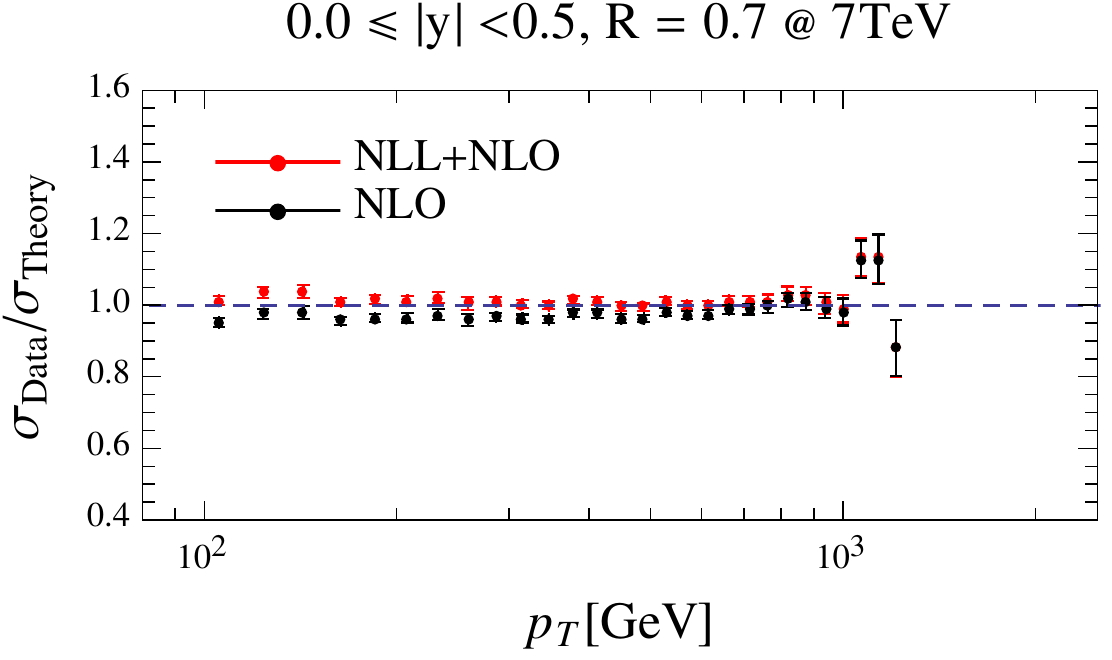}
  \end{subfigure}
  \caption{
    The ratio $\sigma_{\rm Data}/\sigma_{\rm Theory}$ 
    for the CMS data $\sqrt{S}=7$~TeV of~\cite{Chatrchyan:2014gia} with
    $R= 0.5$ (left) and $R=0.7$ (right) to the theoretical results 
    at NLO (black) and at NLO $+$ NLL (red) accuracy 
    using the MMHT PDF set~\cite{Harland-Lang:2014zoa} at NLO.}    
 \label{fig:7tevRscandata}
 \end{figure*}
 \begin{figure*}
 \begin{subfigure}{.5\textwidth}
  \includegraphics[width=0.95\linewidth]{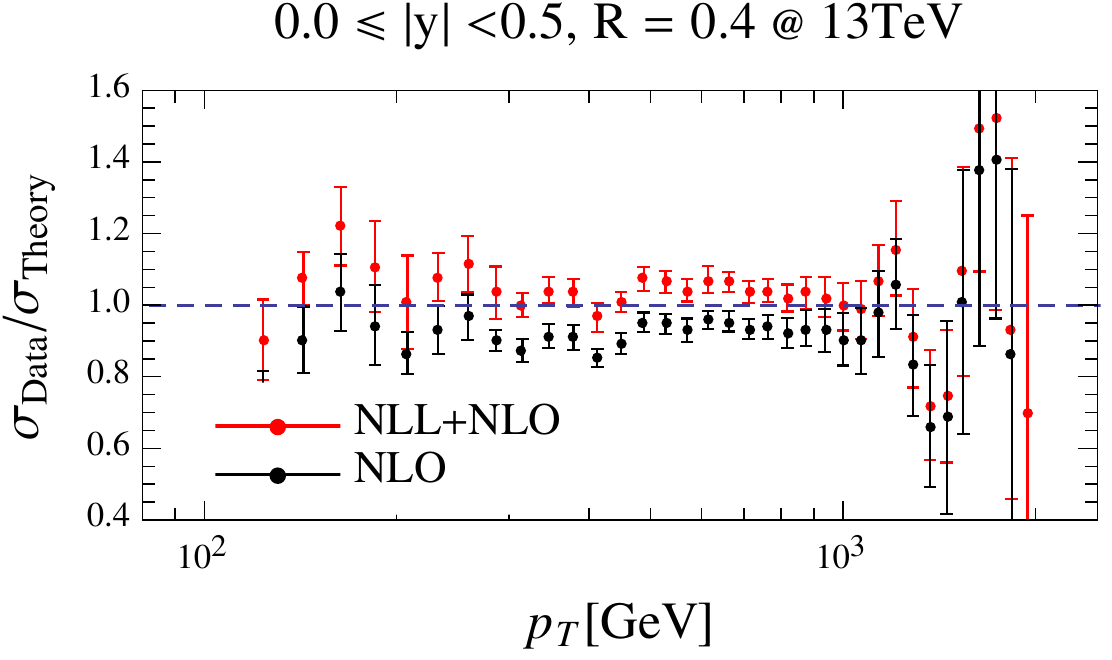}
 \end{subfigure}%
 \begin{subfigure}{.5\textwidth}
  \includegraphics[width=0.95\linewidth]{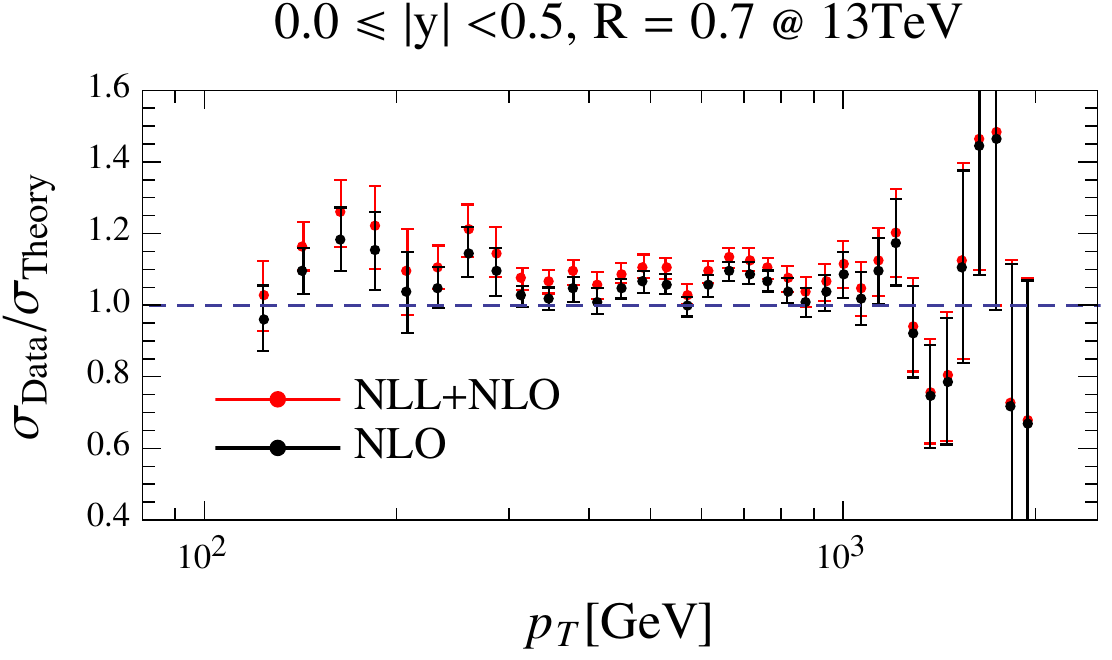}
  \end{subfigure}
  \caption{Same as Fig.~\ref{fig:7tevRscandata} for the CMS data
    of~\cite{Khachatryan:2016wdh} at $\sqrt{S}=13$~TeV 
    with $R= 0.4$  (left) and $R=0.7$ (right).}
  \label{fig:13tevRscandata}
 \end{figure*}

\subsection{Numerical impact of the joint resummation}

In Fig.~\ref{fig:Rscan}, we plot the ratio $K_R$
 \bea
 \label{eq:KR}
 K_R \,=\, \frac{\sigma_{\rm NLL+NLO}(R)}{\sigma_{\rm NLO}(R)} \,,
 \eea
of the NLO $+$ NLL and the NLO cross sections for different jet radii 
as a function of the signal-jet $p_T$ and $|y| < 0.5$ at both $\sqrt{S}=7$~TeV (left) and $13$ TeV (right). 
Results for selected values of $p_T$ are listed in Tab.~\ref{tab:KR}.
For all numerical calculations in this section we use the MMHT14
PDFs~\cite{Harland-Lang:2014zoa} at NLO as an example. 
We find that for a large range of the jet $p_T$, the joint resummation leads to a decrease of the NLO cross section. 
The effect is more pronounced for smaller values of $R$, 
where the impact of the $\ln(R)$ resummation becomes more noticeable and leads
to a significant decrease of the cross section. 
For larger values of the jet $p_T$, the threshold enhancement will compensate the $\ln(R)$ contributions 
and can eventually result in an enhancement. 
This effect is most clearly illustrated in Fig.~\ref{fig:Rscan} and Tab.~\ref{tab:KR} 
for $R = 0.9$ for $p_T$ around ${\cal O}(1 \> {\rm TeV})$. 
A similar trend has been observed in recent work on threshold summation 
with a parton shower event generator including quantum interference~\cite{Nagy:2017dxh}.

To illustrate the dependence of the cross section on the jet radius parameter $R$ 
we plot in Figs.~\ref{fig:Rscan-ratio-7tev} and~\ref{fig:Rscan-ratio-13tev}
for both, fixed order NLO and NLO $+$ NLL resummed predictions, 
the respective ratios $D_R$ 
 \bea
 \label{eq:DR}
 D_R \,=\, \frac{\sigma(R)}{\sigma(R_{\rm fixed})} \,,
 \eea
as a function of the signal-jet $p_T$ and $|y| < 0.5$ 
for the energies $\sqrt{S}=7$ and $13$~TeV. 
Nonperturbative (NP) correction factors are taken from~\cite{Chatrchyan:2014gia} 
and results for selected values of $p_T$ are listed in Tab.~\ref{tab:DR}.
The predicted dependence of the cross section on the jet radius parameter $R$ 
tends to be more pronounced when the effects of the joint resummation are considered. 
For the smaller value of $R=0.2$, the ratio with resummation $D^{{\rm NLO} + {\rm NLL}}_{R}$ is 
smaller than the corresponding fixed order result $D^{{\rm NLO}}_{R}$, while
the opposite trend is observed for larger radii $R \ge 0.7$, cf. Tab.~\ref{tab:DR}.
We emphasize that the ratios $D_R$ are quite insensitive to the chosen input PDFs.
This leads to precise predictions of the NLL $+$ NLO and the NLO calculations
which are experimentally well testable.
In Fig.~\ref{fig:Rscan-ratio-7tev}, we have performed a comparison of
predictions with the inclusive jet data for $D_{0.7}$ shown in red dots 
measured by CMS~\cite{Chatrchyan:2014gia} at the LHC with at $\sqrt{S}=7$~TeV.
The error bars represent the experimental uncertainties which are small due to the
cancellation of systematic errors in the ratio. 
After considering the NP effects, the NLO $+$ NLL resummed result
in Fig.~\ref{fig:Rscan-ratio-7tev} (left) 
agrees very well with the measurement while the fixed order NLO prediction 
in Fig.~\ref{fig:Rscan-ratio-7tev} (right) 
fails to describe the data for the entire range of jet $p_T$ considered. 
We note that those CMS data for $D_{0.7}$ are even larger than the 
fixed order NLO prediction $D_{0.9}$, i.e. a situation which intuitively should be reversed.  

As a further illustration of the resummation effects we compare the
predicted central values at fixed NLO and at NLO $+$ NLL
accuracy in Eq.~(\ref{eq:sigma-NLLres}) 
with the LHC data~\cite{Chatrchyan:2014gia, Khachatryan:2016wdh}, 
collected at $\sqrt{S}=7$ and $13$~TeV, respectively. 
In Figs.~\ref{fig:7tevRscandata} and~\ref{fig:13tevRscandata}, 
the LHC data in the rapidity bin $|y| < 0.5$ is normalized to the theoretical
predictions and displayed as a function of the signal-jet $p_T$.  
For $\sqrt{S}=7$~TeV the NP effects have been included in the predictions in
Fig.~\ref{fig:7tevRscandata}. 
It is clearly visible how the resummed predictions lead to an increase of the
ratio of cross sections $\sigma_{\rm Data}/\sigma_{\rm Theory}$ compared to
the NLO result for all choices of jet radii, so that 
the NLO $+$ NLL results of Eq.~(\ref{eq:sigma-NLLres}) are in perfect agreement
with the $\sqrt{S}=7$~TeV data~\cite{Chatrchyan:2014gia}. 
For the $\sqrt{S}=13$~TeV data~\cite{Khachatryan:2016wdh} 
with the choice of $R = 0.4$, the resummation improves the theory description as well, 
whereas the data for $R = 0.7$ slightly overshoots the theory predictions.  
However, the experimental uncertainties of those data sets are still relatively large.

The observations presented here do neither depend significantly on the scales 
chosen as the leading jet transverse momentum $\mu_F = \mu_R = p_T^{\rm max}$ nor on the PDFs. 
This will be quantified in detail in the following sections.

\subsection{Scale dependence}

\begin{figure}[H]
  \includegraphics[width=1.0\linewidth]{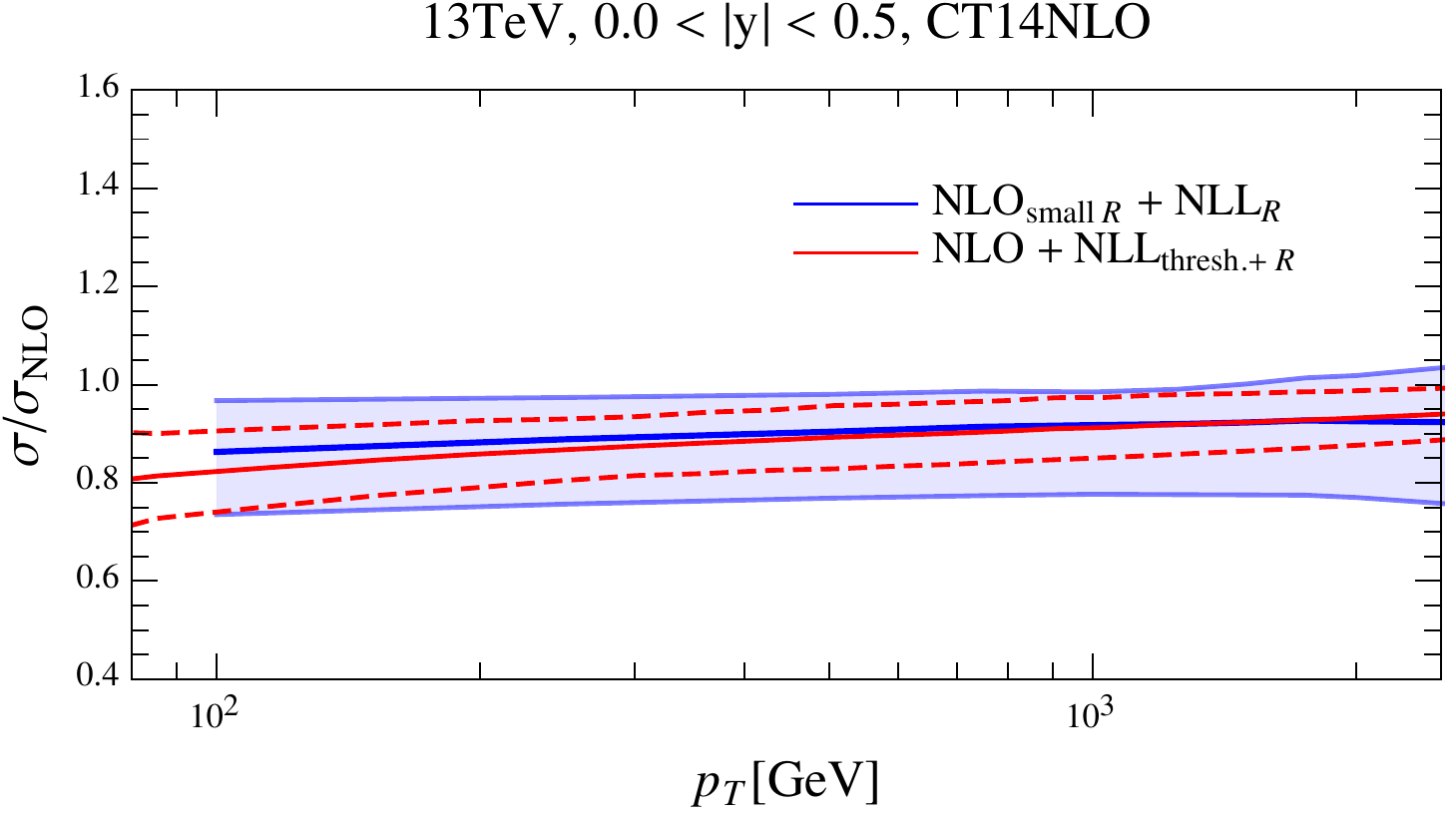}
 \caption{
   Comparison of the scale dependence of the jointly resummed cross section (red) and the case where only small-$R$ logarithms are resummed to all orders (blue)~\cite{Kang:2016mcy}. As an example, we use the CT14 PDF set~\cite{Dulat:2015mca} at NLO with $R = 0.4$ and both results are normalized to NLO. See text for more details.
    }
 \label{fig:nll-comparison}
 \end{figure}

In this section we analyze in detail the scale dependence of the jointly
resummed single-inclusive jet cross section. It is instructive to compare the
obtained scale dependence to the case where only the logarithms in the jet
size parameter $\ln(R)$ are resummed~\cite{Kang:2016mcy}. In
Fig.~\ref{fig:nll-comparison}, we show the residual scale uncertainty of the
jointly resummed cross section normalized to NLO. The scale band is obtained
as discussed in the section~\ref{sec:th} above. In addition, we show the cross
section where only $\ln(R)$ terms are resummed. In this case the scale band is
obtained by varying only the hard scale $\mu_h$ and the jet scale $\mu_J$ by
factors of two around their canonical choices. One observes a significant
reduction of the residual scale dependence once also threshold resummation is
taken into account. This observation holds true even though for the small-$R$
resummed calculation there are only two scales that are varied in order to
estimate the QCD uncertainty whereas there are three separate scales that are
all varied independently for the jointly resummed result. 
One also notices that the reduction of the scale uncertainty gets more pronounced at higher jet
transverse momenta where threshold resummation is more relevant. In fact this
behavior is generally expected for threshold resummed calculations and has
been analyzed in more detail before in many instances, see for example 
the studies for Higgs boson hadro-production~\cite{Catani:2003zt,Ahrens:2008nc}. When approaching large $p_T$, the joint resummation surpasses the small-$R$ resummed cross section due to the threshold enhancement.

An important caveat here is that the $\ln(R)$ resummed calculation
of~\cite{Kang:2016mcy} can currently be performed only with the scale choice
$\mu=p_T$ whereas in the threshold limit we always have
$\mu=p_T^{\text{max}}$. This difference is most relevant at small values of
the jet transverse momentum and likely explains the difference of the central
values of the two curves at small $p_T$. On the other hand, it is interesting
to note that for both scale choices, the resummation consistently leads to a
suppression relative to the respective NLO calculation. In addition, the
jointly resummed calculation is matched and normalized to the full
NLO. Instead, the $\ln(R)$ resummed calculation is using the narrow jet
approximation. However, the differences are of order ${\cal O}(R^2)$ which are
negligible for $R=0.4$~\cite{Jager:2004jh,Mukherjee:2012uz}. 

\subsection{Comparison to LHC data}

Now we move on to the comparison of the theoretical predictions with 
the CMS inclusive jet analyses for both $\sqrt{S}=7$ and $13$~TeV~\cite{Chatrchyan:2014gia, Khachatryan:2016wdh}.
Other data sets, such as those by CMS collected at $\sqrt{S}=2.76$~TeV~\cite{Khachatryan:2016jfl} and 
the one ATLAS at $\sqrt{S}=13$~TeV~\cite{ATLAS-CONF-2017-048} have already been considered in~\cite{Liu:2017pbb}.

We start with $\sqrt{S}=7$~TeV following the CMS analysis~\cite{Chatrchyan:2014gia} and focus on 
the cross section data with the anti-$k_T$ jet radius $R = 0.5$, 
which we bin into 4 different rapidity regions: 
$0.0 \le |y| < 0.5$, $0.5 \le |y| < 1.0$, $1.0 \le |y| < 1.5$ and $1.5 \le |y| < 2.0$. 
For each rapidity bin, we present the pQCD predictions $\sigma_{\rm NLO +
  NLL}$ at NLO $+$ NLL and $\sigma_{\rm NLO}$ at NLO accuracy based on the CT10 PDFs~\cite{Lai:2010vv} at NLO as in the original CMS analysis~\cite{Chatrchyan:2014gia}.

\begin{figure*}
 \begin{subfigure}{.5\textwidth}
  \includegraphics[width=0.95\linewidth]{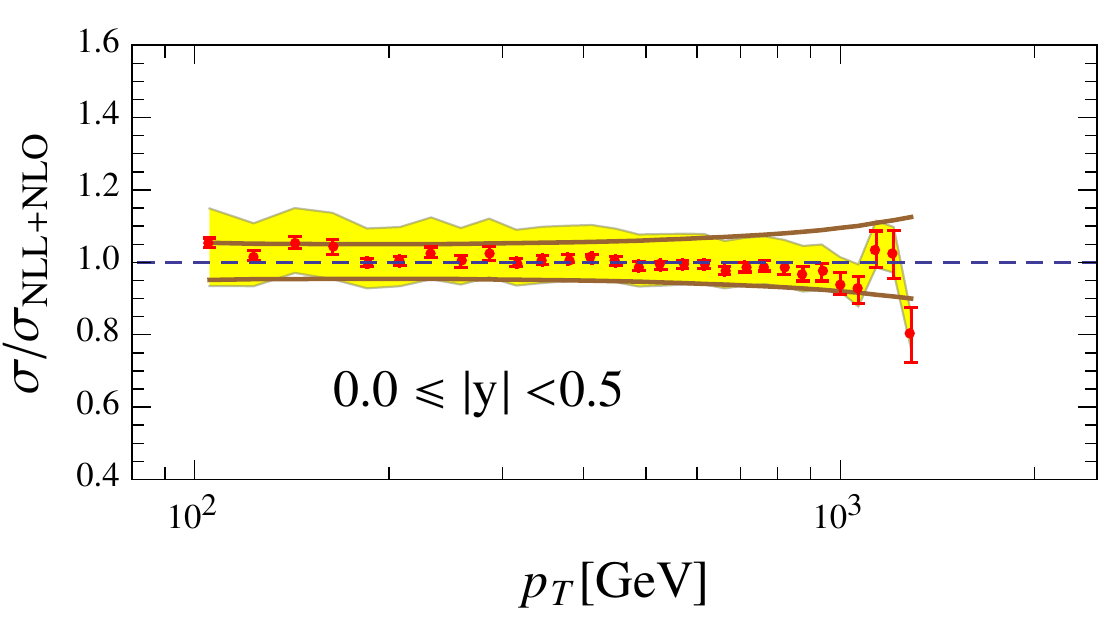}
 \end{subfigure}%
 \begin{subfigure}{.5\textwidth}
  \includegraphics[width=0.95\linewidth]{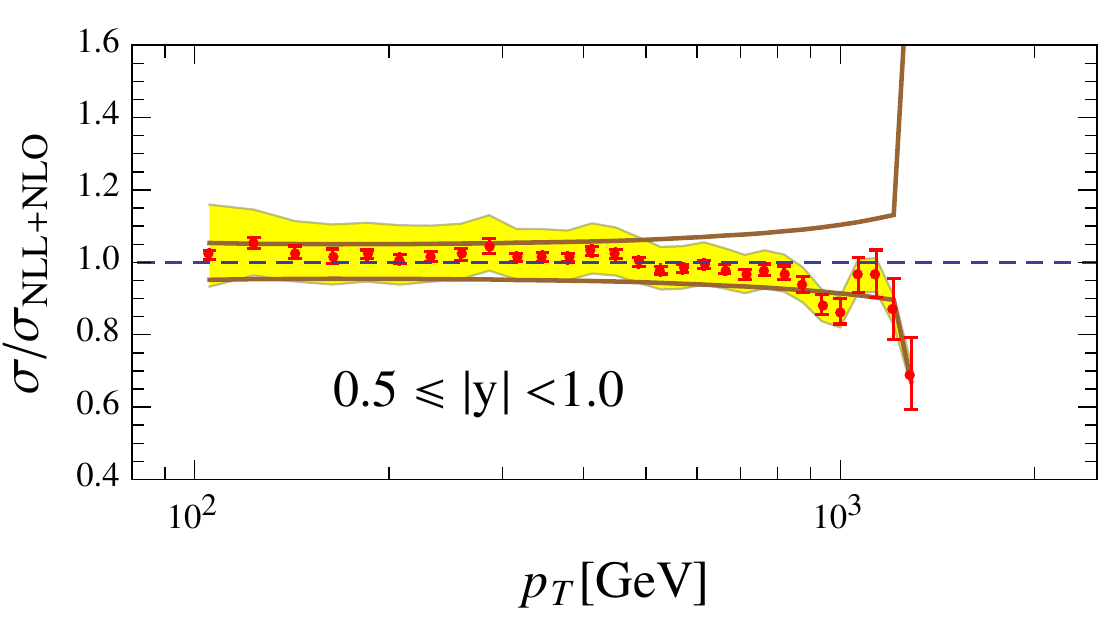}
  \end{subfigure}
   \begin{subfigure}{.5\textwidth}
  \includegraphics[width=0.95\linewidth]{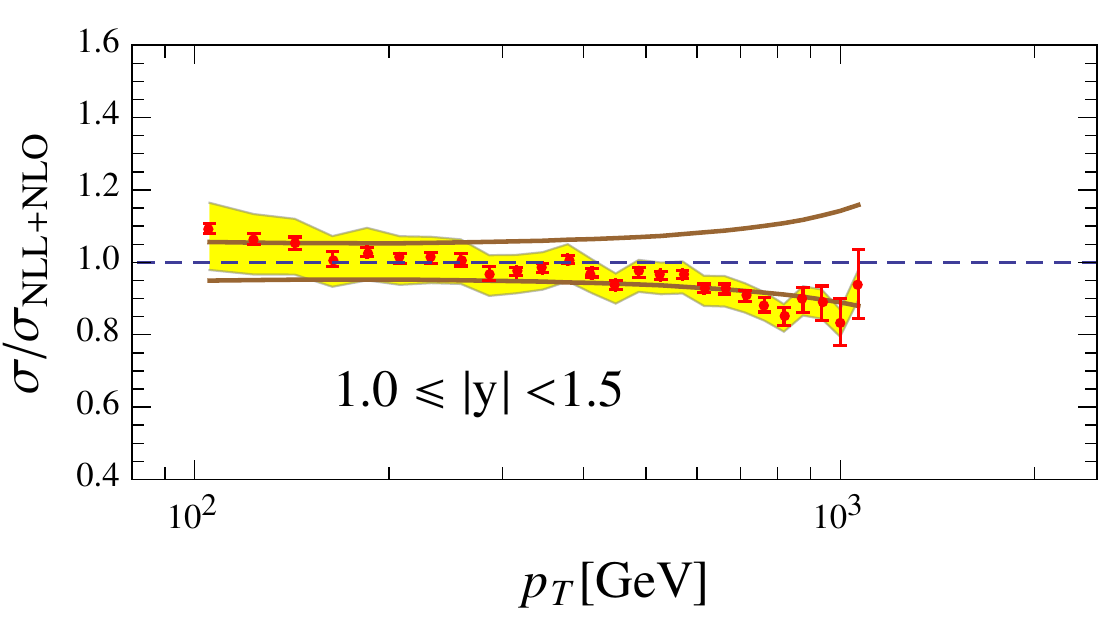}
   \end{subfigure}%
    \begin{subfigure}{.5\textwidth}
  \includegraphics[width=0.95\linewidth]{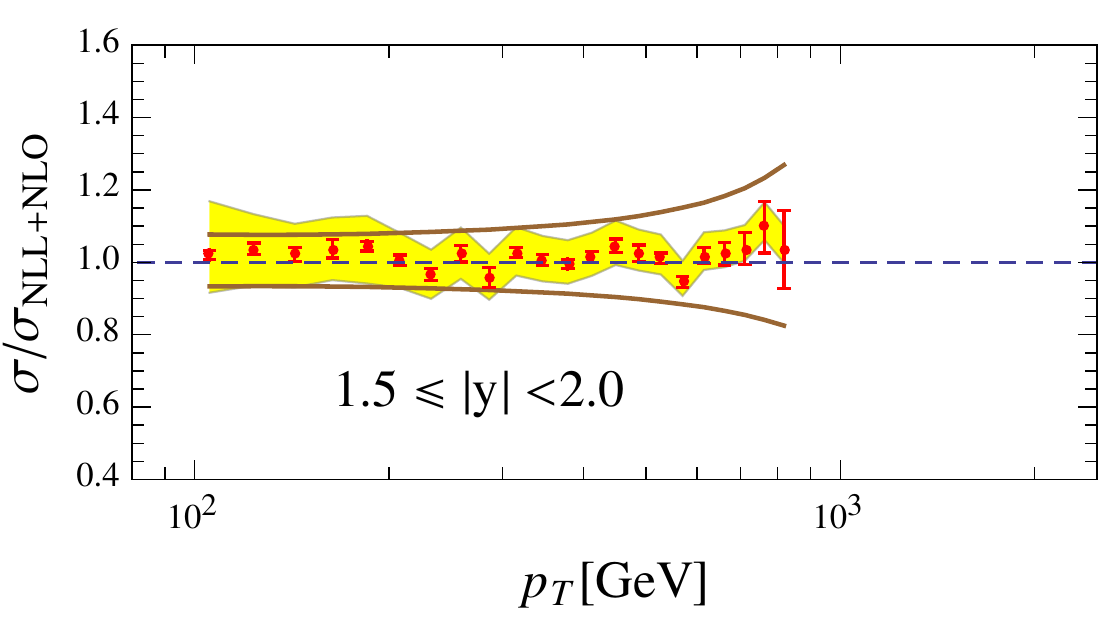}
     \end{subfigure}
 \caption{
   The ratio $\sigma_{\rm Data}/\sigma_{\rm NLO + NLL}$ 
   for the CMS data collected at $\sqrt{S}=7$~TeV~\cite{Chatrchyan:2014gia} with
   $R= 0.5$ to the theoretical results 
   using the CT10 PDF set~\cite{Lai:2010vv} at NLO.
   The error bars represent the experimental statistical errors and 
   the solid (brown) lines the systematic ones.
   The band (yellow) indicates theoretical scale uncertainties. 
   The NP corrections from~\cite{Chatrchyan:2014gia} have been included.
  }
 \label{fig:7tevnll}
 \end{figure*}

\begin{figure*}
 \begin{subfigure}{.5\textwidth}
  \includegraphics[width=0.95\linewidth]{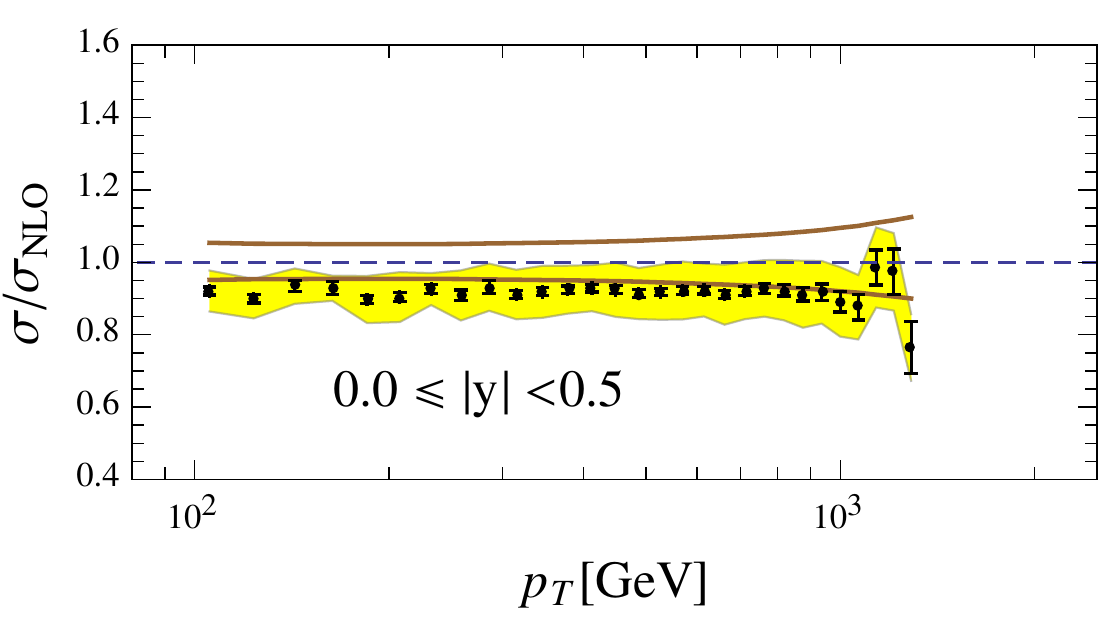}
 \end{subfigure}%
 \begin{subfigure}{.5\textwidth}
  \includegraphics[width=0.95\linewidth]{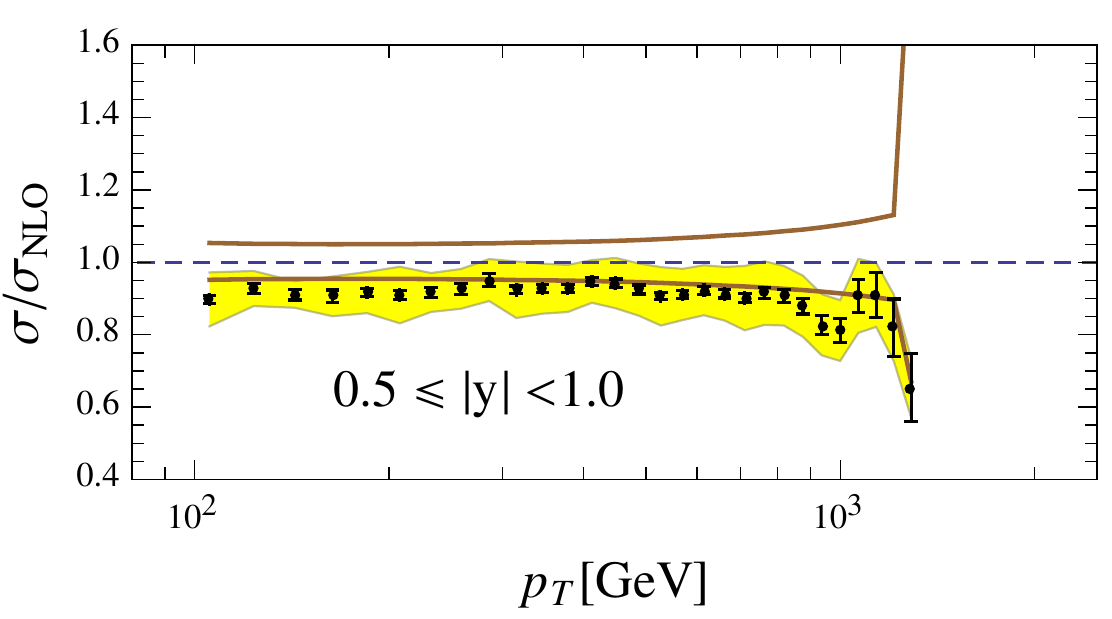}
  \end{subfigure}
   \begin{subfigure}{.5\textwidth}
  \includegraphics[width=0.95\linewidth]{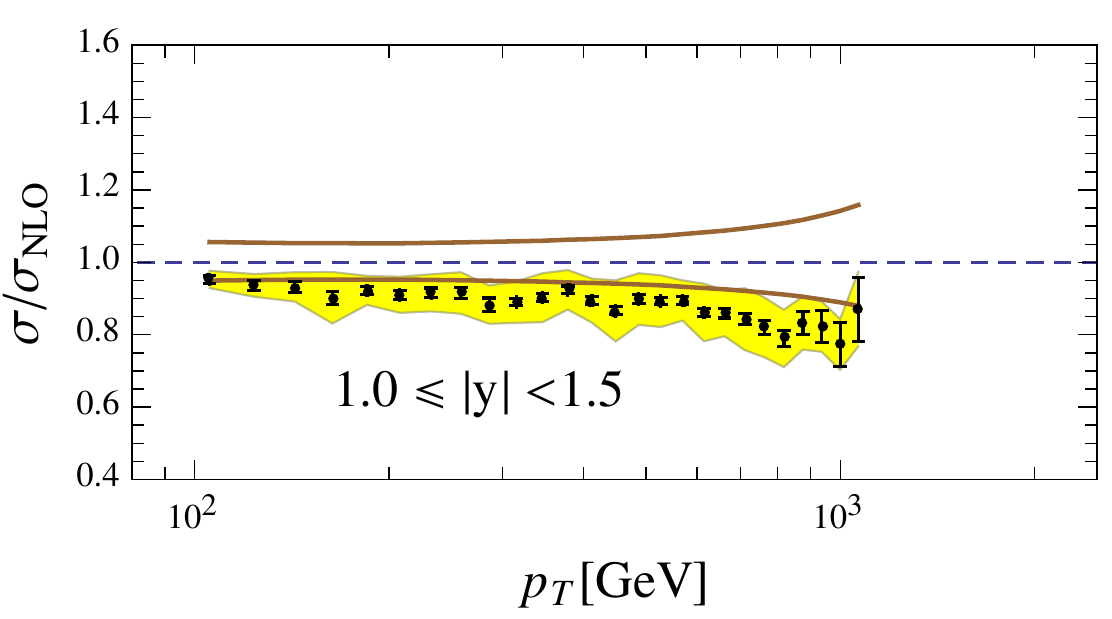}
   \end{subfigure}%
    \begin{subfigure}{.5\textwidth}
  \includegraphics[width=0.95\linewidth]{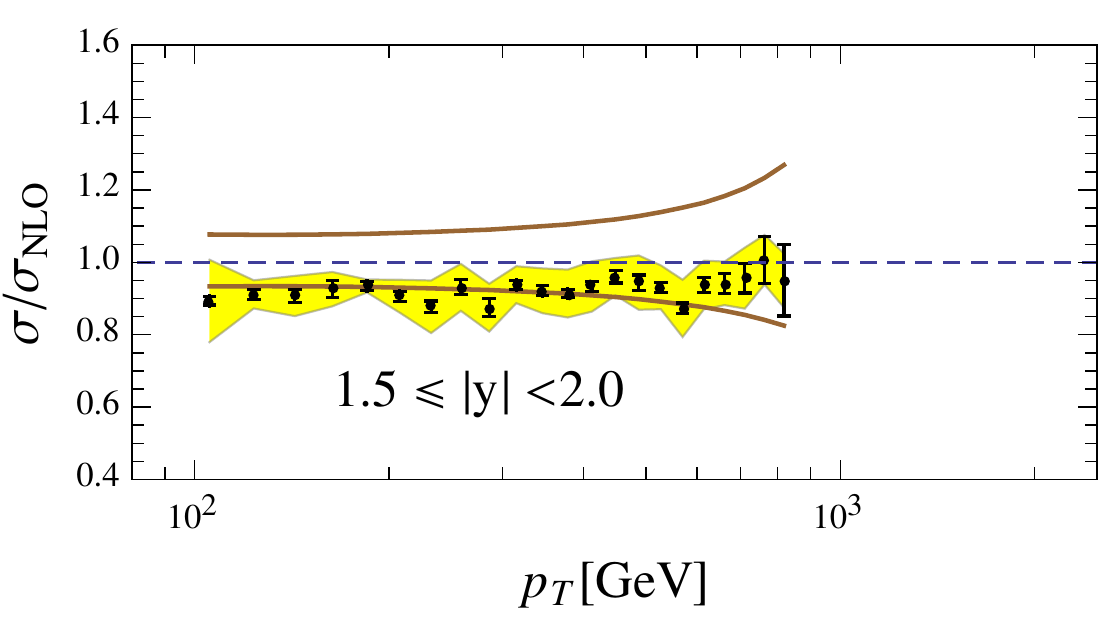}
     \end{subfigure}
 \caption{
   Same as Fig.~\ref{fig:7tevnll} for the ratio $\sigma_{\rm Data}/\sigma_{\rm NLO}$.}    
 \label{fig:7tevnlo}
 \end{figure*}

Figs.~\ref{fig:7tevnll} and~\ref{fig:7tevnlo} show the ratio of the CMS data to 
the theoretical predictions, that is $\sigma_{\rm Data}/\sigma_{\rm Theory}$ 
for both NLO $+$ NLL and NLO accuracy. 
For both cases, also the NP effects as provided by
CMS~\cite{Chatrchyan:2014gia} have been included in the perturbative
calculations to convert the predictions from the parton level to the particle
level.
The yellow bands in Figs.~\ref{fig:7tevnll} and~\ref{fig:7tevnlo} indicate the
theoretical uncertainties from scale variations obtained as discussed 
in the previous section with the hard scale chosen as $\mu=p_T^{\text{max}}$.
The solid brown lines on the other hand indicate the experimental systematic
errors, whereas the error bars on the data represent the experimental statistical errors~\cite{Chatrchyan:2014gia}.  

In Fig.~\ref{fig:7tevnll} we observe very good agreement with the data 
in all rapidity regions for the NLO $+$ NLL predictions where the NLL joint
resummation is taken into account. 
In the high-$p_T$ region, the NLO $+$ NLL calculations still somewhat overestimate
the CMS data. 
However, this can be further improved by switching from the CT10 PDFs~\cite{Lai:2010vv}
to more recent PDF sets, as we will detail in the next section.
In contrast, the NLO results in Fig.~\ref{fig:7tevnlo} are consistently larger
than the inclusive jet data by an amount of $10\%$ in all rapidity bins.
Thus, all predictions are lying along the lower boundary of 
the systematic errors (brown lines) in Fig.~\ref{fig:7tevnlo}, 
except for the high $p_T$ tail region of the rapidity bin $1.5\le |y| < 2.0$.

The theoretical uncertainty of the NLL $+$ NLO predictions in Fig.~\ref{fig:7tevnll}
is still large and comparable with the experimental errors.
However, this can be reduced further in the future with the help improved accuracy
for the resummation, i.e., upon resumming the relevant logarithms to NNLL
accuracy together with matching to the available NNLO calculations.

\begin{figure*}
 \begin{subfigure}{.5\textwidth}
  \includegraphics[width=0.95\linewidth]{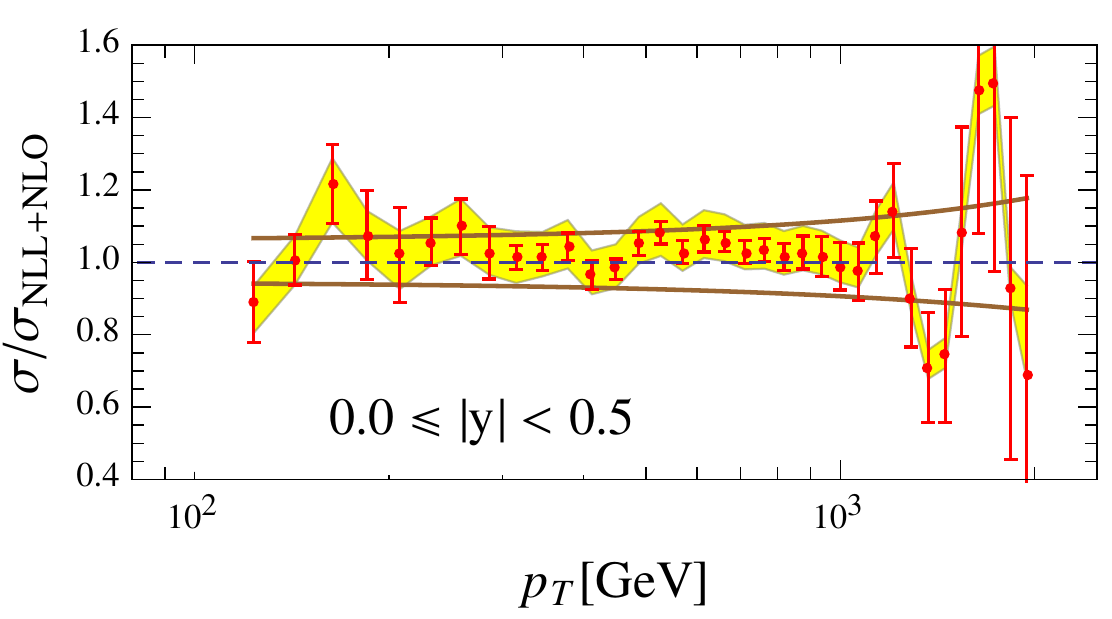}
 \end{subfigure}%
 \begin{subfigure}{.5\textwidth}
  \includegraphics[width=0.95\linewidth]{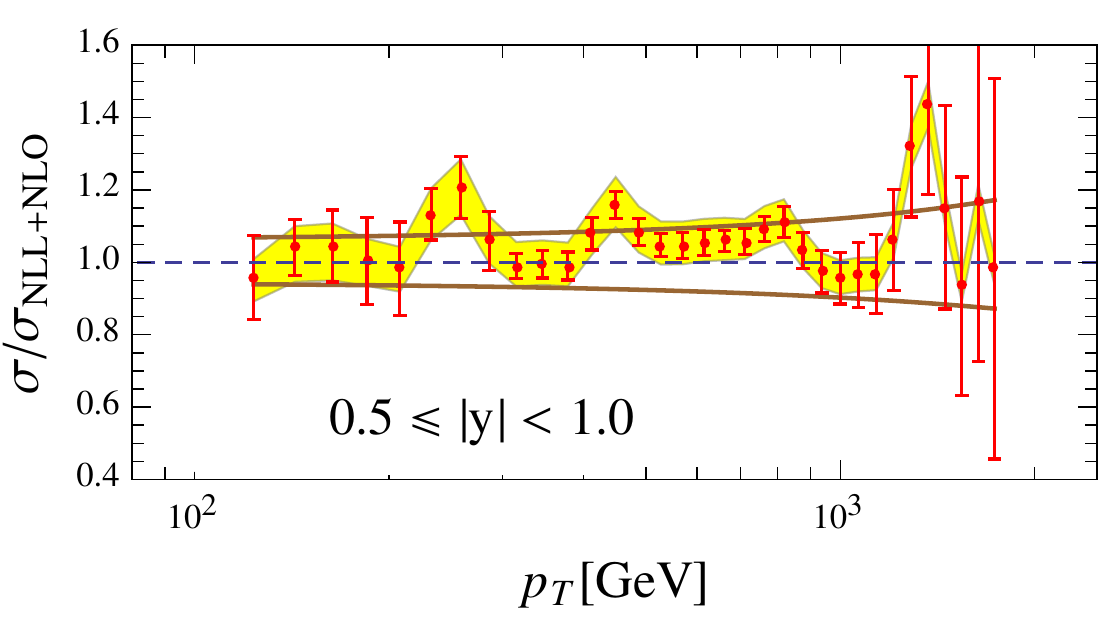}
  \end{subfigure}
   \begin{subfigure}{.5\textwidth}
  \includegraphics[width=0.95\linewidth]{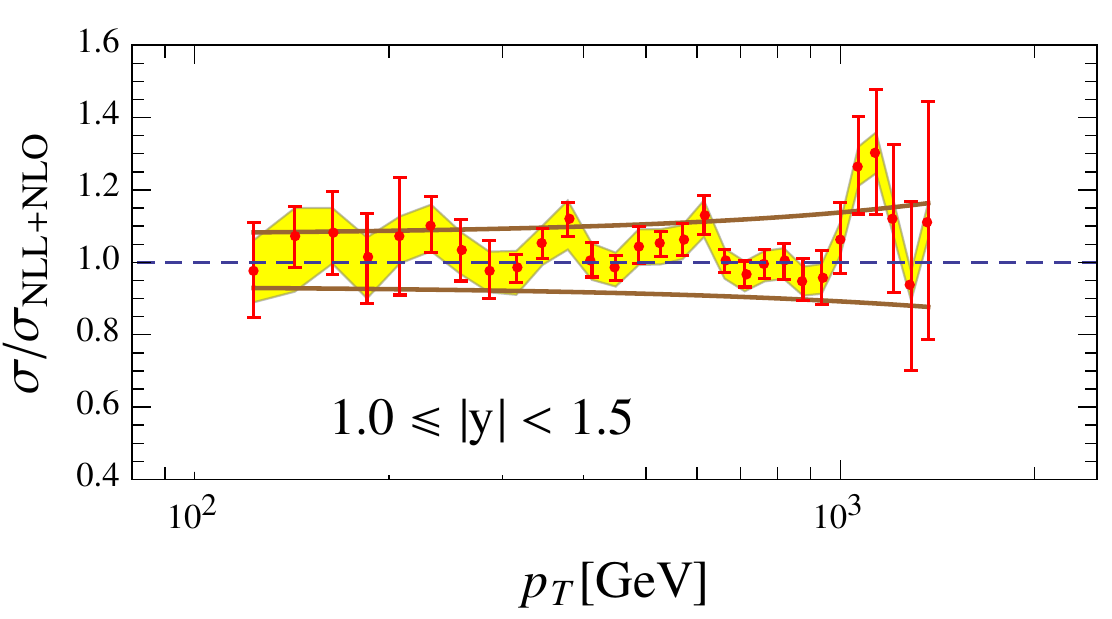}
   \end{subfigure}%
    \begin{subfigure}{.5\textwidth}
  \includegraphics[width=0.95\linewidth]{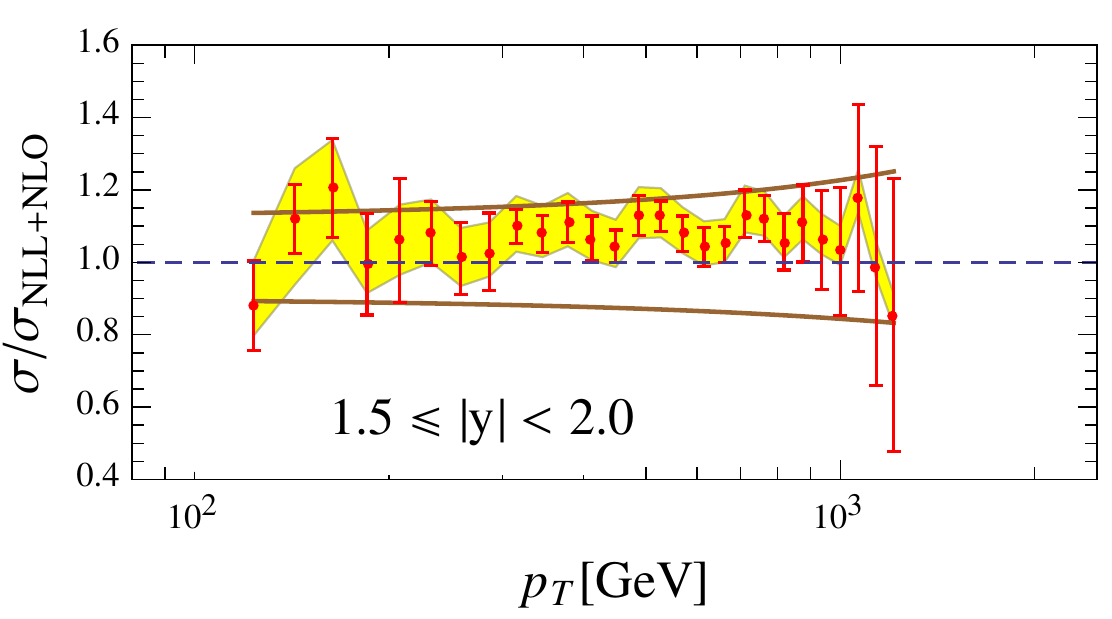}
     \end{subfigure}
 \caption{
   The ratio $\sigma_{\rm Data}/\sigma_{\rm NLO + NLL}$ 
   for CMS data collected at $\sqrt{S}=13$~TeV~\cite{Khachatryan:2016wdh} with
   $R= 0.4$ to the theoretical results 
   using the PDFs of CT14~\cite{Dulat:2015mca} at NLO.
   The error bars represent the experimental statistical errors and 
   the solid (brown) lines the systematic ones.
   The band (yellow) indicates theoretical scale uncertainties.
 } 
 \label{fig:13tevnll}
 \end{figure*}

\begin{figure*}
 \begin{subfigure}{.5\textwidth}
  \includegraphics[width=0.95\linewidth]{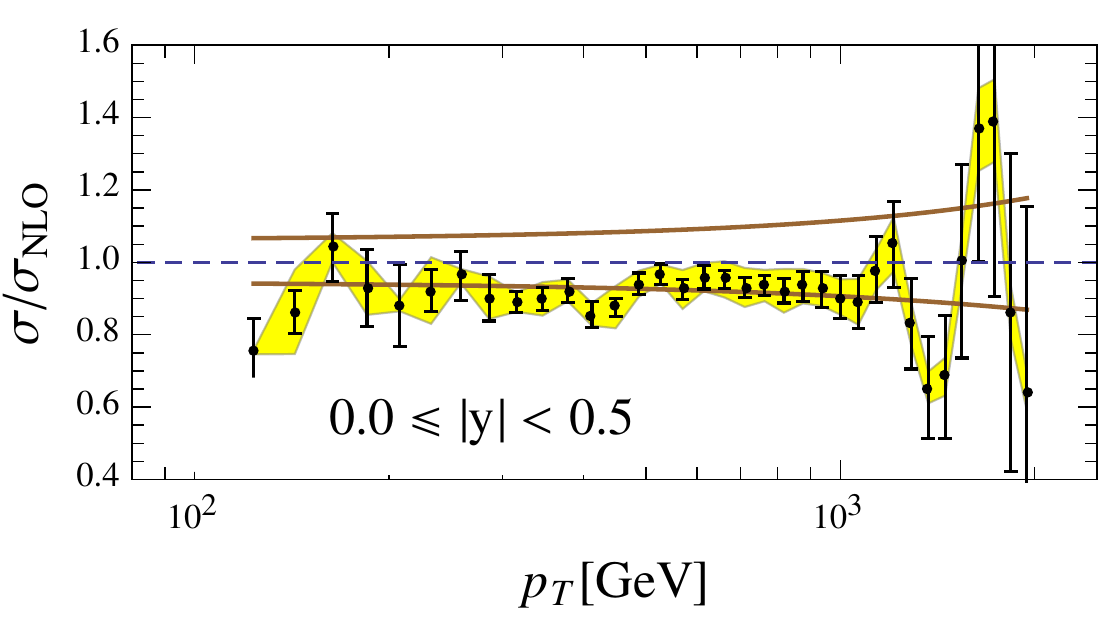}
 \end{subfigure}%
 \begin{subfigure}{.5\textwidth}
  \includegraphics[width=0.95\linewidth]{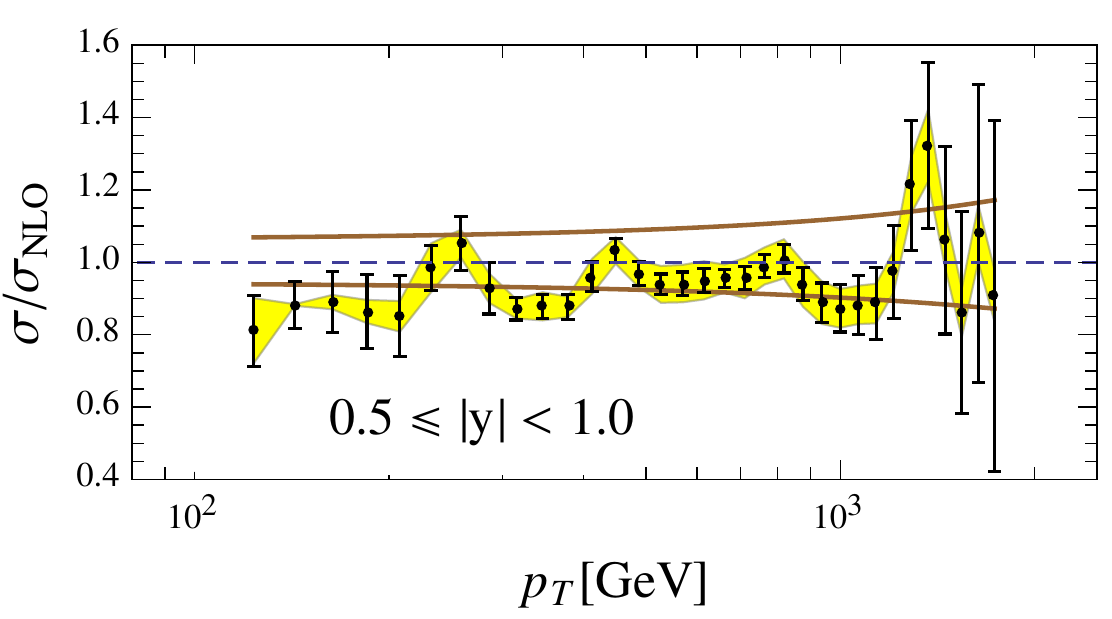}
  \end{subfigure}
   \begin{subfigure}{.5\textwidth}
  \includegraphics[width=0.95\linewidth]{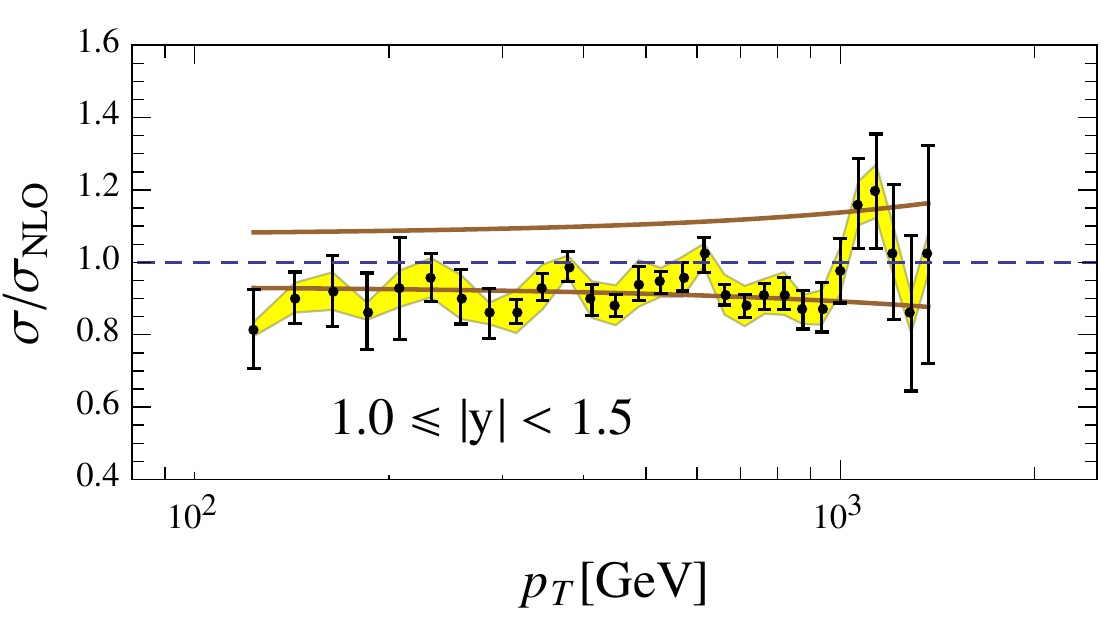}
   \end{subfigure}%
    \begin{subfigure}{.5\textwidth}
  \includegraphics[width=0.95\linewidth]{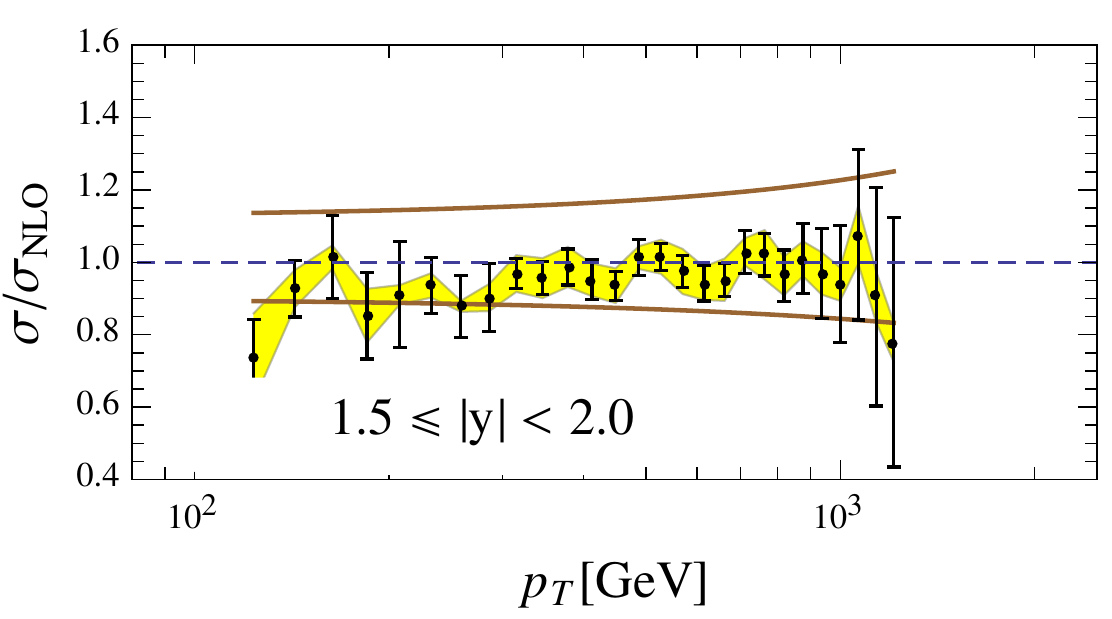}
     \end{subfigure}
 \caption{
   Same as Fig.~\ref{fig:13tevnll} for the ratio $\sigma_{\rm Data}/\sigma_{\rm NLO}$.}    
 \label{fig:13tevnlo}
 \end{figure*}

Next we study the inclusive jet production cross section with $R = 0.4$ at
$\sqrt{S}=13$~TeV. The results are shown in Figs.~\ref{fig:13tevnll} and~\ref{fig:13tevnlo}, 
in which the ratio of the CMS data~\cite{Khachatryan:2016wdh} 
to the cross sections $\sigma_{\rm NLO + NLL}$ and $\sigma_{\rm NLO}$ are displayed, respectively. 
Here, we have applied the CT14 PDF set~\cite{Dulat:2015mca} at NLO for both
predictions and we note that the NP and the electroweak effects have not been
included in this analysis. 
Again, the yellow band in Figs.~\ref{fig:13tevnll} and~\ref{fig:13tevnlo} 
represents the theoretical scale uncertainties whereas the experimental systematic and
statistical errors are shown as solid brown lines and the error bars, respectively. 
At present, the published CMS data at $\sqrt{S}=13$~TeV have larger statistical
errors, since they are based on data samples corresponding to a relativley 
small integrated luminosities of 71 and 44 inverse picobarns, 
whereas the $\sqrt{S}=7$ TeV data~\cite{Chatrchyan:2014gia} 
discussed above correspond to an integrated luminosity of 5.0 inverse femtobarns.

As shown in Fig.~\ref{fig:13tevnll}, the NLO $+$ NLL calculation leads to a good
agreement of the ratio $\sigma_{\rm Data}/\sigma_{\rm Theory}$ with unity 
in the region of central rapidities, but slightly overshoots it 
in the rapidity bin $1.5 \le |y|< 2.0$, although still being compatible
within the errors.
On the contrary, in Fig.~\ref{fig:13tevnlo} 
the ratio $\sigma_{\rm Data}/\sigma_{\rm Theory}$ based on the NLO predictions
systematically undershoots unity in the rapidity region $|y|< 1.5$, 
but it is still compatible within the quoted uncertainties. 
Better consistency of the NLO results with the CMS data is only observed 
in the rapidity region $1.5 \le |y|< 2.0$.


\subsection{Impact of different PDFs}

We now proceed to study the sensitivity of different choices of PDFs in
predicting the inclusive jet $p_T$ distributions. We benchmark our study using
the CMS data at $\sqrt{S}=7$~TeV since the experimental errors of those data
are relatively small. 
Besides the CT10 PDFs used above, we also consider the following alternative 
PDF extractions which are available in the literature to NLO and NNLO accuracy in pQCD:
ABMP16~\cite{Alekhin:2018toappear,Alekhin:2017kpj}, CT14~\cite{Dulat:2015mca},
HERAPDF2.0~\cite{Abramowicz:2015mha}, MMHT2014~\cite{Harland-Lang:2014zoa} and
NNPDF3.1~\cite{Ball:2017nwa}. 
In addition, we use the PDF set of~\cite{Bonvini:2015ira} 
obtained within the framework of NNPDF by fitting only data for the Drell-Yan (DY) process, 
deep-inelastic scattering (DIS) and top-quark hadro-production but including
threshold resummation in all theory predictions for the hard scattering.

 \begin{figure*}
 \begin{subfigure}{.5\textwidth}
  \includegraphics[width=0.95\linewidth]{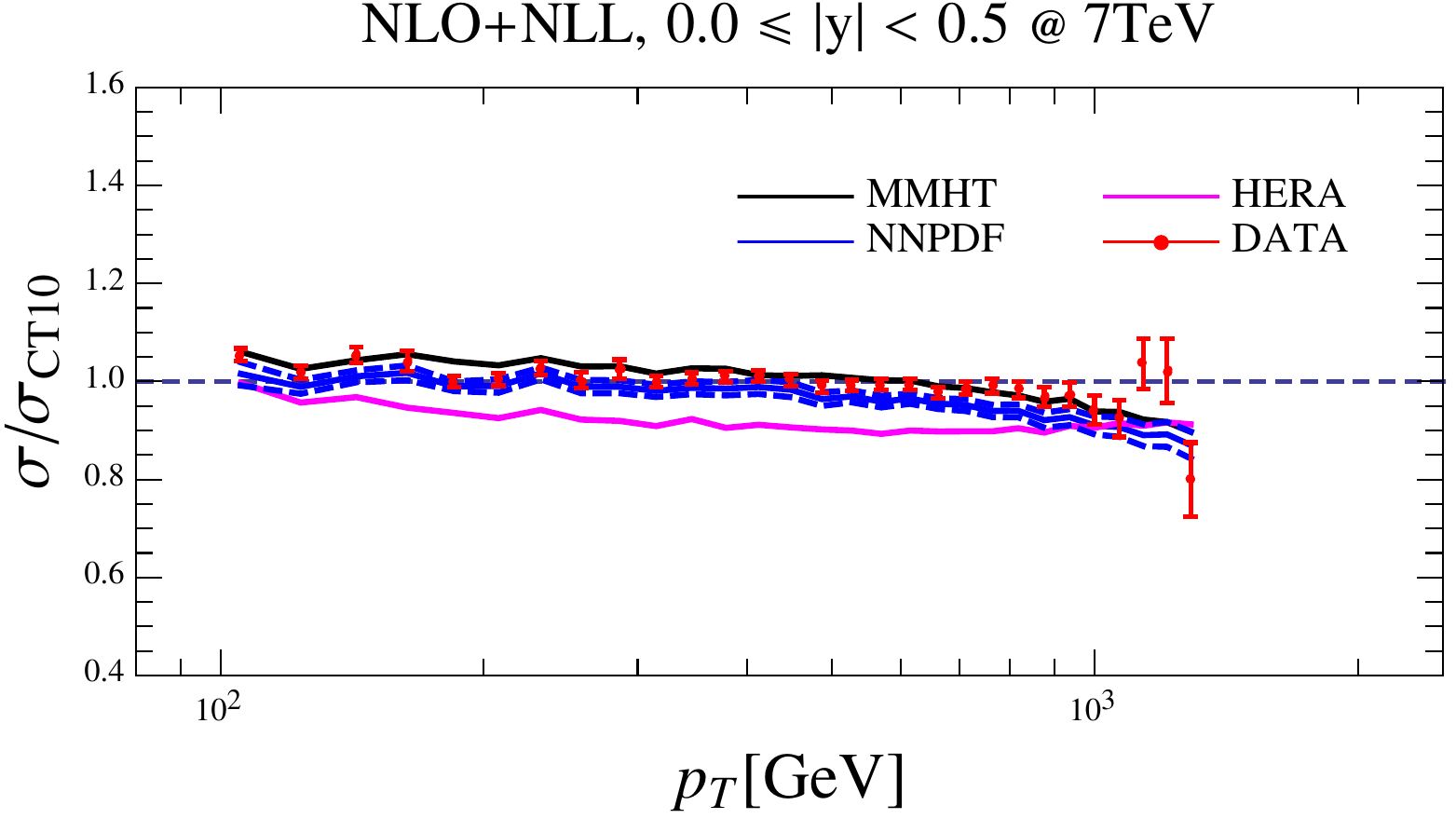}
 \end{subfigure}%
 \begin{subfigure}{.5\textwidth}
  \includegraphics[width=0.95\linewidth]{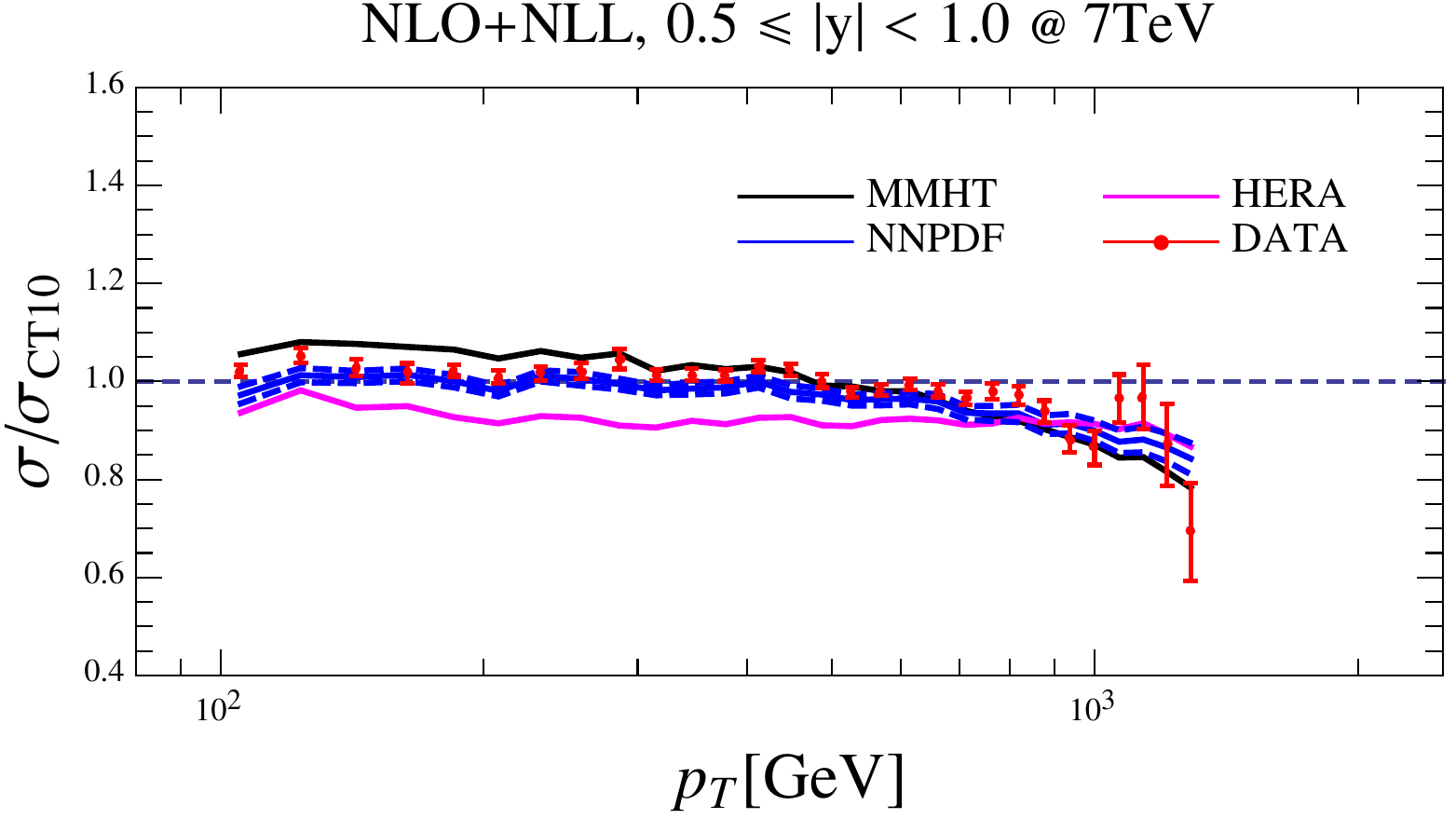}
  \end{subfigure}
   \begin{subfigure}{.5\textwidth}
  \includegraphics[width=0.95\linewidth]{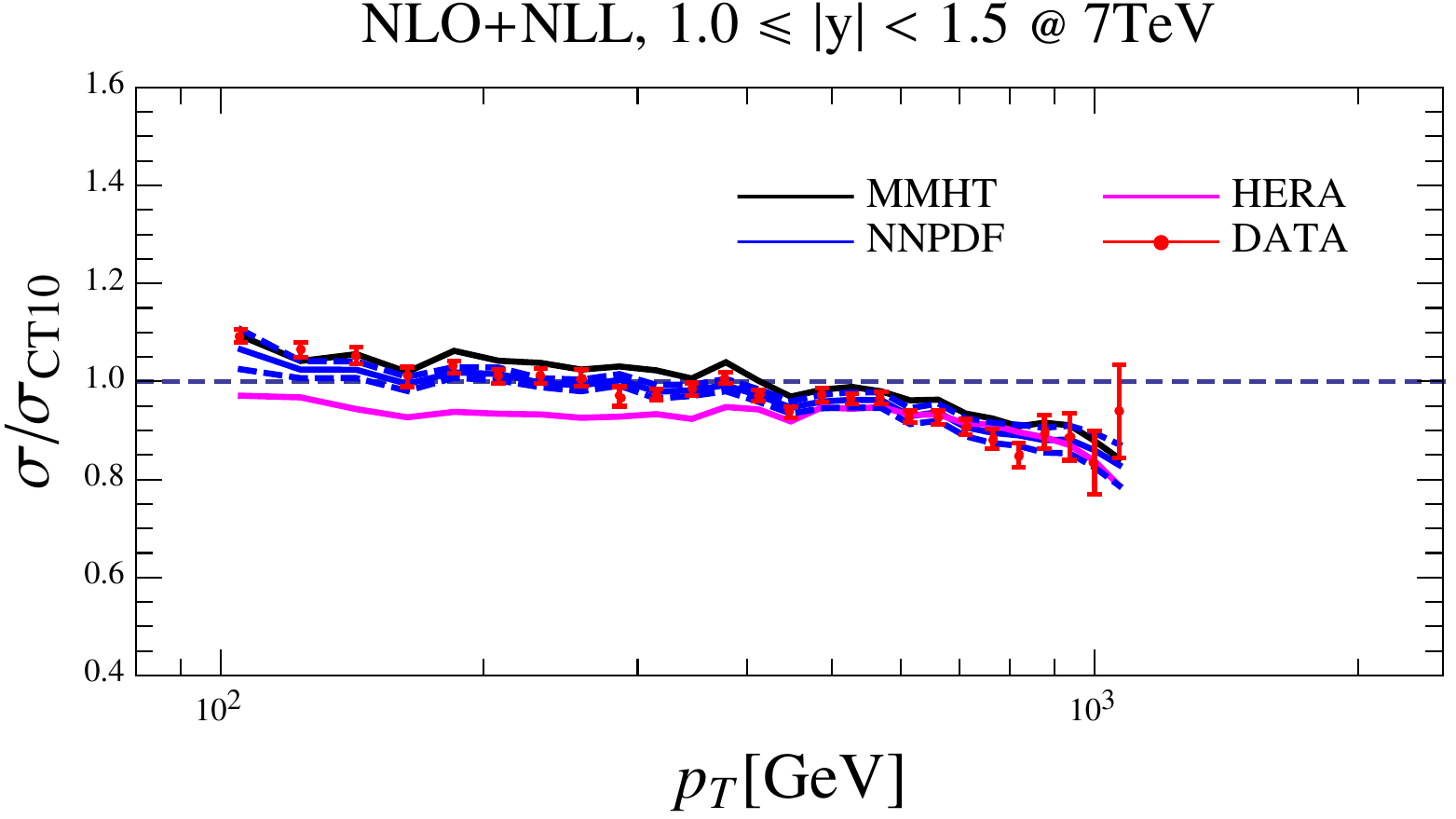}
   \end{subfigure}%
    \begin{subfigure}{.5\textwidth}
  \includegraphics[width=0.95\linewidth]{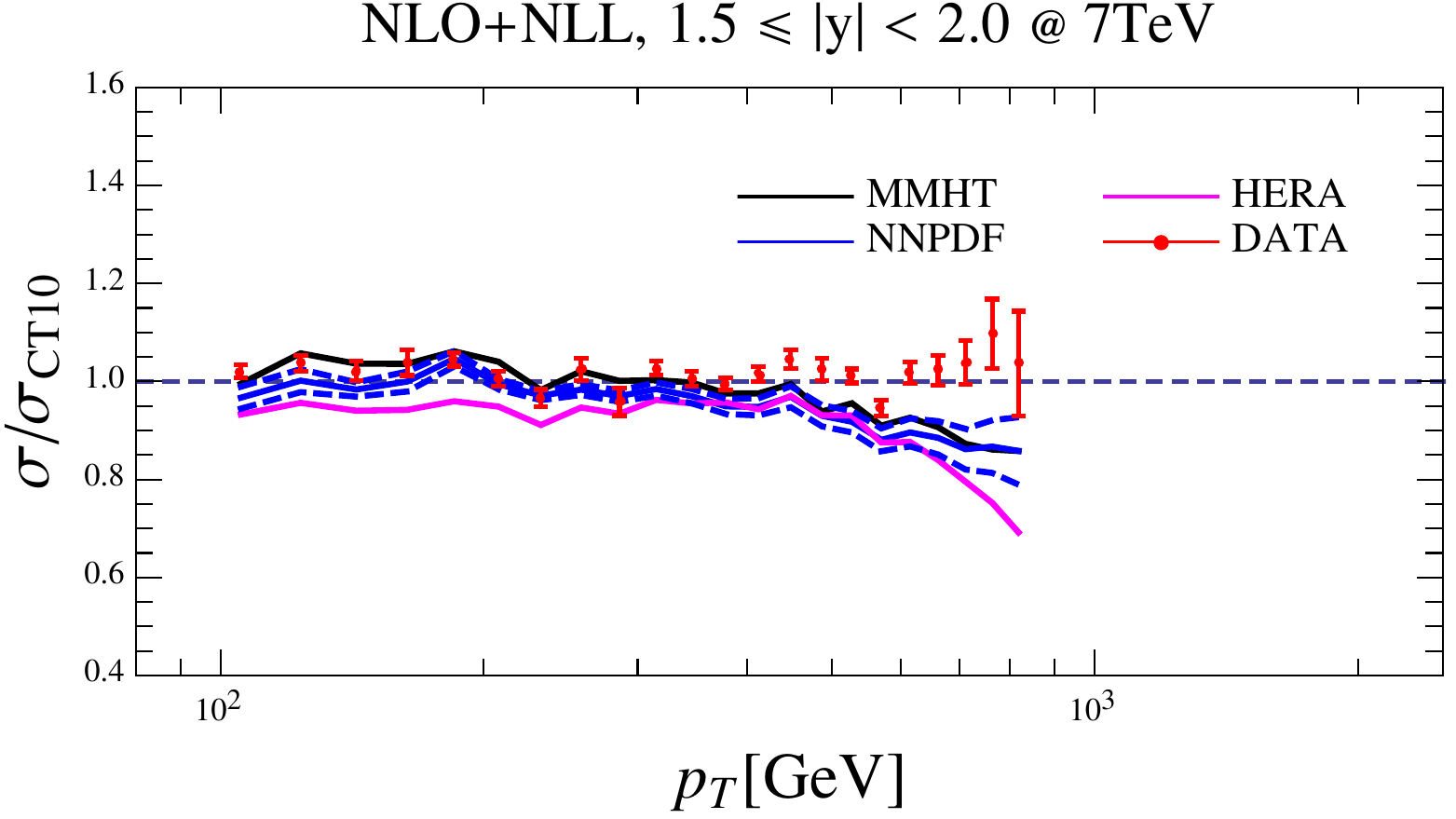}
     \end{subfigure}
 \caption{
   Cross sections $\sigma_{\rm NLO + NLL}$ at NLL $+$ NLO accuracy with $R= 0.5$
   using the central PDF sets HERAPDF2.0~\cite{Abramowicz:2015mha}, MMHT2014~\cite{Harland-Lang:2014zoa} and
   NNPDF3.1~\cite{Ball:2017nwa}
   normalized to the one with CT10 PDFs~\cite{Lai:2010vv} at NLO.
   The dashed lines (blue) indicate the PDF uncertainties for the NNPDF3.1 set.
   The CMS data collected at $\sqrt{S}=7$ TeV~\cite{Chatrchyan:2014gia} with $R= 0.5$
   with their experimental statistical errors are displayed as dots (red).
 }
 \label{fig:7tevpdfscan-sets1}
 \end{figure*}
 \begin{figure*}
 \begin{subfigure}{.5\textwidth}
  \includegraphics[width=0.95\linewidth]{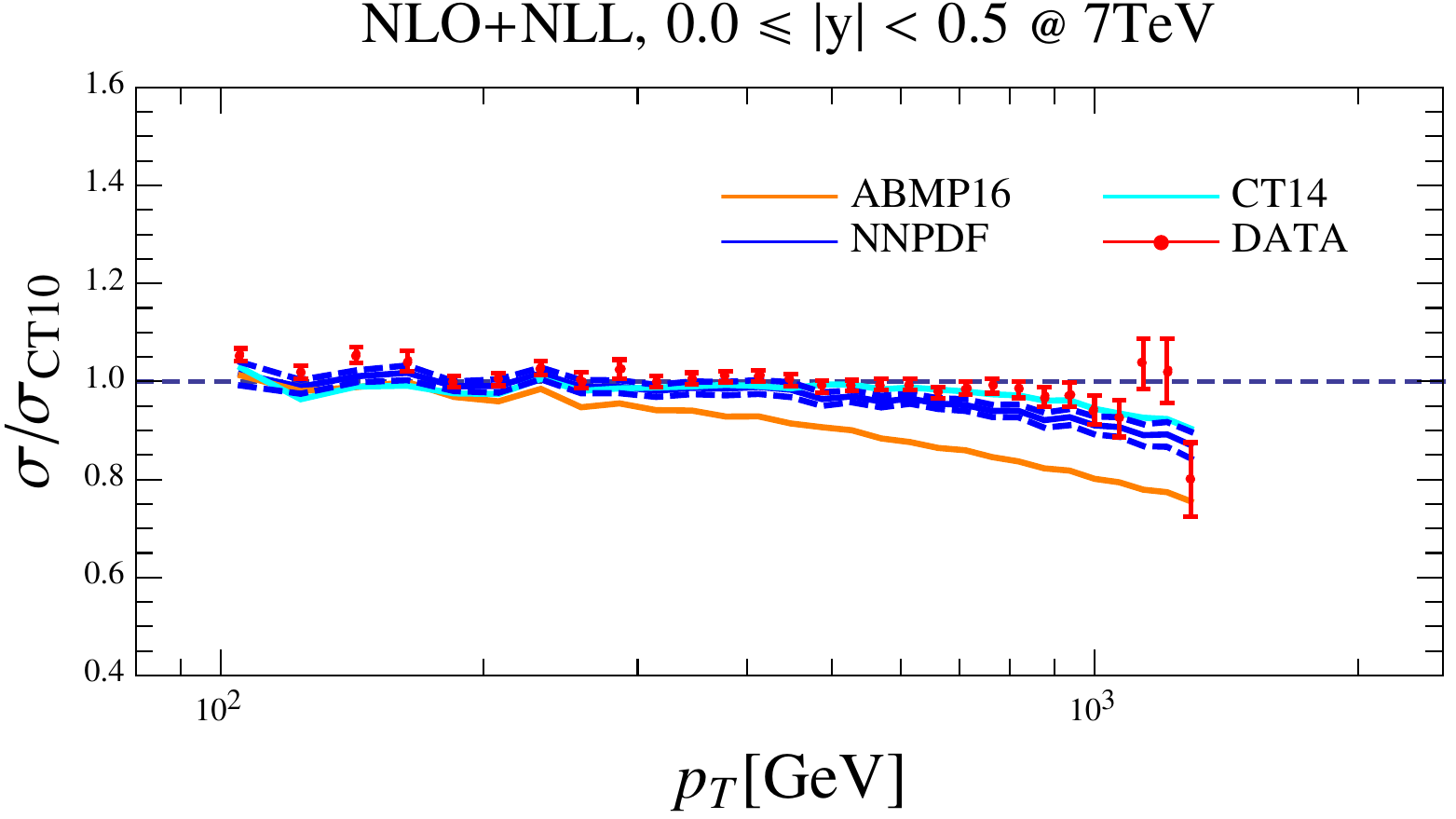}
 \end{subfigure}%
 \begin{subfigure}{.5\textwidth}
  \includegraphics[width=0.95\linewidth]{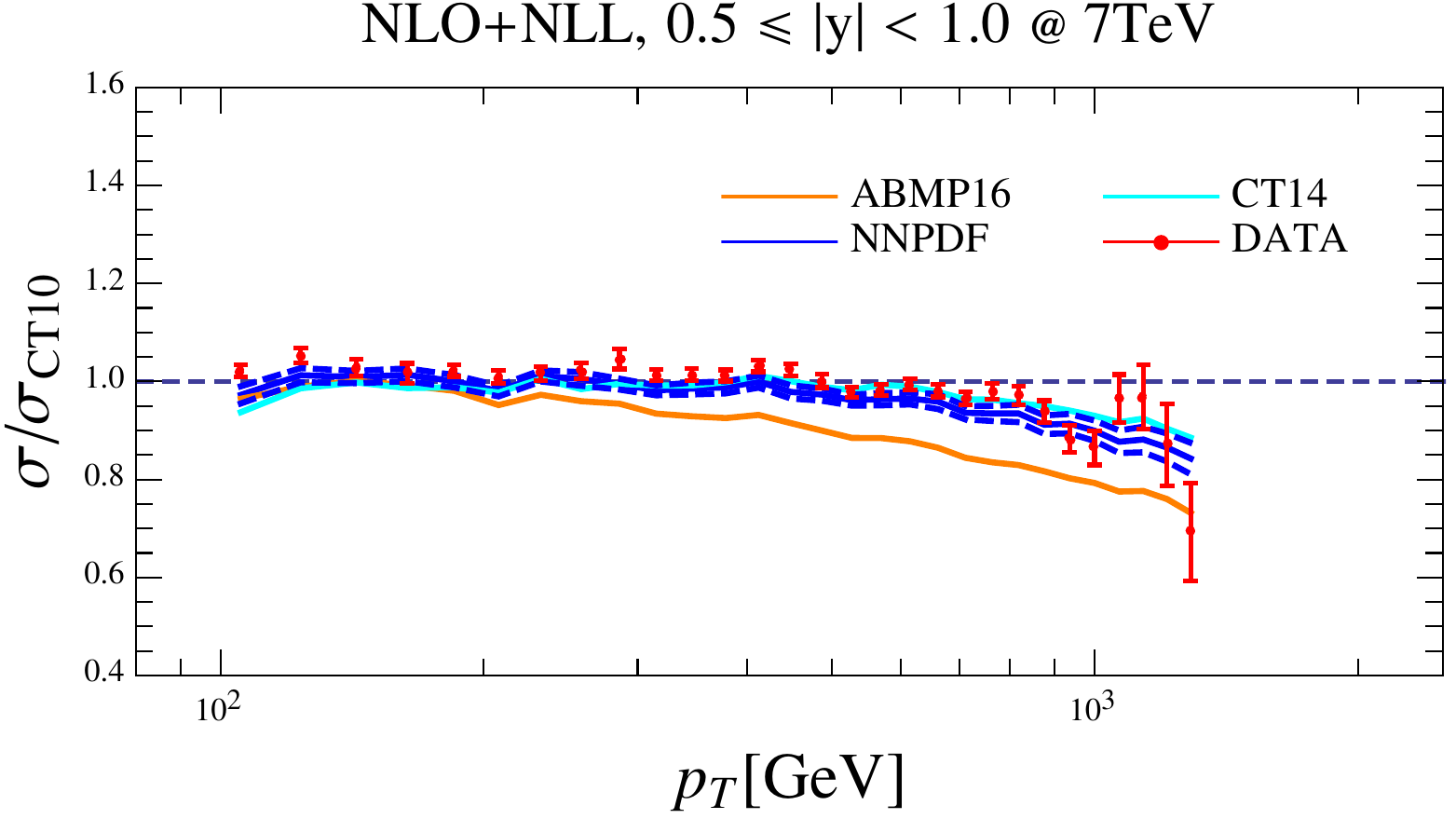}
  \end{subfigure}
   \begin{subfigure}{.5\textwidth}
  \includegraphics[width=0.95\linewidth]{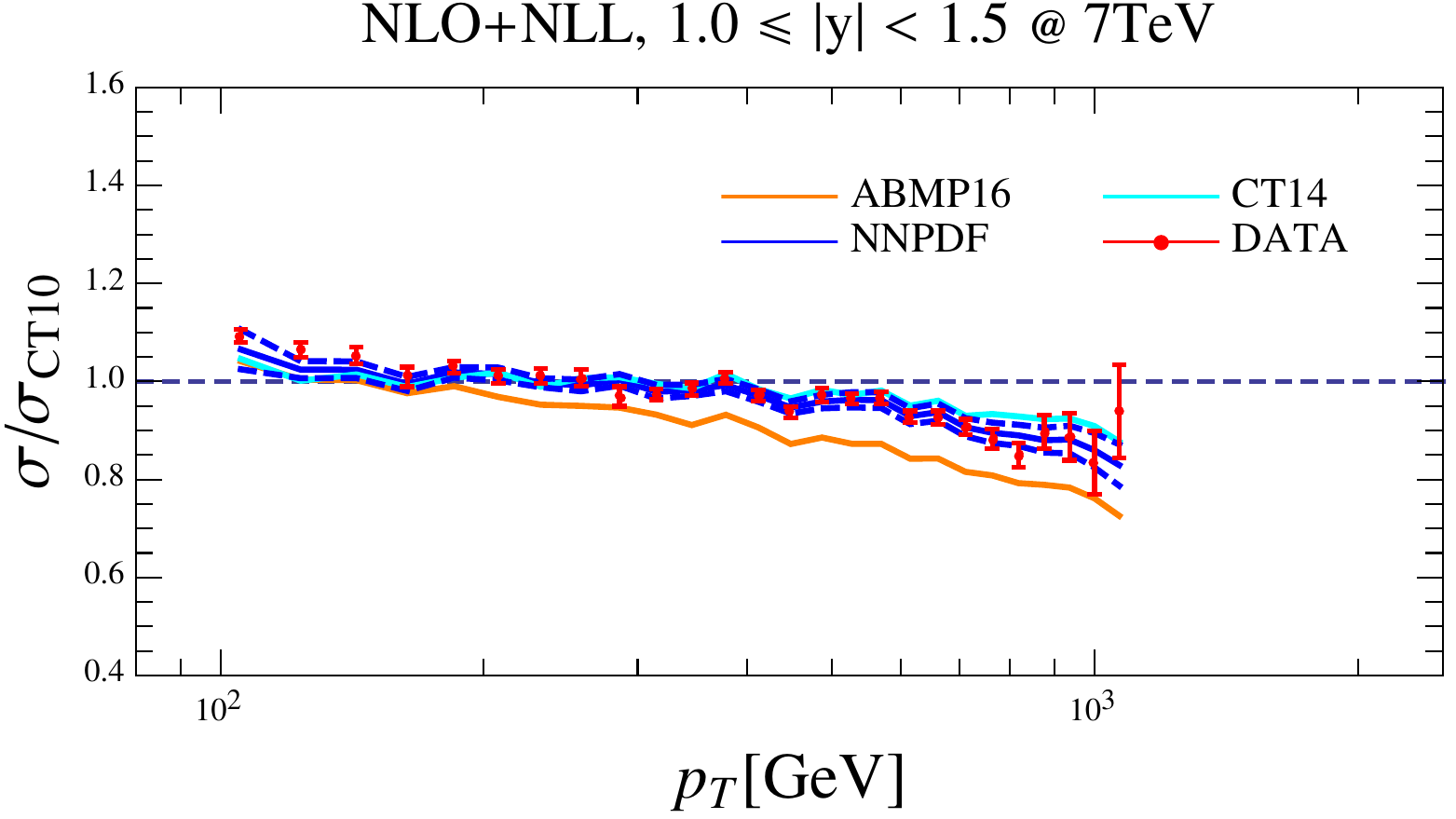}
   \end{subfigure}%
    \begin{subfigure}{.5\textwidth}
  \includegraphics[width=0.95\linewidth]{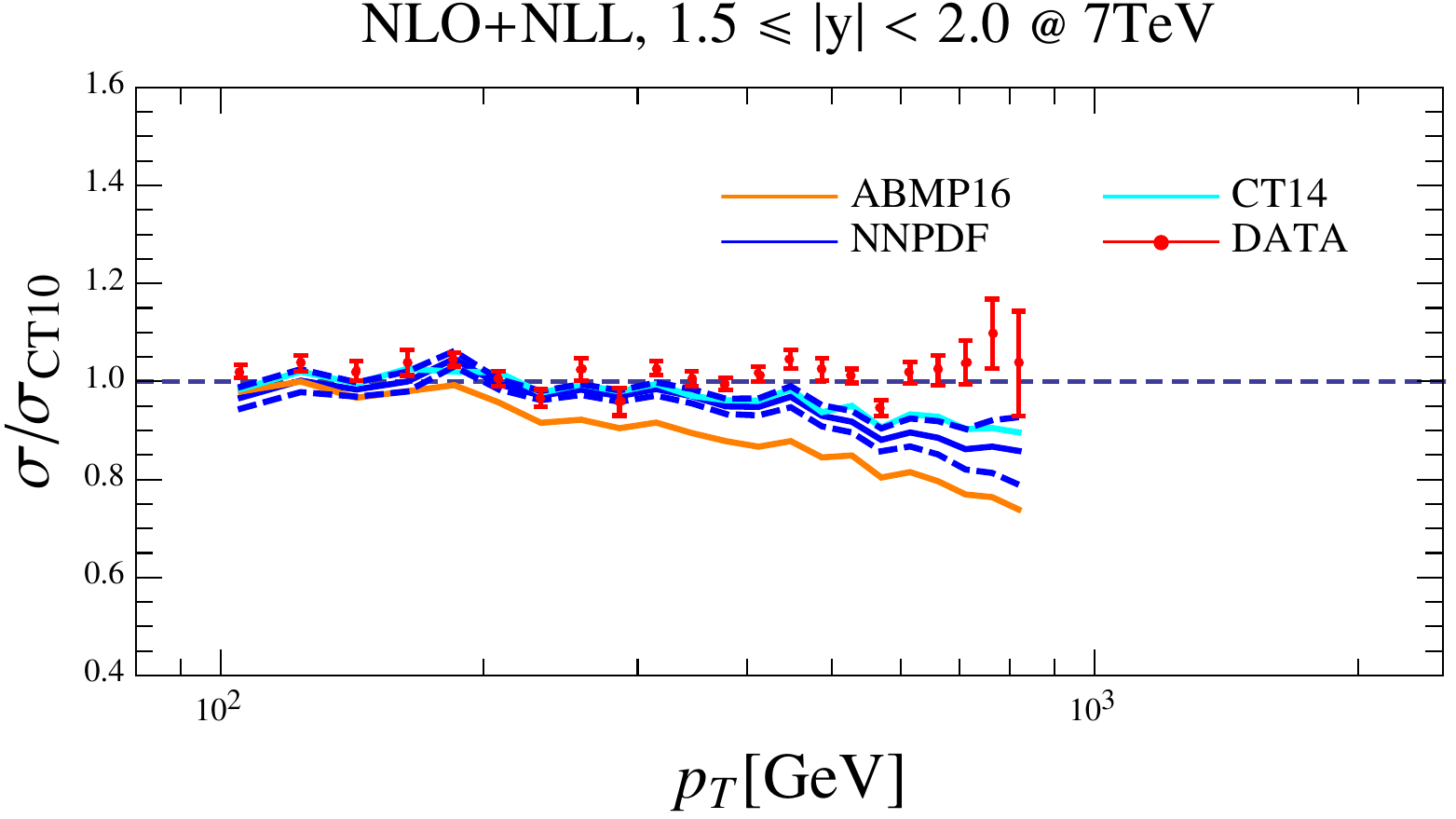}
     \end{subfigure}
 \caption{
   Same as Fig.~\ref{fig:7tevpdfscan-sets1} for the PDF sets 
   ABMP16~\cite{Alekhin:2018toappear}, CT14~\cite{Dulat:2015mca} and
   NNPDF3.1~\cite{Ball:2017nwa}.}    
 \label{fig:7tevpdfscan-sets2}
 \end{figure*}
 \begin{figure*}
 \begin{subfigure}{.5\textwidth}
  \includegraphics[width=0.95\linewidth]{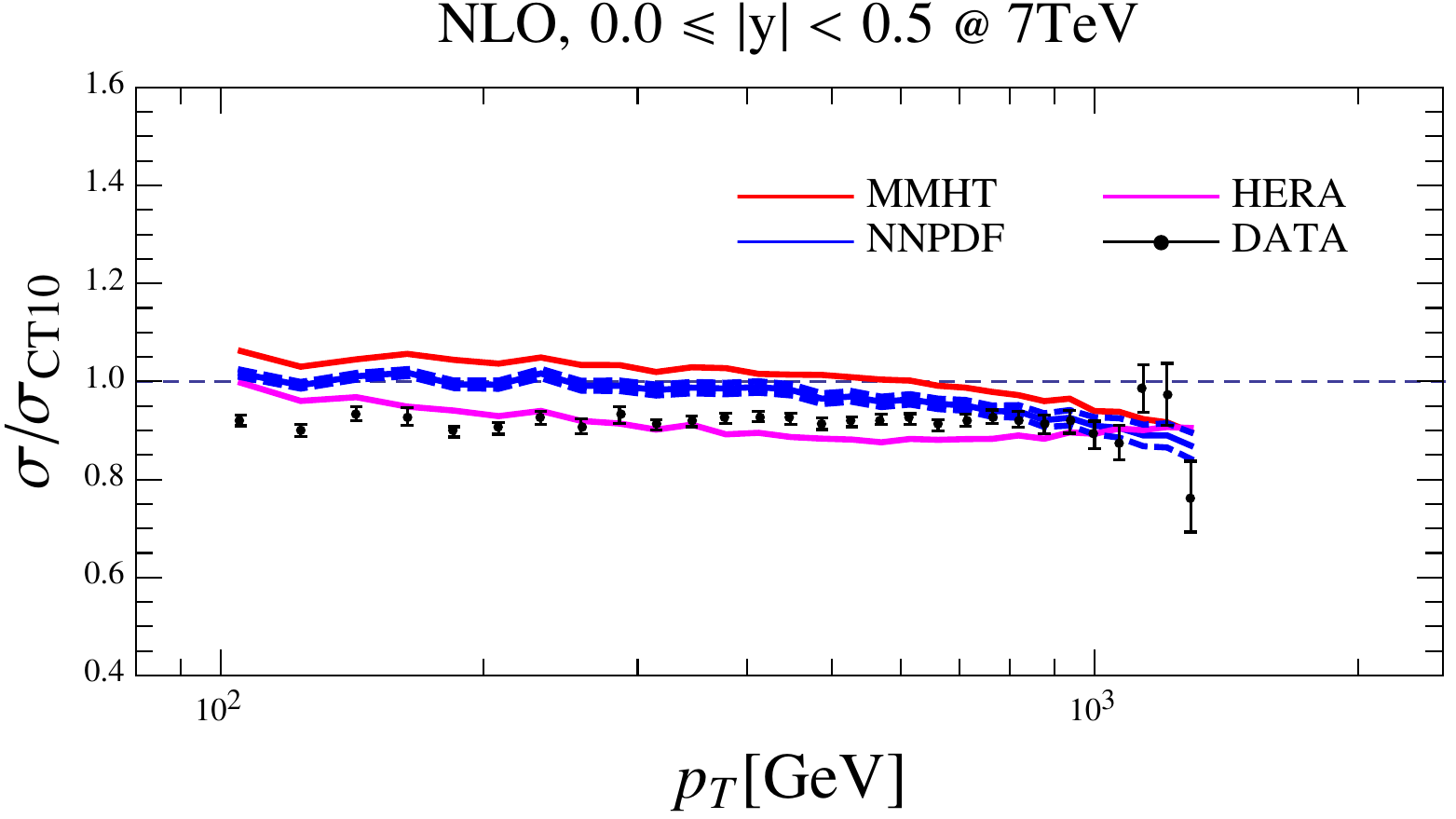}
 \end{subfigure}%
 \begin{subfigure}{.5\textwidth}
  \includegraphics[width=0.95\linewidth]{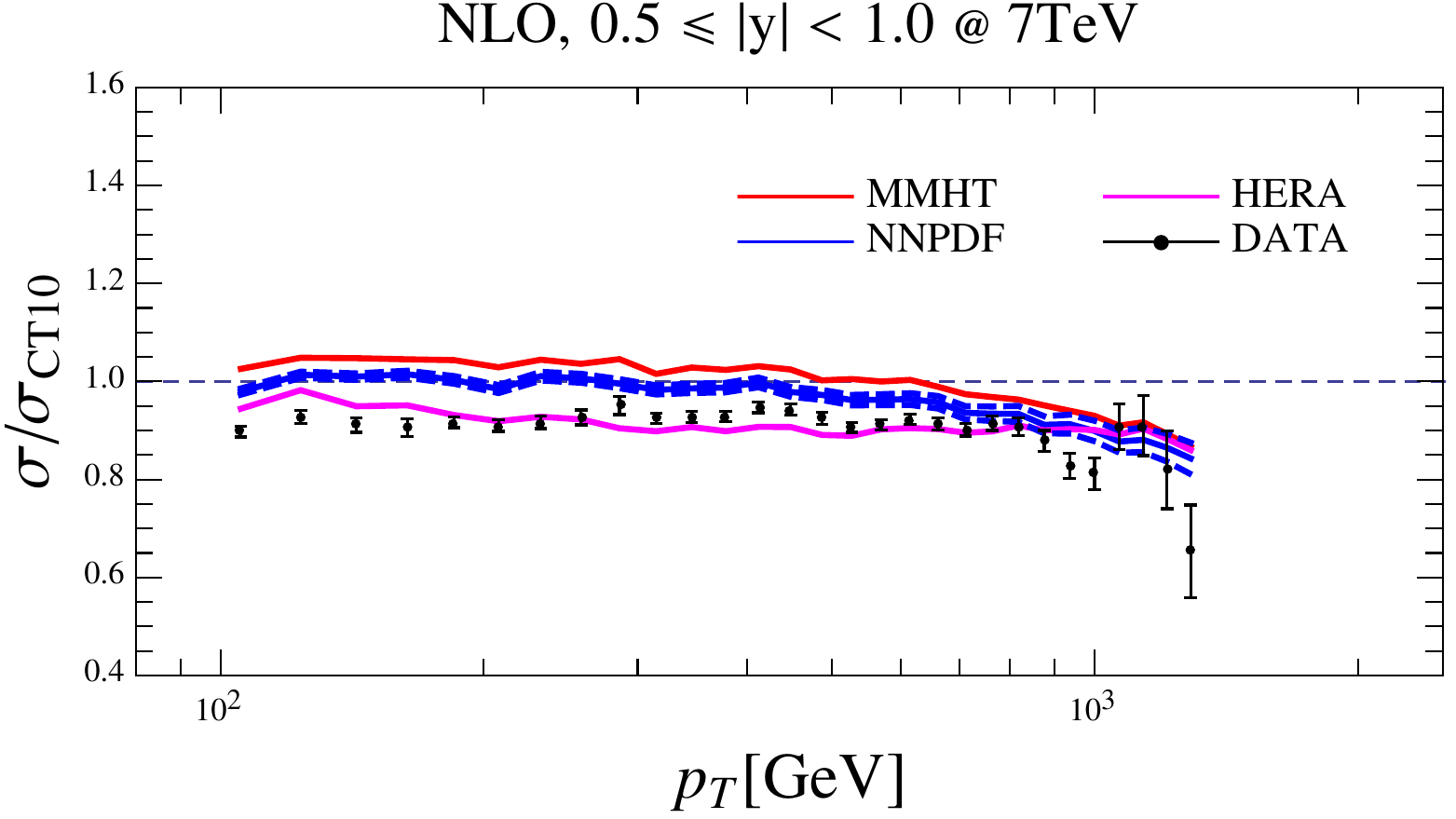}
  \end{subfigure}
   \begin{subfigure}{.5\textwidth}
  \includegraphics[width=0.95\linewidth]{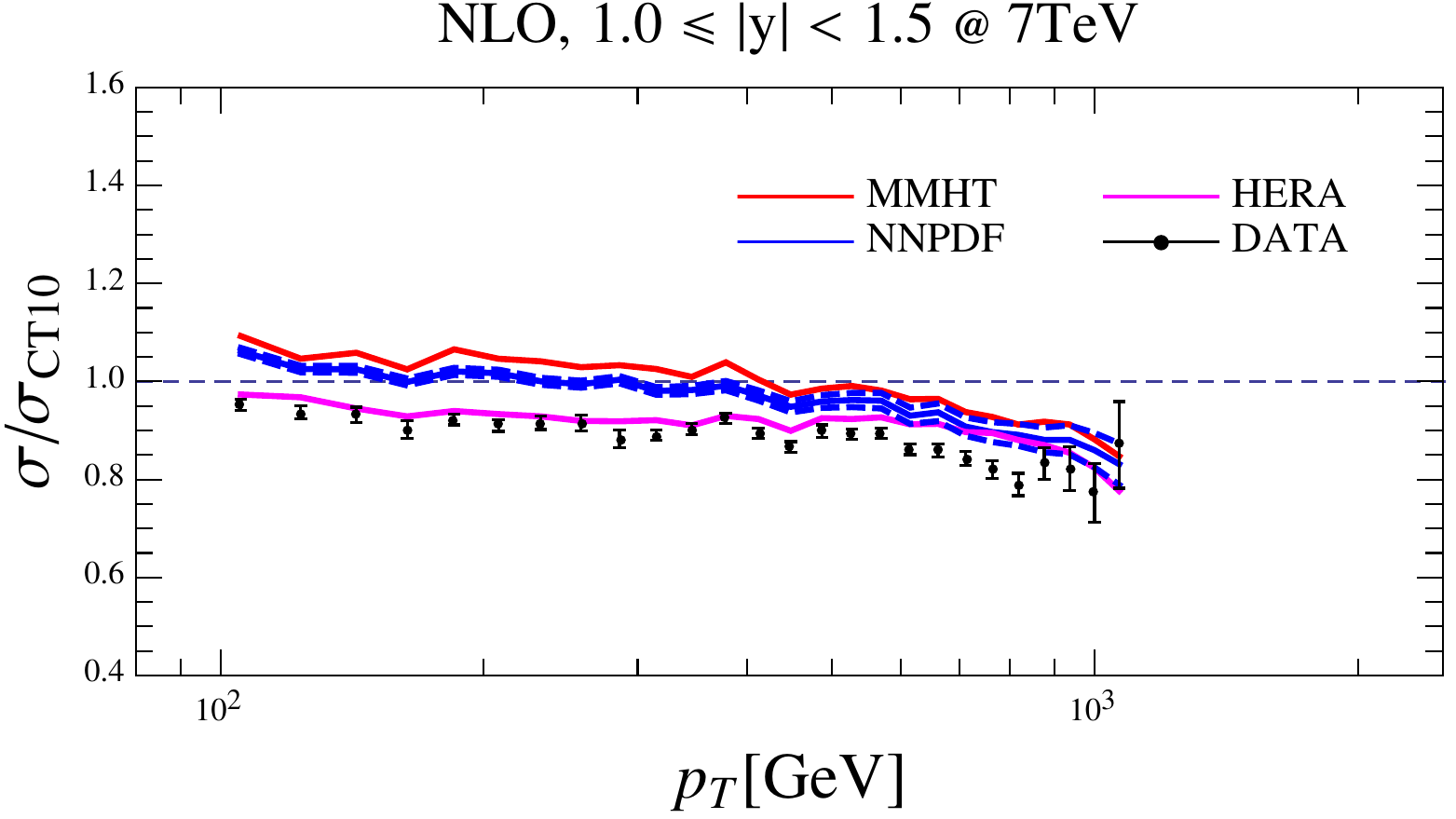}
   \end{subfigure}%
    \begin{subfigure}{.5\textwidth}
  \includegraphics[width=0.95\linewidth]{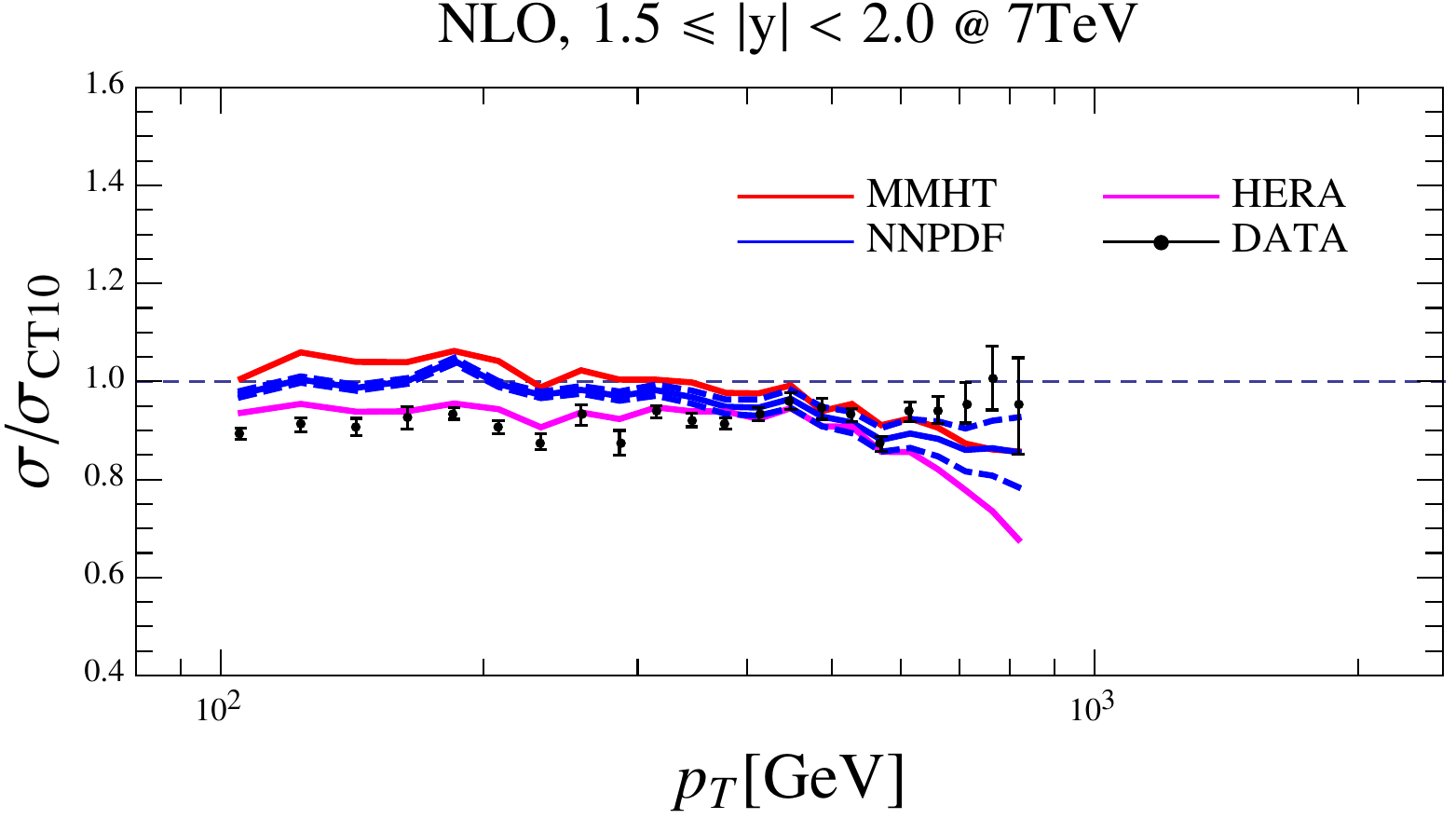}
     \end{subfigure}
 \caption{
   Same as Fig.~\ref{fig:7tevpdfscan-sets1} for cross sections $\sigma_{\rm NLO}$ at NLO in pQCD.}
 \label{fig:7tevpdfscan-nlo-sets1}
 \end{figure*}
 \begin{figure*}
 \begin{subfigure}{.5\textwidth}
  \includegraphics[width=0.95\linewidth]{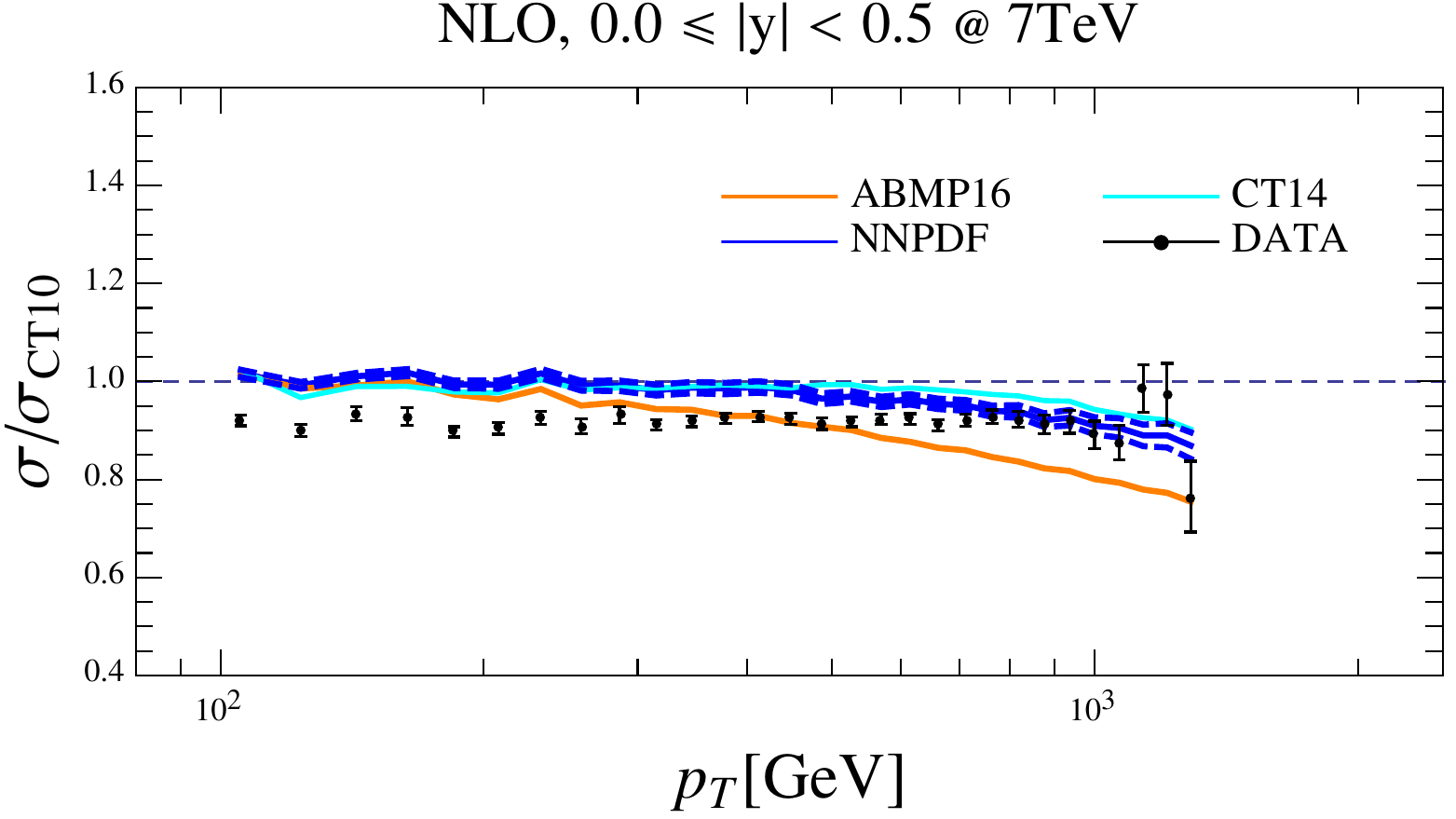}
 \end{subfigure}%
 \begin{subfigure}{.5\textwidth}
  \includegraphics[width=0.95\linewidth]{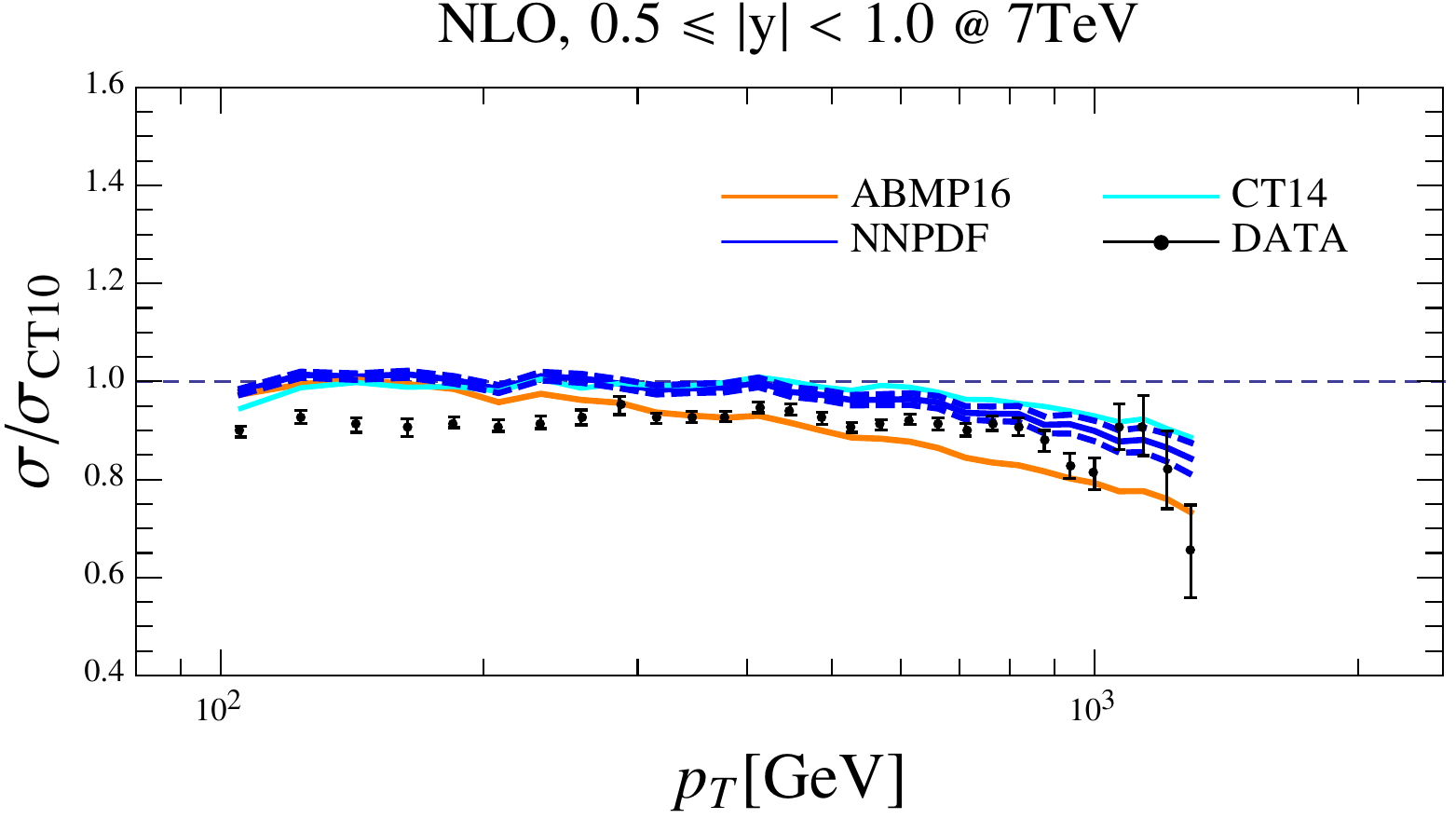}
  \end{subfigure}
   \begin{subfigure}{.5\textwidth}
  \includegraphics[width=0.95\linewidth]{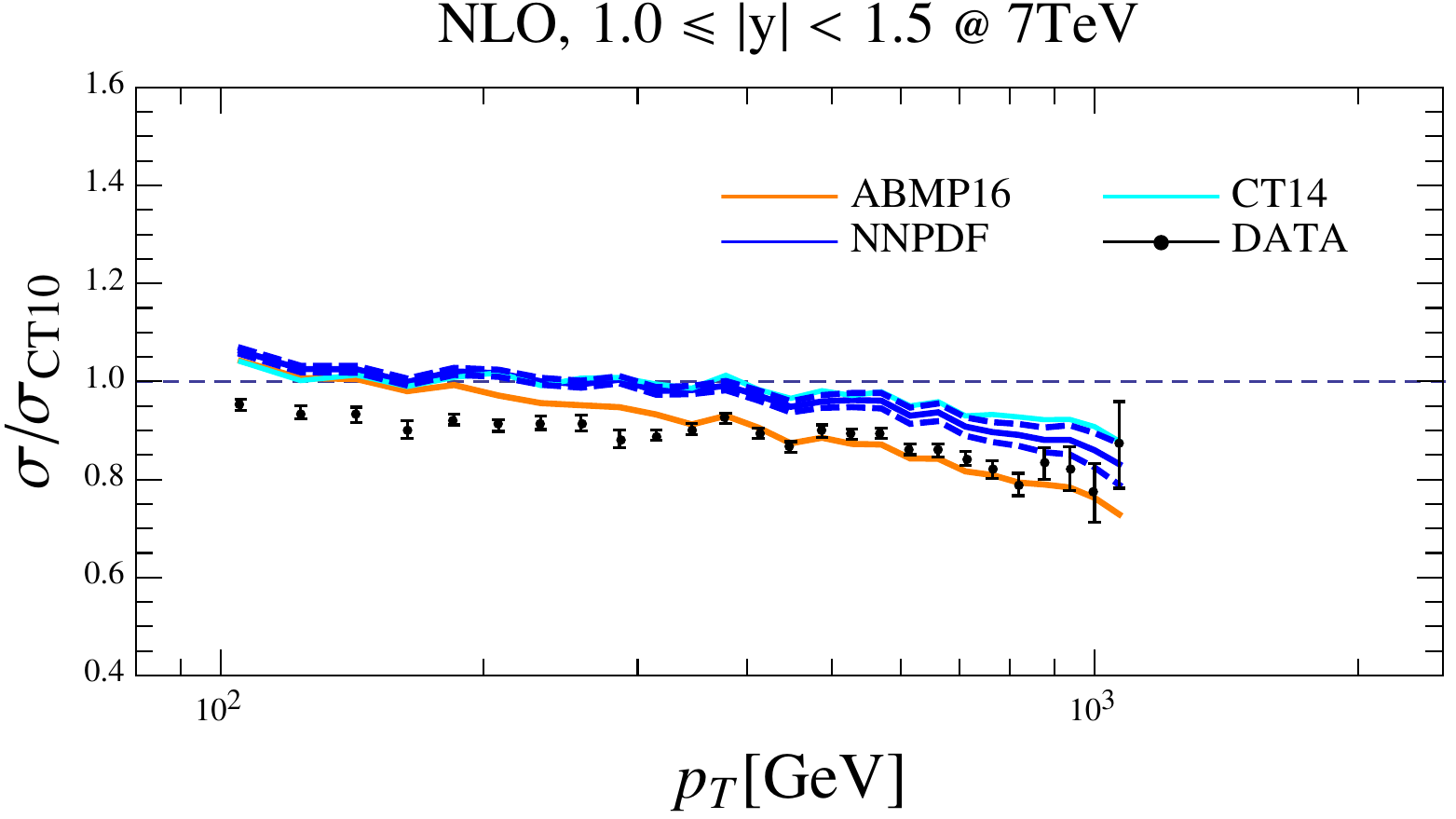}
   \end{subfigure}%
    \begin{subfigure}{.5\textwidth}
  \includegraphics[width=0.95\linewidth]{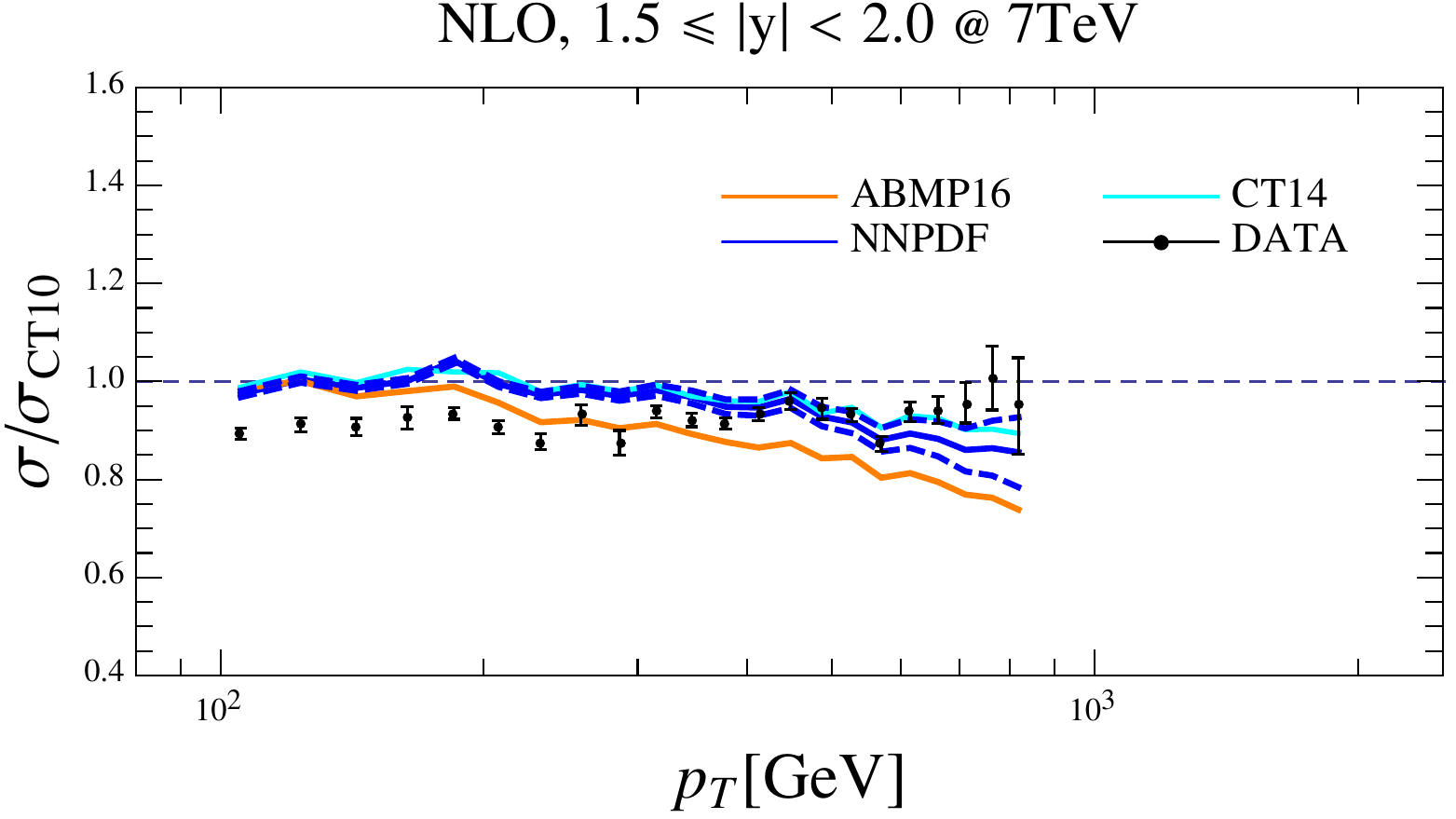}
     \end{subfigure}
 \caption{
   Same as Fig.~\ref{fig:7tevpdfscan-sets2} for cross sections $\sigma_{\rm NLO}$ at NLO in pQCD.}
  \label{fig:7tevpdfscan-nlo-sets2}
 \end{figure*}
 \begin{figure*}
 \begin{subfigure}{.5\textwidth}
  \includegraphics[width=0.95\linewidth]{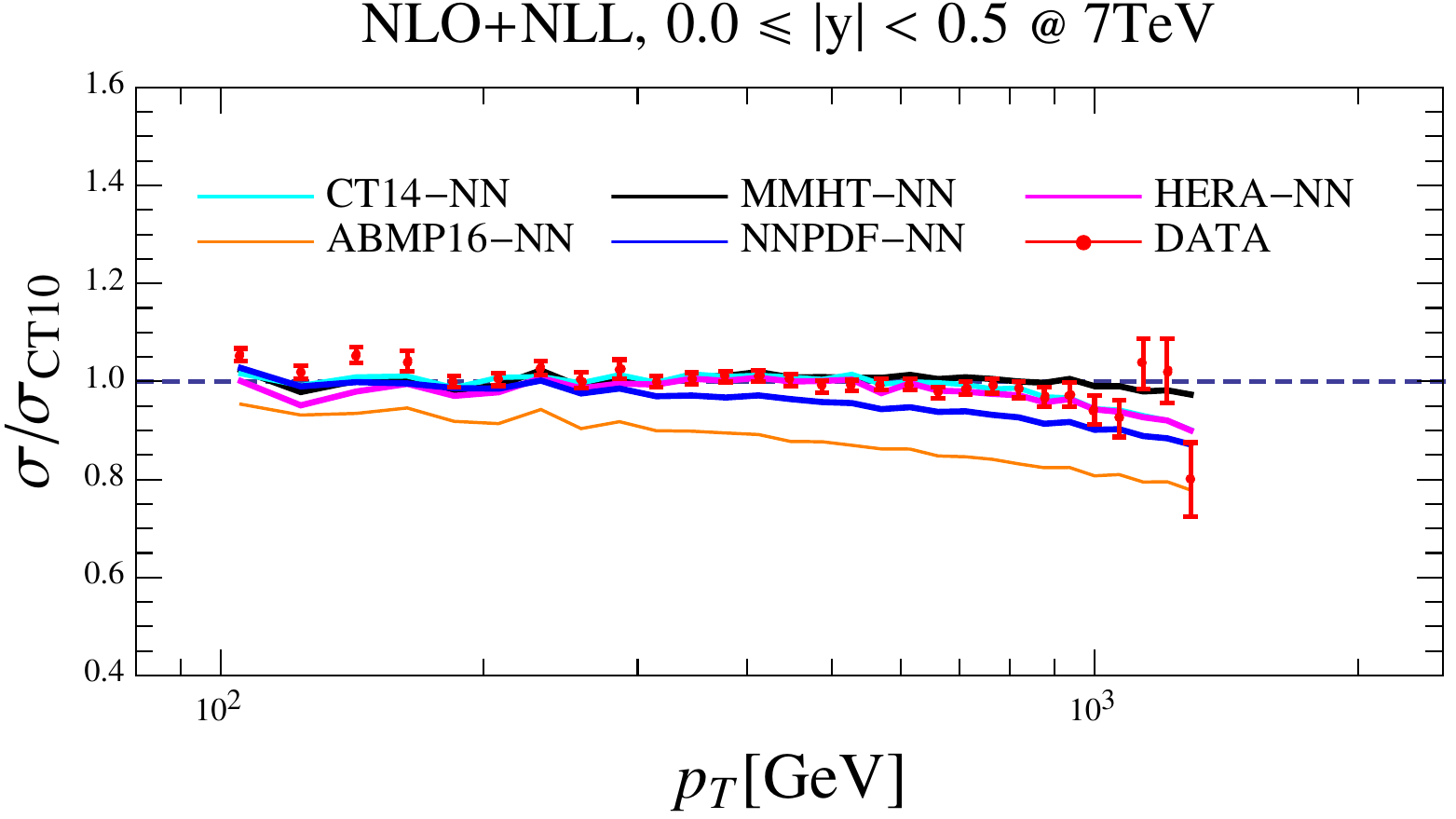}
 \end{subfigure}%
 \begin{subfigure}{.5\textwidth}
  \includegraphics[width=0.95\linewidth]{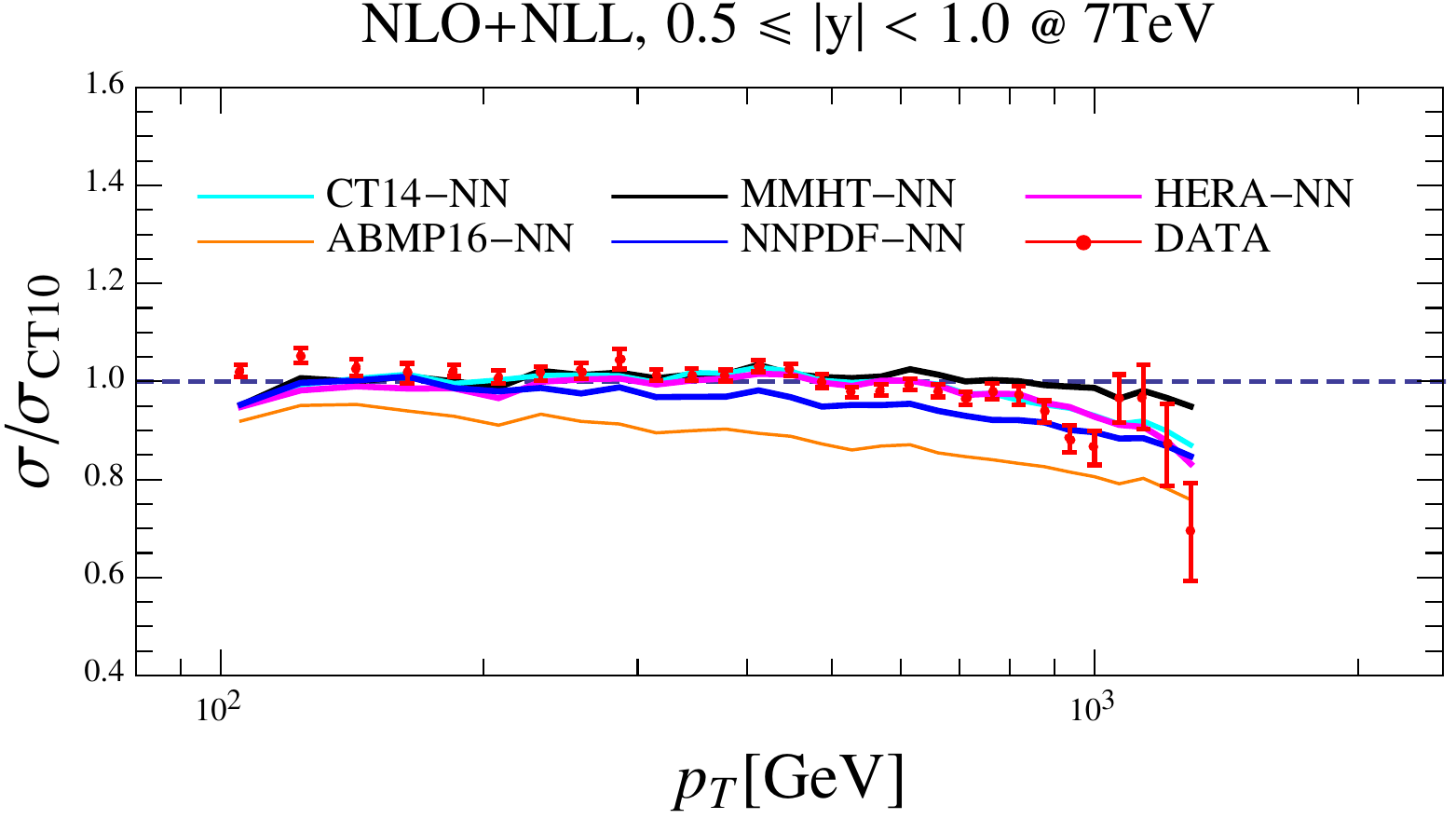}
  \end{subfigure}
   \begin{subfigure}{.5\textwidth}
  \includegraphics[width=0.95\linewidth]{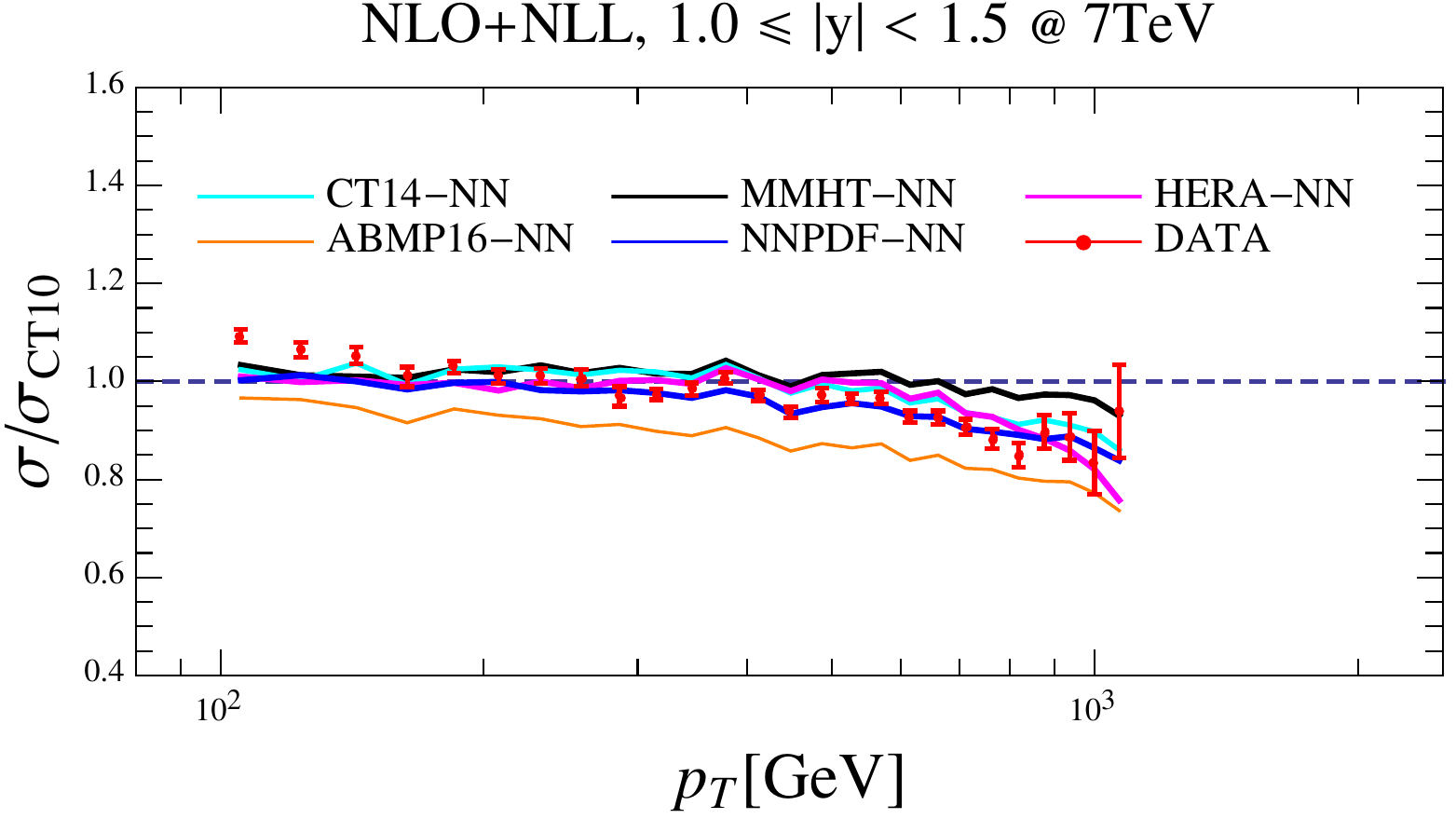}
   \end{subfigure}%
    \begin{subfigure}{.5\textwidth}
  \includegraphics[width=0.95\linewidth]{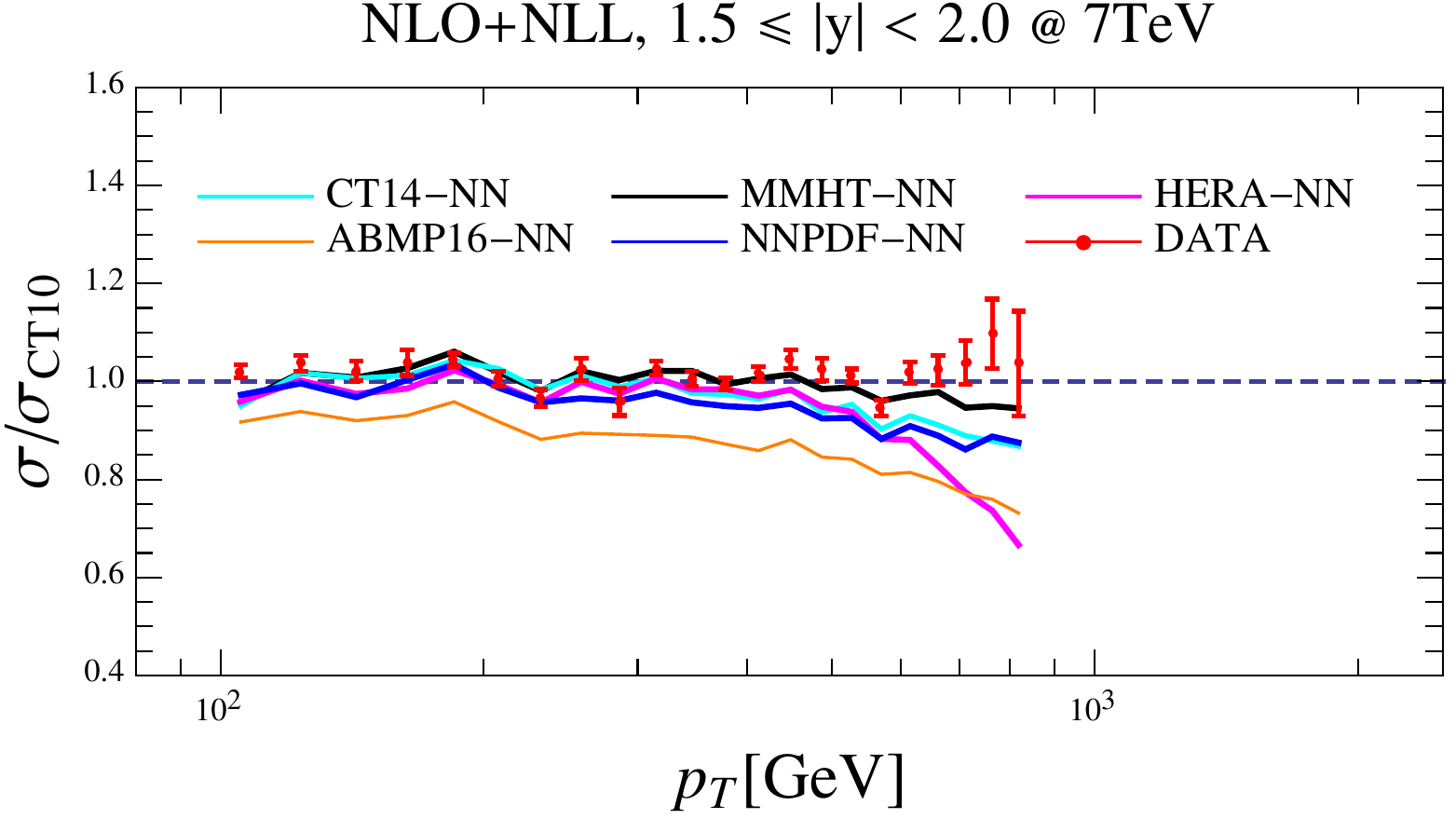}
     \end{subfigure}
 \caption{
   Same as Fig.~\ref{fig:7tevpdfscan-sets1} with the NNLO variant taken for all PDF sets 
   ABMP16~\cite{Alekhin:2017kpj}, CT14~\cite{Dulat:2015mca},
   HERAPDF2.0~\cite{Abramowicz:2015mha}, MMHT2014~\cite{Harland-Lang:2014zoa} and 
   NNPDF3.1~\cite{Ball:2017nwa}.
 }    
 \label{fig:7tevpdfscan-nn}
 \end{figure*}

The NLO $+$ NLL studies are presented 
in Figs.~\ref{fig:7tevpdfscan-sets1} and~\ref{fig:7tevpdfscan-sets2},
and the NLO ones 
in Figs.~\ref{fig:7tevpdfscan-nlo-sets1} and~\ref{fig:7tevpdfscan-nlo-sets2}, respectively. 
To maintain consistency, the NLO variants for all PDFs are used here,
the value of the strong coupling $\alpha_s(M_Z)$ is taken as provided by the respective PDF sets and the predictions are normalized to the one with CT10 PDFs at NLO to
allow for comparisons with the CMS analysis~\cite{Chatrchyan:2014gia}.
In Figs.~\ref{fig:7tevpdfscan-sets1} and~\ref{fig:7tevpdfscan-sets2} we see 
that the predictions based on the NLL joint resummation achieve excellent agreement
with the experimental data for the PDF sets CT14, MMHT2014 and NNPDF3.1, 
while the PDF set ABMP16 slightly undershoots the data for large values of $p_T$.
Only in the high-$p_T$ region of the bin with rapidities $1.5 \le |y| < 2.0$, the
data are in excess of the theoretical predictions, but the discrepancy is still 
within the experimental error, see Fig.~\ref{fig:7tevnll}.
The predictions with HERAPDF2.0 show a somewhat different trend.
They are lower than the data in most of the $p_T$ region and only tend to agree with the data in
the high-$p_T$ regime around $p_T \simeq 1$~TeV. 
To illustrate the robustness of the observations further, we also plot the uncertainties
for the NNPDF3.1 PDFs, which turn out to be quite small 
compared with the theoretical error from the scale uncertainty, see again Fig.~\ref{fig:7tevnll}.
The uncertainties for other PDFs are of similar size.

The situation deteriorates substantially when only the NLO corrections are taken into account. 
For $\sigma_{\rm NLO}$ at NLO, as shown in Figs.~\ref{fig:7tevpdfscan-nlo-sets1} and~\ref{fig:7tevpdfscan-nlo-sets2}, 
none of the PDFs do a good job in describing the CMS inclusive jet data.
The predictions with ABMP16, CT14, MMHT2014 and NNPDF3.1 are all higher than
the data in all rapidity bins in most of the jet $p_T$ regions. For $p_T \gtrsim 800$~GeV 
those predictions tend to agree with or slightly undershoot the data for rapidities $|y|< 1.5$.
For HERAPDF2.0 the NLO predictions are within the experimental errors of the inclusive jet data,
but the shape of the $p_T$ dependence of the cross section differs slightly from the one of the data. Like in the NLO $+$ NLL case above, we also display the uncertainties for the
NNPDF3.1 PDF sets in Figs.~\ref{fig:7tevpdfscan-nlo-sets1} and~\ref{fig:7tevpdfscan-nlo-sets2}.
As the PDF errors are found to be small, the theory predictions 
are rather stable against the uncertainties of current PDFs from global fits 
and the strong tension between the NLO theory and the data persists. 
We also note, that recent studies~\cite{Nocera:2017zge,Harland-Lang:2017ytb} found 
it to be impossible to re-constrain the PDFs within a global analysis including inclusive jet
data from the LHC when all current cross-correlations among different rapidity
bins are consistently taken into account.

Fig.~\ref{fig:7tevpdfscan-nn} shows again the comparison of the NLO $+$ NLL
calculations with the CMS data at $\sqrt{S}=7$ TeV but now using the NNLO
variants of the PDF sets under study.
This choice is reasonable to a certain extent, since 
the NLL resummation includes a dominant part of the full NNLO contributions. 
On the other hand, potentially large NNLO corrections, 
for instance possible large corrections from the complete two-loop virtual 
corrections are still missing in $\sigma_{\rm NLO + NLL}$.
We can see from Fig.~\ref{fig:7tevpdfscan-nn} that the NNLO variants of CT14, 
MMHT2014, NNPDF3.1 get slightly shifted, but are well consistent with the CMS data, 
again except for the highest values of $p_T$ in the rapidity bin $1.5 \le |y| < 2.0$.
In contrast, the predictions with the NNLO variant of HERAPDF2.0 are
significantly shifted compared to the NLO one,
cf. Fig.~\ref{fig:7tevpdfscan-sets1}, and display now also good consistency
with the CMS data. 
The NNLO variant of the AMBP16 PDFs predicts the correct shape, 
but it is lower than the data for all rapidity bins as a consequence 
of the lower value of $\alpha_s(M_Z)=0.1147$ compared to  
$\alpha_s(M_Z)=0.1180$ used by CT14, HERAPDF2.0, MMHT2014 or NNPDF3.1.
This sensitivity to $\alpha_s(M_Z)$ confirms again the great potential 
of inclusive jet cross section data for the determination of the strong
coupling constant~\cite{Britzger:2017maj}.

\begin{figure*}
 \begin{subfigure}{.5\textwidth}
  \includegraphics[width=0.95\linewidth]{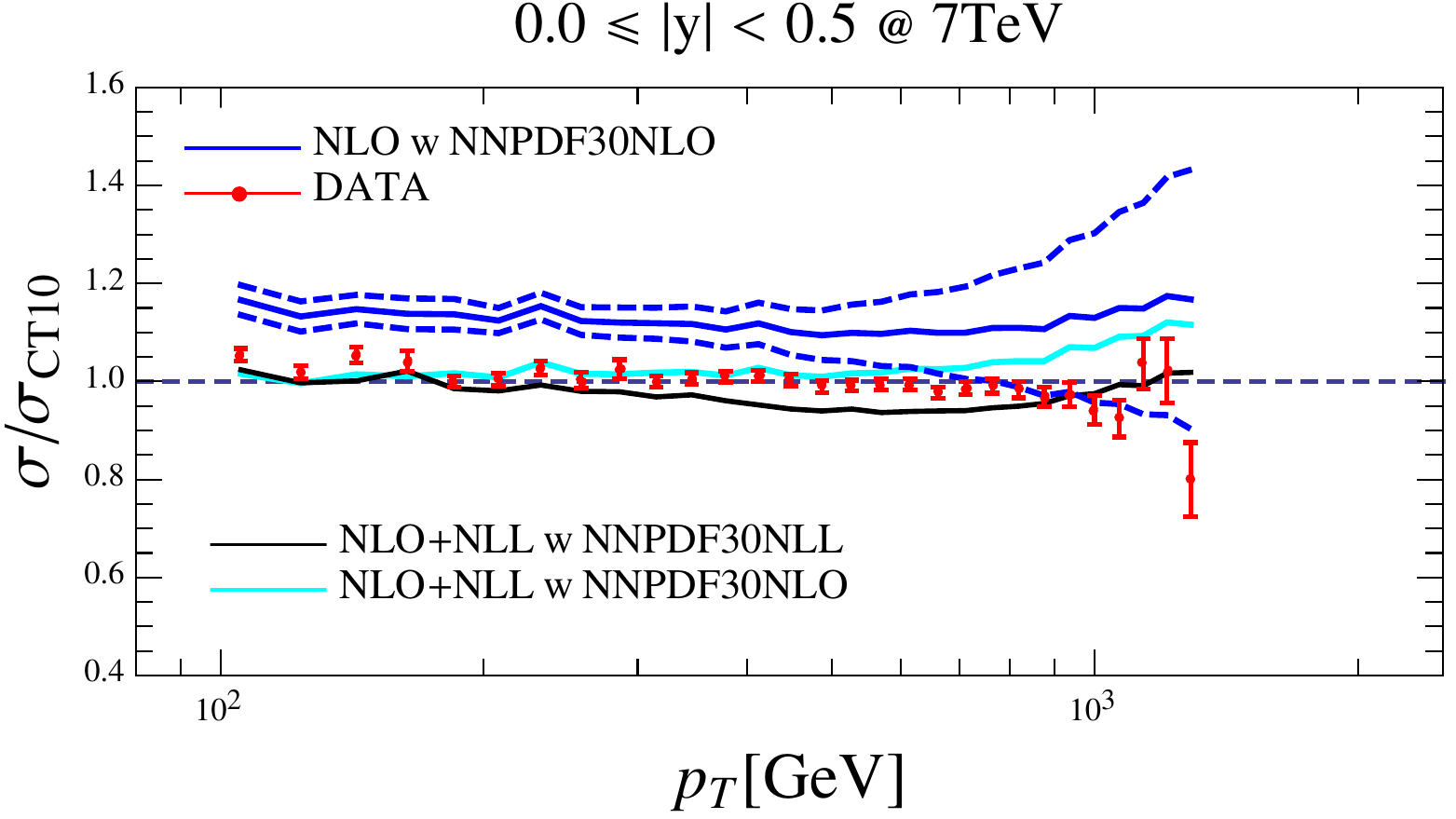}
 \end{subfigure}%
 \begin{subfigure}{.5\textwidth}
  \includegraphics[width=0.95\linewidth]{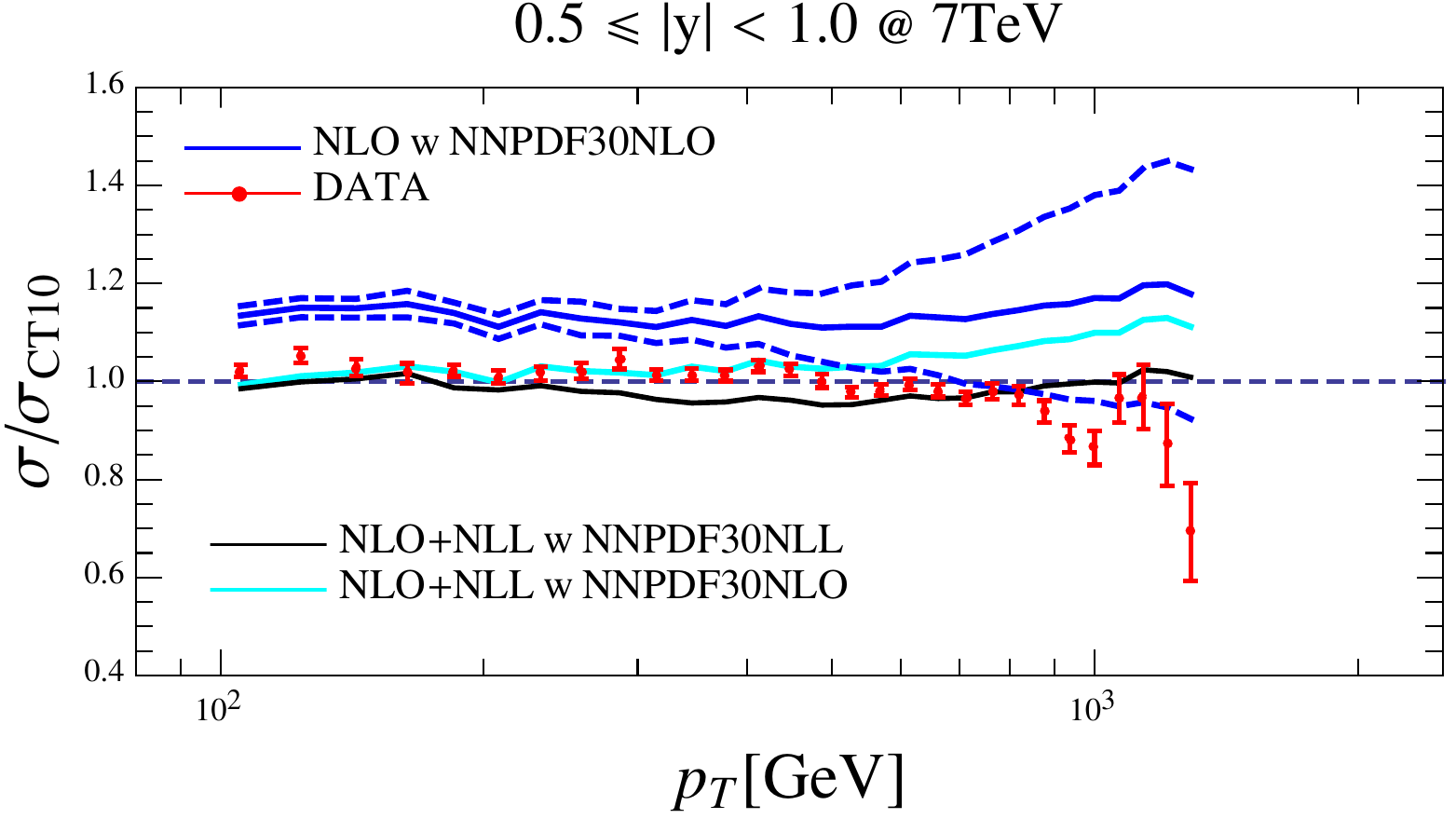}
  \end{subfigure}
   \begin{subfigure}{.5\textwidth}
  \includegraphics[width=0.95\linewidth]{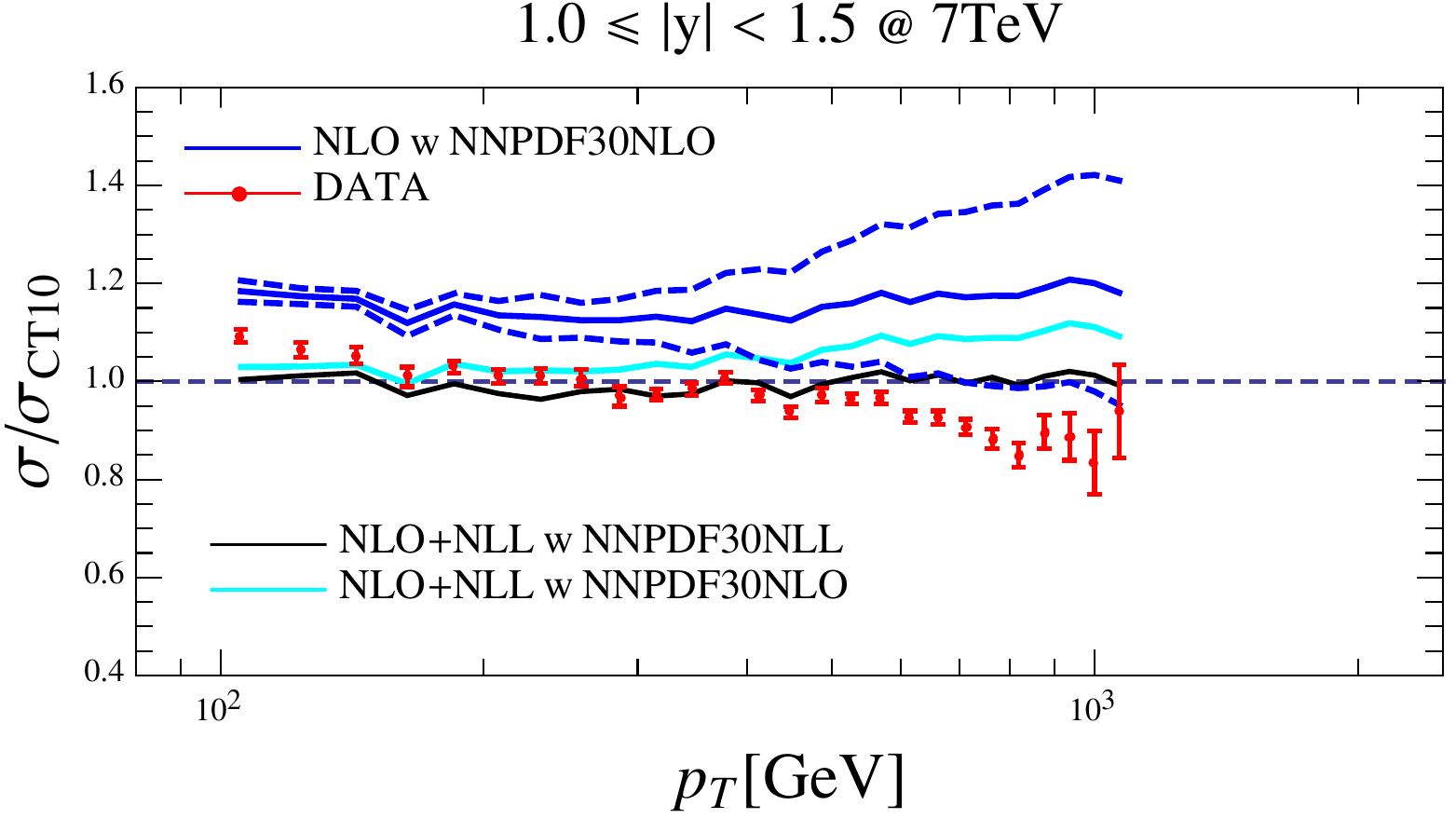}
   \end{subfigure}%
    \begin{subfigure}{.5\textwidth}
  \includegraphics[width=0.95\linewidth]{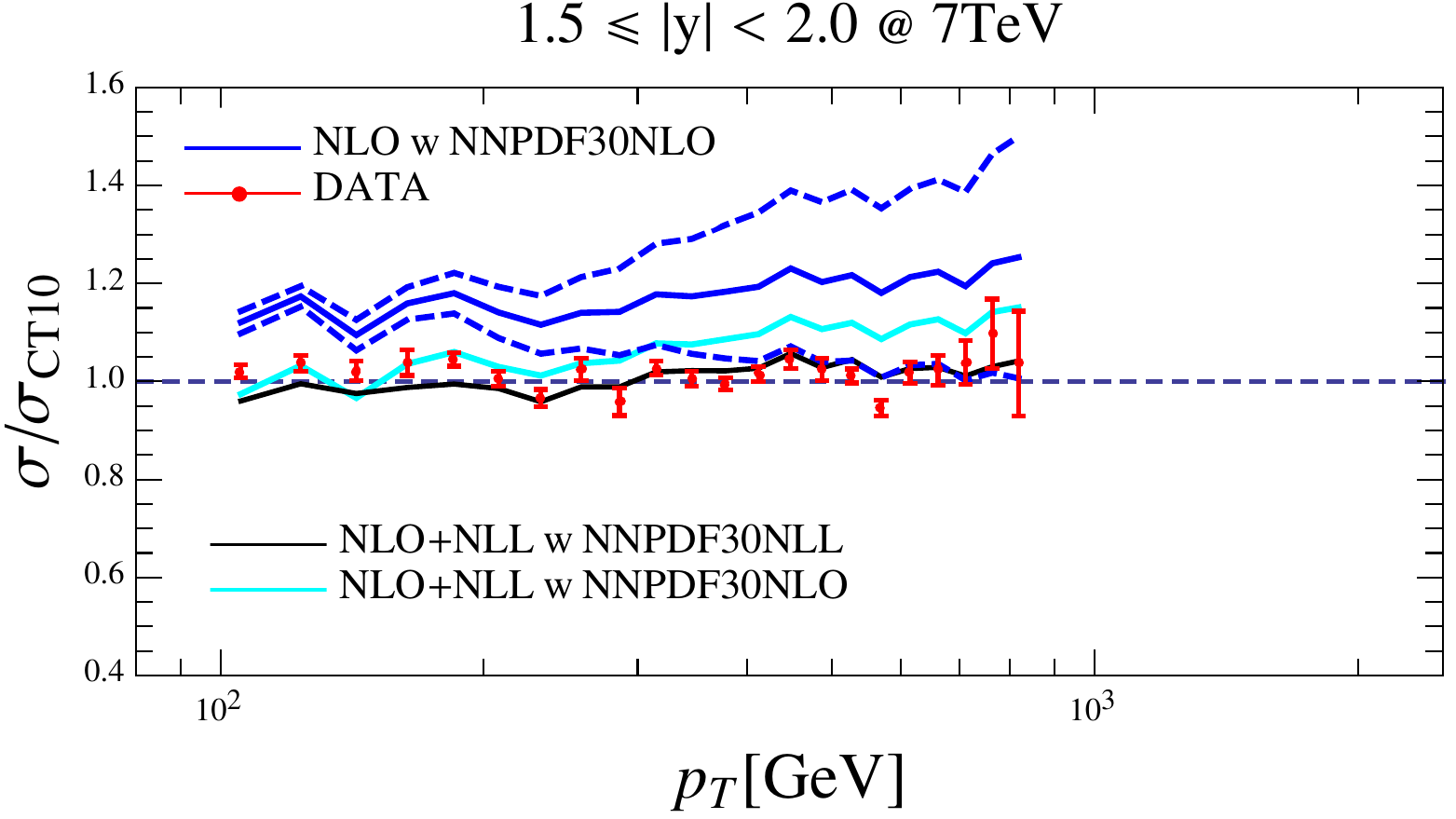}
     \end{subfigure}
 \caption{
   Same as in Figs.~\ref{fig:7tevpdfscan-sets1} and~\ref{fig:7tevpdfscan-nlo-sets1}
   for the cross sections $\sigma_{\rm NLO + NLL}$ and $\sigma_{\rm NLO}$ 
   using the PDF sets NNPDF3.0 at NLO without and with
   resummation~\cite{Bonvini:2015ira}, labeled as 
   {\tt NNPDF30NLO} and {\tt NNPDF30NLL} in the plots, respectively.
   The dashed lines (blue) indicate the PDF uncertainties for $\sigma_{\rm NLO}$ with the set {\tt NNPDF30NLO}.
 }
 \label{fig:7tevpdfscan-nl}
 \end{figure*}

Finally, in Fig.~\ref{fig:7tevpdfscan-nl}, we display the results with the
PDFs of~\cite{Bonvini:2015ira} which have been extracted from data for 
the DY process, DIS and top-quark hadro-production within the NNPDF framework.
These PDFs are subject to improvements at large-$x$, 
since the theory predictions for DIS and DY as well as for top-quark
hadro-production include threshold resummation.
For the PDF variant without threshold resummation (labeled as {\tt NNPDF30NLO}
in Fig.~\ref{fig:7tevpdfscan-nl}) both cross sections at NLO $+$ NLL and NLO
accuracy, $\sigma_{\rm NLO + NLL}$ and $\sigma_{\rm NLO}$, respectively, are shown.
In addition to that, the NLO $+$ NLL results $\sigma_{\rm NLO + NLL}$ 
for the PDF variant with threshold resummation (labeled as {\tt NNPDF30NLL}
in Fig.~\ref{fig:7tevpdfscan-nl}) are presented as well.
Overall, the NLO $+$ NLL predictions exhibit better agreement with the data 
compared to the NLO results, although the PDFs uncertainties of~\cite{Bonvini:2015ira}
are substantially larger than the ones of global fits.
Those large PDF uncertainties at large-$x$ and relevant scales of 
$p_T \simeq \mu \simeq 0.5 \dots 1$~TeV originate from the gluon PDF at $x \gtrsim 0.1$ 
and the light flavor PDFs at lower $x$ through the standard parton evolution.
The findings in Fig.~\ref{fig:7tevpdfscan-nl} underpin the necessity to carefully examine and analyze data 
which constrain those PDFs, including the need to delineate resummation effects from
power corrections in the kinematic regions.  
It will be interesting to observe to what extend improvements can be made 
in the future in extractions of PDFs with threshold resummation 
when the inclusive jet data are included.

\section{Summary and Conclusions \label{sec:conc}}

In this manuscript, we have provided a detailed study of pQCD calculations from first principles 
for cross sections of single-inclusive jet production at the LHC.
We have performed comprehensive comparisons between the fixed NLO results 
and the NLL threshold and small-$R$ joint resummation improved calculations obtained recently, 
and we have achieved remarkable advances in perturbative predictions upon
using the latter.
In our studies, significant differences between the NLO and the NLO $+$ NLL joint resummation 
predictions have been observed in the kinematic regions of interest for the LHC analyses
and we have found that these differences account for the discrepancy between the NLO predictions 
and the LHC data for the jet $p_T$ spectrum in various
rapidity bins collected by the CMS experiment at various center-of-mass energies.
Once the joint resummation has been included, 
a remarkable agreement was found between the QCD theory predictions and the
LHC data in a large range of jet rapidities. 

We have illustrated the impact of the joint resummation in a study of the jet
radius ratios $D_R$ at both, $\sqrt{S} = 7$ and $13$~TeV, which have the advantage of
being largely independent of the PDFs and other residual theory uncertainties.
At $\sqrt{S} = 7$~TeV these jet radius ratios between $R = 0.5$ and $R = 0.7$, 
i.e., $D_{0.7}$, 
have been compared with a CMS analysis in Fig.~\ref{fig:Rscan-ratio-7tev} 
and overall we have found a significant improvement in the theoretical
description of those data.
While the predicted double-differential cross sections in $p_T$ and $y$ at NLO
in pQCD are, for a given value of $R$, systematically higher than the central values of those LHC data 
in all rapidity bins, arguably they still agree within the theoretical and experimental uncertainties.
For the jet radius ratios $D_R$, however, such consistency is definitely not
the case due to the much reduced experimental uncertainties.
The NLO predictions for $D_{0.7}$ completely miss those LHC data and also 
cannot be changed by considering different PDF sets since those effects largely cancel
out in the jet radius ratios $D_R$. 
Therefore, we conclude that the NLO $+$ NLL joint resummation 
is a crucial ingredient in order to achieve a good description of the $\sqrt{S} = 7$~TeV jet data within pQCD.
We have also presented predictions for the jet radius ratios $D_R$ 
at $\sqrt{S} = 13$~TeV in Fig.~\ref{fig:Rscan-ratio-13tev}. 
using different jet radii with a jet $p_T$ up to $2 \> {\rm TeV}$. 
These results will be useful for future experimental analyses of inclusive jet data.  

Due to the great importance of the inclusive jet data for constraints on PDFs
and determinations of the strong coupling $\alpha_s(M_Z)$ 
we have also investigated in this study the impact of different PDF sets on the
theoretical predictions.
We have found that the NLO $+$ NLL predictions at $\sqrt{S} = 7$~TeV 
based on the NLO variants of the PDF sets ABMP16, CT14, MMHT2014 and NNPDF3.1 
or on the NNLO variant of HERAPDF2.0, respectively,  
describe the $p_T$ distributions remarkably well for the various rapidity bins.
On the other hand, the tension with the CMS inclusive jet data and 
the pure NLO predictions $\sigma_{\rm NLO}$ persists for all those PDF sets 
and cannot be removed or relieved by selecting a specific PDF set. 
Therefore, global PDFs which also fit inclusive jet data from the LHC 
need to be based on theory predictions using the joint resummation 
for the single-inclusive jet production in order to avoid a possible bias 
in the PDF extraction due to missing large logarithms in the hard cross
sections beyond NLO.
We have also noticed in our studies that PDFs extracted with account of 
threshold resummation but without inclusive jet data have significantly increased PDF uncertainties 
compared to the ones from the global fits. We suggest to use the joint resummed 
calculations of the present article in the on-going efforts to refine those PDFs.

Finally, we note that although the NLO $+$ NLL calculations greatly improve
the theoretical predictions, the associated scale uncertainties 
are still large and comparable with the current experimental errors. 
Therefore, in future studies it will be necessary to go beyond 
the currently achieved accuracy by matching the resummed results 
with the now available inclusive jet calculations at NNLO 
and by improving the logarithmic accuracy of the joint threshold and small-$R$ joint resummation to NNLL.
Both these tasks are feasible within the present framework for resummation
and will be subject of future work.  

\acknowledgments 
We would like to thank Jun Gao and Katerina Lipka 
for helpful discussions and Marco Bonvini for providing the PDF set of~\cite{Bonvini:2015ira}.
X.L. would like to thank Zhong-bo Kang and UCLA department of Physics and Astronomy for the hospitality. 
X.L. is supported by the National Natural Science Foundation of China under
Grants No. 11775023 and the Fundamental Research Funds for the Central Universities. 
S.M. acknowledges contract 05H15GUCC1 by BMBF. 
F.R. is supported by the U.S. Department of Energy under Contract No. DE-AC02-05CH11231 and by the LDRD Program of Lawrence Berkeley National Laboratory. This research used resources of the Argonne Leadership Computing Facility, which is a DOE Office of Science User Facility supported under Contract DE-AC02-06CH11357.

\appendix
\renewcommand{\theequation}{\ref{sec:appA}.\arabic{equation}}
\setcounter{equation}{0}
\renewcommand{\thetable}{\ref{sec:appA}.\arabic{table}}
\setcounter{table}{0}
\section{Cross sections at $\sqrt{S}=13$~TeV}
\label{sec:appA}

We present the cross sections for the LHC in Tabs.~\ref{tab:cms13-ct14-r04-nll}--\ref{tab:cms13-ct14-r07},
respectively, including the theory uncertainty arising from scale variations.
At NLO the scale uncertainties have been obtained from the envelope of the variation 
around $\mu_R=\mu_F=p_T^{\text{max}}$ up and down by a factor of two, 
while the scale uncertainties at NLO $+$ NLL have been computed as described
in section~\ref{sec:th}.
The values for the cross sections contain an additional error of 
${\cal O}(1.5\%)$ not shown explicitly from the numerical integration of the NLO corrections, 
which dominates both and is correlated between $\sigma_{\rm NLO}$ and $\sigma_{\rm NLO + NLL}$.
The PDF sets used and other parameters are given in the table captions.
We also note that for the small $p_T$ bins ($p_T \lesssim 200$~GeV) 
at $\sqrt{S}=13$~TeV, the threshold resummation may not be applicable anymore.

\onecolumngrid

\begin{table}[h!]
\renewcommand{\arraystretch}{1.3}
\begin{center}
\begin{tabular}{|r|r|r|r|r|r|r|}
  \hline
  \multicolumn{7}{l}{$\mathbf{\sigma_{\rm NLO + NLL}}$ \bf{for LHC at} $\mathbf{\sqrt{S}=13~\mbox{\bf TeV,}}$ $\mathbf{R=0.4}$}\\ 
  \hline
  \hline
\multicolumn{2}{|c|}{$p_T$ [GeV]} &
\multicolumn{5}{|c|}{$d\sigma/dp_T/d|y|$ [pb] \quad ($\pm \Delta \sigma_{\rm th.~scale}~\mbox{in}~\%) $} \\
\multicolumn{1}{|c|}{$p_T^{\min}$} &
\multicolumn{1}{|c|}{$p_T^{\max}$} &
\multicolumn{1}{|c|}{$0 \le |y| < 0.5$} &
\multicolumn{1}{|c|}{$0.5 \le |y| < 1.0$} &
\multicolumn{1}{|c|}{$1.0 \le |y| < 1.5$} &
\multicolumn{1}{|c|}{$1.5 \le |y| < 2.0$} &
\multicolumn{1}{|c|}{$2.0 \le |y| < 2.5$} \\
\hline
  56 &   74 & 1.20$\,\cdot\, 10^{+5}$  ($\pm\!\!$ 15.06\%) & 1.17$\,\cdot\, 10^{+5}$  ($\pm\!\!$ 13.72\%) & 1.07$\,\cdot\, 10^{+5}$  ($\pm\!\!$ 14.10\%) & 8.99$\,\cdot\, 10^{+4}$  ($\pm\!\!$ 19.41\%) & 7.25$\,\cdot\, 10^{+4}$  ($\pm\!\!$ 25.95 \%)\\
  74 &   97 & 3.49$\,\cdot\, 10^{+4}$  ($\pm\!\!$ 10.61\%) & 3.28$\,\cdot\, 10^{+4}$  ($\pm\!\!$ 11.83\%) & 2.95$\,\cdot\, 10^{+4}$  ($\pm\!\!$ 12.10\%) & 2.47$\,\cdot\, 10^{+4}$  ($\pm\!\!$ 15.22\%) & 1.93$\,\cdot\, 10^{+4}$  ($\pm\!\!$ 21.42 \%)\\
  97 &  133 & 8.76$\,\cdot\, 10^{+3}$  ($\pm\!\!$  9.53\%) & 8.23$\,\cdot\, 10^{+3}$  ($\pm\!\!$ 10.41\%) & 7.31$\,\cdot\, 10^{+3}$  ($\pm\!\!$ 10.91\%) & 6.04$\,\cdot\, 10^{+3}$  ($\pm\!\!$ 12.44\%) & 4.71$\,\cdot\, 10^{+3}$  ($\pm\!\!$ 17.60 \%)\\
 133 &  174 & 2.05$\,\cdot\, 10^{+3}$  ($\pm\!\!$  8.50\%) & 1.93$\,\cdot\, 10^{+3}$  ($\pm\!\!$  8.34\%) & 1.69$\,\cdot\, 10^{+3}$  ($\pm\!\!$  8.24\%) & 1.37$\,\cdot\, 10^{+3}$  ($\pm\!\!$  9.75\%) & 1.02$\,\cdot\, 10^{+3}$  ($\pm\!\!$ 15.06 \%)\\
 174 &  220 & 5.66$\,\cdot\, 10^{+2}$  ($\pm\!\!$  7.96\%) & 5.27$\,\cdot\, 10^{+2}$  ($\pm\!\!$  8.06\%) & 4.55$\,\cdot\, 10^{+2}$  ($\pm\!\!$  8.41\%) & 3.63$\,\cdot\, 10^{+2}$  ($\pm\!\!$  8.38\%) & 2.60$\,\cdot\, 10^{+2}$  ($\pm\!\!$ 13.37 \%)\\
 220 &  272 & 1.75$\,\cdot\, 10^{+2}$  ($\pm\!\!$  7.24\%) & 1.63$\,\cdot\, 10^{+2}$  ($\pm\!\!$  7.04\%) & 1.39$\,\cdot\, 10^{+2}$  ($\pm\!\!$  7.10\%) & 1.08$\,\cdot\, 10^{+2}$  ($\pm\!\!$  7.46\%) & 7.26$\,\cdot\, 10^{+1}$  ($\pm\!\!$ 12.22 \%)\\
 272 &  330 & 5.91$\,\cdot\, 10^{+1}$  ($\pm\!\!$  6.86\%) & 5.42$\,\cdot\, 10^{+1}$  ($\pm\!\!$  6.60\%) & 4.57$\,\cdot\, 10^{+1}$  ($\pm\!\!$  6.22\%) & 3.43$\,\cdot\, 10^{+1}$  ($\pm\!\!$  7.08\%) & 2.14$\,\cdot\, 10^{+1}$  ($\pm\!\!$ 11.40 \%)\\
 330 &  395 & 2.11$\,\cdot\, 10^{+1}$  ($\pm\!\!$  7.07\%) & 1.93$\,\cdot\, 10^{+1}$  ($\pm\!\!$  6.25\%) & 1.59$\,\cdot\, 10^{+1}$  ($\pm\!\!$  5.80\%) & 1.15$\,\cdot\, 10^{+1}$  ($\pm\!\!$  6.65\%) & 6.42$\,\cdot\, 10^{+0}$  ($\pm\!\!$ 10.81 \%)\\
 395 &  468 & 7.91$\,\cdot\, 10^{+0}$  ($\pm\!\!$  6.93\%) & 7.16$\,\cdot\, 10^{+0}$  ($\pm\!\!$  6.10\%) & 5.79$\,\cdot\, 10^{+0}$  ($\pm\!\!$  5.18\%) & 3.95$\,\cdot\, 10^{+0}$  ($\pm\!\!$  6.48\%) & 1.90$\,\cdot\, 10^{+0}$  ($\pm\!\!$ 10.41 \%)\\
 468 &  548 & 3.07$\,\cdot\, 10^{+0}$  ($\pm\!\!$  7.22\%) & 2.75$\,\cdot\, 10^{+0}$  ($\pm\!\!$  6.00\%) & 2.17$\,\cdot\, 10^{+0}$  ($\pm\!\!$  4.45\%) & 1.37$\,\cdot\, 10^{+0}$  ($\pm\!\!$  6.35\%) & 5.44$\,\cdot\, 10^{-1}$  ($\pm\!\!$ 10.00 \%)\\
 548 &  638 & 1.23$\,\cdot\, 10^{+0}$  ($\pm\!\!$  7.03\%) & 1.09$\,\cdot\, 10^{+0}$  ($\pm\!\!$  5.96\%) & 8.28$\,\cdot\, 10^{-1}$  ($\pm\!\!$  3.99\%) & 4.78$\,\cdot\, 10^{-1}$  ($\pm\!\!$  6.29\%) & 1.48$\,\cdot\, 10^{-1}$  ($\pm\!\!$  9.61 \%)\\
 638 &  737 & 4.98$\,\cdot\, 10^{-1}$  ($\pm\!\!$  7.02\%) & 4.36$\,\cdot\, 10^{-1}$  ($\pm\!\!$  5.73\%) & 3.17$\,\cdot\, 10^{-1}$  ($\pm\!\!$  3.51\%) & 1.62$\,\cdot\, 10^{-1}$  ($\pm\!\!$  6.31\%) & 3.67$\,\cdot\, 10^{-2}$  ($\pm\!\!$  9.15 \%)\\
 737 &  846 & 2.07$\,\cdot\, 10^{-1}$  ($\pm\!\!$  6.86\%) & 1.77$\,\cdot\, 10^{-1}$  ($\pm\!\!$  5.49\%) & 1.22$\,\cdot\, 10^{-1}$  ($\pm\!\!$  3.35\%) & 5.32$\,\cdot\, 10^{-2}$  ($\pm\!\!$  6.33\%) & 8.05$\,\cdot\, 10^{-3}$  ($\pm\!\!$  8.77 \%)\\
 846 &  967 & 8.63$\,\cdot\, 10^{-2}$  ($\pm\!\!$  6.93\%) & 7.23$\,\cdot\, 10^{-2}$  ($\pm\!\!$  5.12\%) & 4.61$\,\cdot\, 10^{-2}$  ($\pm\!\!$  3.45\%) & 1.66$\,\cdot\, 10^{-2}$  ($\pm\!\!$  6.34\%) & 1.50$\,\cdot\, 10^{-3}$  ($\pm\!\!$  8.41 \%)\\
 967 & 1101 & 3.58$\,\cdot\, 10^{-2}$  ($\pm\!\!$  6.73\%) & 2.92$\,\cdot\, 10^{-2}$  ($\pm\!\!$  4.81\%) & 1.70$\,\cdot\, 10^{-2}$  ($\pm\!\!$  3.59\%) & 4.79$\,\cdot\, 10^{-3}$  ($\pm\!\!$  6.34\%) & 2.19$\,\cdot\, 10^{-4}$  ($\pm\!\!$  8.04 \%)\\
1101 & 1248 & 1.49$\,\cdot\, 10^{-2}$  ($\pm\!\!$  6.69\%) & 1.17$\,\cdot\, 10^{-2}$  ($\pm\!\!$  4.58\%) & 6.05$\,\cdot\, 10^{-3}$  ($\pm\!\!$  3.84\%) & 1.25$\,\cdot\, 10^{-3}$  ($\pm\!\!$  6.39\%) & 2.21$\,\cdot\, 10^{-5}$  ($\pm\!\!$  8.57 \%)\\
1248 & 1410 & 6.09$\,\cdot\, 10^{-3}$  ($\pm\!\!$  6.53\%) & 4.57$\,\cdot\, 10^{-3}$  ($\pm\!\!$  4.23\%) & 2.05$\,\cdot\, 10^{-3}$  ($\pm\!\!$  4.03\%) & 2.86$\,\cdot\, 10^{-4}$  ($\pm\!\!$  6.42\%) & 1.18$\,\cdot\, 10^{-6}$  ($\pm\!\!$  9.93 \%)\\
1410 & 1588 & 2.45$\,\cdot\, 10^{-3}$  ($\pm\!\!$  6.35\%) & 1.73$\,\cdot\, 10^{-3}$  ($\pm\!\!$  3.84\%) & 6.49$\,\cdot\, 10^{-4}$  ($\pm\!\!$  4.22\%) & 5.50$\,\cdot\, 10^{-5}$  ($\pm\!\!$  6.48\%) & 1.75$\,\cdot\, 10^{-8}$  ($\pm\!\!$ 12.79 \%)\\
1588 & 1784 & 9.60$\,\cdot\, 10^{-4}$  ($\pm\!\!$  6.23\%) & 6.30$\,\cdot\, 10^{-4}$  ($\pm\!\!$  3.50\%) & 1.88$\,\cdot\, 10^{-4}$  ($\pm\!\!$  4.41\%) & 8.30$\,\cdot\, 10^{-6}$  ($\pm\!\!$  6.55\%) & 5.45$\,\cdot\, 10^{-12}$  ($\pm\!\!$ 23.09 \%)\\
1784 & 2000 & 3.60$\,\cdot\, 10^{-4}$  ($\pm\!\!$  6.03\%) & 2.16$\,\cdot\, 10^{-4}$  ($\pm\!\!$  3.16\%) & 4.85$\,\cdot\, 10^{-5}$  ($\pm\!\!$  4.61\%) & 8.57$\,\cdot\, 10^{-7}$  ($\pm\!\!$  6.64\%) & -----\\
2000 & 2238 & 1.28$\,\cdot\, 10^{-4}$  ($\pm\!\!$  5.89\%) & 6.84$\,\cdot\, 10^{-5}$  ($\pm\!\!$  2.86\%) & 1.07$\,\cdot\, 10^{-5}$  ($\pm\!\!$  4.81\%) & 4.69$\,\cdot\, 10^{-8}$  ($\pm\!\!$  7.07\%) & -----\\
2238 & 2500 & 4.25$\,\cdot\, 10^{-5}$  ($\pm\!\!$  5.70\%) & 1.96$\,\cdot\, 10^{-5}$  ($\pm\!\!$  2.59\%) & 1.91$\,\cdot\, 10^{-6}$  ($\pm\!\!$  5.02\%) & 7.58$\,\cdot\, 10^{-10}$  ($\pm\!\!$  8.68\%) & -----\\
2500 & 2787 & 1.29$\,\cdot\, 10^{-5}$  ($\pm\!\!$  5.49\%) & 4.91$\,\cdot\, 10^{-6}$  ($\pm\!\!$  2.79\%) & 2.57$\,\cdot\, 10^{-7}$  ($\pm\!\!$  5.23\%) & 4.71$\,\cdot\, 10^{-13}$  ($\pm\!\!$ 12.81\%) & -----\\
2787 & 3103 & 3.51$\,\cdot\, 10^{-6}$  ($\pm\!\!$  5.22\%) & 1.04$\,\cdot\, 10^{-6}$  ($\pm\!\!$  3.03\%) & 2.25$\,\cdot\, 10^{-8}$  ($\pm\!\!$  5.47\%) & ----- & -----\\
\hline
\end{tabular}
\caption{\label{tab:cms13-ct14-r04-nll}
   The double-differential cross sections $\sigma_{\rm NLO + NLL}$ 
   in bins of $p_T$ and $y$ at $\sqrt{S}=13$~TeV 
   with $R= 0.4$ using the CT14~\cite{Dulat:2015mca} PDFs at NLO.
   Theoretical uncertainties from the scale variation are given in
   parenthesis.
 }
\end{center}
\end{table}

\begin{table}[h!]
\renewcommand{\arraystretch}{1.3}
\begin{center}
\begin{tabular}{|r|r|r|r|r|r|r|}
  \hline
  \multicolumn{7}{l}{$\mathbf{\sigma_{\rm NLO}}$ \bf{for LHC at} $\mathbf{\sqrt{S}=13~\mbox{\bf TeV,}}$ $\mathbf{R=0.4}$}\\ 
  \hline
  \hline
\multicolumn{2}{|c|}{$p_T$ [GeV]} &
\multicolumn{5}{|c|}{$d\sigma/dp_T/d|y|$ [pb] \quad ($\pm \Delta \sigma_{\rm th.~scale}~\mbox{in}~\%) $} \\
\multicolumn{1}{|c|}{$p_T^{\min}$} &
\multicolumn{1}{|c|}{$p_T^{\max}$} &
\multicolumn{1}{|c|}{$0 \le |y| < 0.5$} &
\multicolumn{1}{|c|}{$0.5 \le |y| < 1.0$} &
\multicolumn{1}{|c|}{$1.0 \le |y| < 1.5$} &
\multicolumn{1}{|c|}{$1.5 \le |y| < 2.0$} &
\multicolumn{1}{|c|}{$2.0 \le |y| < 2.5$} \\
\hline
  56 &   74 & 1.52$\,\cdot\, 10^{+5}$  ($\pm\!\!$  2.60\%) & 1.48$\,\cdot\, 10^{+5}$  ($\pm\!\!$  3.96\%) & 1.36$\,\cdot\, 10^{+5}$  ($\pm\!\!$  5.17\%) & 1.17$\,\cdot\, 10^{+5}$  ($\pm\!\!$  4.59\%) & 9.57$\,\cdot\, 10^{+4}$  ($\pm\!\!$  7.17 \%)\\
  74 &   97 & 4.28$\,\cdot\, 10^{+4}$  ($\pm\!\!$  6.03\%) & 4.05$\,\cdot\, 10^{+4}$  ($\pm\!\!$  3.83\%) & 3.66$\,\cdot\, 10^{+4}$  ($\pm\!\!$  5.55\%) & 3.11$\,\cdot\, 10^{+4}$  ($\pm\!\!$  5.34\%) & 2.47$\,\cdot\, 10^{+4}$  ($\pm\!\!$  5.42 \%)\\
  97 &  133 & 1.05$\,\cdot\, 10^{+4}$  ($\pm\!\!$  4.66\%) & 9.94$\,\cdot\, 10^{+3}$  ($\pm\!\!$  4.44\%) & 8.87$\,\cdot\, 10^{+3}$  ($\pm\!\!$  5.48\%) & 7.38$\,\cdot\, 10^{+3}$  ($\pm\!\!$  5.14\%) & 5.79$\,\cdot\, 10^{+3}$  ($\pm\!\!$  5.85 \%)\\
 133 &  174 & 2.42$\,\cdot\, 10^{+3}$  ($\pm\!\!$  5.02\%) & 2.28$\,\cdot\, 10^{+3}$  ($\pm\!\!$  5.83\%) & 2.00$\,\cdot\, 10^{+3}$  ($\pm\!\!$  5.45\%) & 1.63$\,\cdot\, 10^{+3}$  ($\pm\!\!$  5.80\%) & 1.22$\,\cdot\, 10^{+3}$  ($\pm\!\!$  6.05 \%)\\
 174 &  220 & 6.59$\,\cdot\, 10^{+2}$  ($\pm\!\!$  5.65\%) & 6.15$\,\cdot\, 10^{+2}$  ($\pm\!\!$  5.93\%) & 5.32$\,\cdot\, 10^{+2}$  ($\pm\!\!$  4.82\%) & 4.24$\,\cdot\, 10^{+2}$  ($\pm\!\!$  5.31\%) & 3.05$\,\cdot\, 10^{+2}$  ($\pm\!\!$  7.15 \%)\\
 220 &  272 & 2.02$\,\cdot\, 10^{+2}$  ($\pm\!\!$  5.26\%) & 1.88$\,\cdot\, 10^{+2}$  ($\pm\!\!$  5.25\%) & 1.60$\,\cdot\, 10^{+2}$  ($\pm\!\!$  4.75\%) & 1.24$\,\cdot\, 10^{+2}$  ($\pm\!\!$  4.24\%) & 8.37$\,\cdot\, 10^{+1}$  ($\pm\!\!$  4.44 \%)\\
 272 &  330 & 6.76$\,\cdot\, 10^{+1}$  ($\pm\!\!$  4.80\%) & 6.20$\,\cdot\, 10^{+1}$  ($\pm\!\!$  5.21\%) & 5.21$\,\cdot\, 10^{+1}$  ($\pm\!\!$  5.07\%) & 3.91$\,\cdot\, 10^{+1}$  ($\pm\!\!$  4.63\%) & 2.43$\,\cdot\, 10^{+1}$  ($\pm\!\!$  5.70 \%)\\
 330 &  395 & 2.40$\,\cdot\, 10^{+1}$  ($\pm\!\!$  4.98\%) & 2.18$\,\cdot\, 10^{+1}$  ($\pm\!\!$  4.35\%) & 1.80$\,\cdot\, 10^{+1}$  ($\pm\!\!$  4.90\%) & 1.30$\,\cdot\, 10^{+1}$  ($\pm\!\!$  4.83\%) & 7.23$\,\cdot\, 10^{+0}$  ($\pm\!\!$  5.90 \%)\\
 395 &  468 & 8.92$\,\cdot\, 10^{+0}$  ($\pm\!\!$  4.87\%) & 8.07$\,\cdot\, 10^{+0}$  ($\pm\!\!$  4.68\%) & 6.50$\,\cdot\, 10^{+0}$  ($\pm\!\!$  3.62\%) & 4.42$\,\cdot\, 10^{+0}$  ($\pm\!\!$  4.66\%) & 2.12$\,\cdot\, 10^{+0}$  ($\pm\!\!$  6.01 \%)\\
 468 &  548 & 3.43$\,\cdot\, 10^{+0}$  ($\pm\!\!$  4.11\%) & 3.08$\,\cdot\, 10^{+0}$  ($\pm\!\!$  4.18\%) & 2.42$\,\cdot\, 10^{+0}$  ($\pm\!\!$  4.23\%) & 1.53$\,\cdot\, 10^{+0}$  ($\pm\!\!$  4.44\%) & 6.05$\,\cdot\, 10^{-1}$  ($\pm\!\!$  6.08 \%)\\
 548 &  638 & 1.37$\,\cdot\, 10^{+0}$  ($\pm\!\!$  4.23\%) & 1.21$\,\cdot\, 10^{+0}$  ($\pm\!\!$  4.19\%) & 9.19$\,\cdot\, 10^{-1}$  ($\pm\!\!$  4.46\%) & 5.29$\,\cdot\, 10^{-1}$  ($\pm\!\!$  4.77\%) & 1.63$\,\cdot\, 10^{-1}$  ($\pm\!\!$  6.91 \%)\\
 638 &  737 & 5.53$\,\cdot\, 10^{-1}$  ($\pm\!\!$  4.43\%) & 4.83$\,\cdot\, 10^{-1}$  ($\pm\!\!$  4.68\%) & 3.50$\,\cdot\, 10^{-1}$  ($\pm\!\!$  4.83\%) & 1.78$\,\cdot\, 10^{-1}$  ($\pm\!\!$  4.87\%) & 4.04$\,\cdot\, 10^{-2}$  ($\pm\!\!$  8.56 \%)\\
 737 &  846 & 2.28$\,\cdot\, 10^{-1}$  ($\pm\!\!$  5.00\%) & 1.96$\,\cdot\, 10^{-1}$  ($\pm\!\!$  5.02\%) & 1.34$\,\cdot\, 10^{-1}$  ($\pm\!\!$  5.30\%) & 5.84$\,\cdot\, 10^{-2}$  ($\pm\!\!$  5.01\%) & 8.86$\,\cdot\, 10^{-3}$  ($\pm\!\!$  9.05 \%)\\
 846 &  967 & 9.48$\,\cdot\, 10^{-2}$  ($\pm\!\!$  5.16\%) & 7.94$\,\cdot\, 10^{-2}$  ($\pm\!\!$  5.36\%) & 5.06$\,\cdot\, 10^{-2}$  ($\pm\!\!$  5.38\%) & 1.82$\,\cdot\, 10^{-2}$  ($\pm\!\!$  5.06\%) & 1.65$\,\cdot\, 10^{-3}$  ($\pm\!\!$ 11.51 \%)\\
 967 & 1101 & 3.92$\,\cdot\, 10^{-2}$  ($\pm\!\!$  5.36\%) & 3.20$\,\cdot\, 10^{-2}$  ($\pm\!\!$  5.65\%) & 1.86$\,\cdot\, 10^{-2}$  ($\pm\!\!$  5.82\%) & 5.24$\,\cdot\, 10^{-3}$  ($\pm\!\!$  5.36\%) & 2.40$\,\cdot\, 10^{-4}$  ($\pm\!\!$ 15.04 \%)\\
1101 & 1248 & 1.62$\,\cdot\, 10^{-2}$  ($\pm\!\!$  5.92\%) & 1.27$\,\cdot\, 10^{-2}$  ($\pm\!\!$  5.94\%) & 6.60$\,\cdot\, 10^{-3}$  ($\pm\!\!$  5.92\%) & 1.37$\,\cdot\, 10^{-3}$  ($\pm\!\!$  5.87\%) & 2.41$\,\cdot\, 10^{-5}$  ($\pm\!\!$ 19.70 \%)\\
1248 & 1410 & 6.62$\,\cdot\, 10^{-3}$  ($\pm\!\!$  6.30\%) & 4.97$\,\cdot\, 10^{-3}$  ($\pm\!\!$  6.35\%) & 2.24$\,\cdot\, 10^{-3}$  ($\pm\!\!$  6.22\%) & 3.14$\,\cdot\, 10^{-4}$  ($\pm\!\!$  6.73\%) & 1.29$\,\cdot\, 10^{-6}$  ($\pm\!\!$ 27.03 \%)\\
1410 & 1588 & 2.65$\,\cdot\, 10^{-3}$  ($\pm\!\!$  6.72\%) & 1.88$\,\cdot\, 10^{-3}$  ($\pm\!\!$  6.81\%) & 7.09$\,\cdot\, 10^{-4}$  ($\pm\!\!$  6.46\%) & 6.05$\,\cdot\, 10^{-5}$  ($\pm\!\!$  8.09\%) & 1.88$\,\cdot\, 10^{-8}$  ($\pm\!\!$ 48.46 \%)\\
1588 & 1784 & 1.04$\,\cdot\, 10^{-3}$  ($\pm\!\!$  7.25\%) & 6.83$\,\cdot\, 10^{-4}$  ($\pm\!\!$  7.25\%) & 2.06$\,\cdot\, 10^{-4}$  ($\pm\!\!$  6.77\%) & 9.17$\,\cdot\, 10^{-6}$  ($\pm\!\!$ 10.21\%) & 5.46$\,\cdot\, 10^{-12}$  ($\pm\!\!$ 111.1\%)\\
1784 & 2000 & 3.87$\,\cdot\, 10^{-4}$  ($\pm\!\!$  7.70\%) & 2.34$\,\cdot\, 10^{-4}$  ($\pm\!\!$  7.62\%) & 5.31$\,\cdot\, 10^{-5}$  ($\pm\!\!$  7.00\%) & 9.53$\,\cdot\, 10^{-7}$  ($\pm\!\!$ 12.01\%) & -----\\
2000 & 2238 & 1.37$\,\cdot\, 10^{-4}$  ($\pm\!\!$  8.16\%) & 7.40$\,\cdot\, 10^{-5}$  ($\pm\!\!$  8.03\%) & 1.17$\,\cdot\, 10^{-5}$  ($\pm\!\!$  7.22\%) & 5.27$\,\cdot\, 10^{-8}$  ($\pm\!\!$ 15.30\%) & -----\\
2238 & 2500 & 4.54$\,\cdot\, 10^{-5}$  ($\pm\!\!$  8.78\%) & 2.12$\,\cdot\, 10^{-5}$  ($\pm\!\!$  8.55\%) & 2.10$\,\cdot\, 10^{-6}$  ($\pm\!\!$  7.51\%) & 8.65$\,\cdot\, 10^{-10}$  ($\pm\!\!$ 23.54\%) & -----\\
2500 & 2787 & 1.37$\,\cdot\, 10^{-5}$  ($\pm\!\!$  9.36\%) & 5.31$\,\cdot\, 10^{-6}$  ($\pm\!\!$  8.94\%) & 2.85$\,\cdot\, 10^{-7}$  ($\pm\!\!$  8.16\%) & 5.59$\,\cdot\, 10^{-13}$  ($\pm\!\!$ 47.42\%) & -----\\
2787 & 3103 & 3.71$\,\cdot\, 10^{-6}$  ($\pm\!\!$  9.96\%) & 1.13$\,\cdot\, 10^{-6}$  ($\pm\!\!$  9.57\%) & 2.52$\,\cdot\, 10^{-8}$  ($\pm\!\!$  8.03\%) & ----- & -----\\
\hline
\end{tabular}
\caption{\label{tab:cms13-ct14-r04}
  Same as Tab.~\ref{tab:cms13-ct14-r04-nll} for the cross sections $\sigma_{\rm NLO}$.
}
\end{center}
\end{table}

\begin{table}[h!]
\renewcommand{\arraystretch}{1.3}
\begin{center}
\begin{tabular}{|r|r|r|r|r|r|r|}
  \hline
  \multicolumn{7}{l}{$\mathbf{\sigma_{\rm NLO + NLL}}$ \bf{for LHC at} $\mathbf{\sqrt{S}=13~\mbox{\bf TeV,}}$ $\mathbf{R=0.7}$}\\ 
  \hline
  \hline
\multicolumn{2}{|c|}{$p_T$ [GeV]} &
\multicolumn{5}{|c|}{$d\sigma/dp_T/d|y|$ [pb] \quad ($\pm \Delta \sigma_{\rm th.~scale}~\mbox{in}~\%) $} \\
\multicolumn{1}{|c|}{$p_T^{\min}$} &
\multicolumn{1}{|c|}{$p_T^{\max}$} &
\multicolumn{1}{|c|}{$0 \le |y| < 0.5$} &
\multicolumn{1}{|c|}{$0.5 \le |y| < 1.0$} &
\multicolumn{1}{|c|}{$1.0 \le |y| < 1.5$} &
\multicolumn{1}{|c|}{$1.5 \le |y| < 2.0$} &
\multicolumn{1}{|c|}{$2.0 \le |y| < 2.5$} \\
\hline
  56 &   74 & 1.69$\,\cdot\, 10^{+5}$  ($\pm\!\!$ 16.06\%) & 1.63$\,\cdot\, 10^{+5}$  ($\pm\!\!$ 13.32\%) & 1.44$\,\cdot\, 10^{+5}$  ($\pm\!\!$ 17.61\%) & 1.29$\,\cdot\, 10^{+5}$  ($\pm\!\!$ 16.04\%) & 1.04$\,\cdot\, 10^{+5}$  ($\pm\!\!$ 18.93 \%)\\
  74 &   97 & 4.70$\,\cdot\, 10^{+4}$  ($\pm\!\!$ 12.44\%) & 4.46$\,\cdot\, 10^{+4}$  ($\pm\!\!$ 13.11\%) & 4.09$\,\cdot\, 10^{+4}$  ($\pm\!\!$ 11.94\%) & 3.40$\,\cdot\, 10^{+4}$  ($\pm\!\!$ 13.50\%) & 2.71$\,\cdot\, 10^{+4}$  ($\pm\!\!$ 15.59 \%)\\
  97 &  133 & 1.17$\,\cdot\, 10^{+4}$  ($\pm\!\!$ 10.46\%) & 1.10$\,\cdot\, 10^{+4}$  ($\pm\!\!$ 12.34\%) & 9.74$\,\cdot\, 10^{+3}$  ($\pm\!\!$ 11.95\%) & 8.14$\,\cdot\, 10^{+3}$  ($\pm\!\!$ 10.86\%) & 6.30$\,\cdot\, 10^{+3}$  ($\pm\!\!$ 13.02 \%)\\
 133 &  174 & 2.69$\,\cdot\, 10^{+3}$  ($\pm\!\!$ 10.72\%) & 2.51$\,\cdot\, 10^{+3}$  ($\pm\!\!$ 10.35\%) & 2.17$\,\cdot\, 10^{+3}$  ($\pm\!\!$ 12.07\%) & 1.79$\,\cdot\, 10^{+3}$  ($\pm\!\!$  9.28\%) & 1.34$\,\cdot\, 10^{+3}$  ($\pm\!\!$ 11.08 \%)\\
 174 &  220 & 7.29$\,\cdot\, 10^{+2}$  ($\pm\!\!$ 10.65\%) & 6.83$\,\cdot\, 10^{+2}$  ($\pm\!\!$ 10.24\%) & 5.89$\,\cdot\, 10^{+2}$  ($\pm\!\!$  8.82\%) & 4.66$\,\cdot\, 10^{+2}$  ($\pm\!\!$  8.06\%) & 3.33$\,\cdot\, 10^{+2}$  ($\pm\!\!$ 10.05 \%)\\
 220 &  272 & 2.25$\,\cdot\, 10^{+2}$  ($\pm\!\!$ 10.59\%) & 2.08$\,\cdot\, 10^{+2}$  ($\pm\!\!$ 10.09\%) & 1.77$\,\cdot\, 10^{+2}$  ($\pm\!\!$  8.66\%) & 1.37$\,\cdot\, 10^{+2}$  ($\pm\!\!$  6.53\%) & 9.08$\,\cdot\, 10^{+1}$  ($\pm\!\!$  9.52 \%)\\
 272 &  330 & 7.49$\,\cdot\, 10^{+1}$  ($\pm\!\!$ 10.71\%) & 6.89$\,\cdot\, 10^{+1}$  ($\pm\!\!$ 10.23\%) & 5.73$\,\cdot\, 10^{+1}$  ($\pm\!\!$  8.30\%) & 4.30$\,\cdot\, 10^{+1}$  ($\pm\!\!$  5.55\%) & 2.65$\,\cdot\, 10^{+1}$  ($\pm\!\!$  8.96 \%)\\
 330 &  395 & 2.66$\,\cdot\, 10^{+1}$  ($\pm\!\!$ 10.60\%) & 2.41$\,\cdot\, 10^{+1}$  ($\pm\!\!$  9.61\%) & 1.98$\,\cdot\, 10^{+1}$  ($\pm\!\!$  7.95\%) & 1.42$\,\cdot\, 10^{+1}$  ($\pm\!\!$  4.93\%) & 7.81$\,\cdot\, 10^{+0}$  ($\pm\!\!$  8.68 \%)\\
 395 &  468 & 9.86$\,\cdot\, 10^{+0}$  ($\pm\!\!$ 10.46\%) & 8.94$\,\cdot\, 10^{+0}$  ($\pm\!\!$  9.53\%) & 7.12$\,\cdot\, 10^{+0}$  ($\pm\!\!$  7.41\%) & 4.81$\,\cdot\, 10^{+0}$  ($\pm\!\!$  4.46\%) & 2.30$\,\cdot\, 10^{+0}$  ($\pm\!\!$  8.38 \%)\\
 468 &  548 & 3.80$\,\cdot\, 10^{+0}$  ($\pm\!\!$ 10.30\%) & 3.39$\,\cdot\, 10^{+0}$  ($\pm\!\!$  9.10\%) & 2.65$\,\cdot\, 10^{+0}$  ($\pm\!\!$  6.86\%) & 1.66$\,\cdot\, 10^{+0}$  ($\pm\!\!$  4.50\%) & 6.54$\,\cdot\, 10^{-1}$  ($\pm\!\!$  8.18 \%)\\
 548 &  638 & 1.51$\,\cdot\, 10^{+0}$  ($\pm\!\!$ 10.19\%) & 1.33$\,\cdot\, 10^{+0}$  ($\pm\!\!$  9.11\%) & 1.00$\,\cdot\, 10^{+0}$  ($\pm\!\!$  6.28\%) & 5.72$\,\cdot\, 10^{-1}$  ($\pm\!\!$  4.64\%) & 1.78$\,\cdot\, 10^{-1}$  ($\pm\!\!$  7.93 \%)\\
 638 &  737 & 6.09$\,\cdot\, 10^{-1}$  ($\pm\!\!$  9.80\%) & 5.30$\,\cdot\, 10^{-1}$  ($\pm\!\!$  8.44\%) & 3.83$\,\cdot\, 10^{-1}$  ($\pm\!\!$  5.73\%) & 1.93$\,\cdot\, 10^{-1}$  ($\pm\!\!$  4.76\%) & 4.39$\,\cdot\, 10^{-2}$  ($\pm\!\!$  7.69 \%)\\
 737 &  846 & 2.51$\,\cdot\, 10^{-1}$  ($\pm\!\!$  9.68\%) & 2.15$\,\cdot\, 10^{-1}$  ($\pm\!\!$  8.04\%) & 1.46$\,\cdot\, 10^{-1}$  ($\pm\!\!$  5.03\%) & 6.33$\,\cdot\, 10^{-2}$  ($\pm\!\!$  4.90\%) & 9.62$\,\cdot\, 10^{-3}$  ($\pm\!\!$  7.48 \%)\\
 846 &  967 & 1.04$\,\cdot\, 10^{-1}$  ($\pm\!\!$  9.35\%) & 8.68$\,\cdot\, 10^{-2}$  ($\pm\!\!$  7.65\%) & 5.50$\,\cdot\, 10^{-2}$  ($\pm\!\!$  4.41\%) & 1.97$\,\cdot\, 10^{-2}$  ($\pm\!\!$  5.03\%) & 1.79$\,\cdot\, 10^{-3}$  ($\pm\!\!$  7.24 \%)\\
 967 & 1101 & 4.32$\,\cdot\, 10^{-2}$  ($\pm\!\!$  9.16\%) & 3.50$\,\cdot\, 10^{-2}$  ($\pm\!\!$  7.18\%) & 2.02$\,\cdot\, 10^{-2}$  ($\pm\!\!$  3.73\%) & 5.66$\,\cdot\, 10^{-3}$  ($\pm\!\!$  5.16\%) & 2.62$\,\cdot\, 10^{-4}$  ($\pm\!\!$  6.95 \%)\\
1101 & 1248 & 1.78$\,\cdot\, 10^{-2}$  ($\pm\!\!$  8.90\%) & 1.39$\,\cdot\, 10^{-2}$  ($\pm\!\!$  6.74\%) & 7.15$\,\cdot\, 10^{-3}$  ($\pm\!\!$  3.09\%) & 1.47$\,\cdot\, 10^{-3}$  ($\pm\!\!$  5.28\%) & 2.64$\,\cdot\, 10^{-5}$  ($\pm\!\!$  6.66 \%)\\
1248 & 1410 & 7.29$\,\cdot\, 10^{-3}$  ($\pm\!\!$  8.70\%) & 5.42$\,\cdot\, 10^{-3}$  ($\pm\!\!$  6.27\%) & 2.41$\,\cdot\, 10^{-3}$  ($\pm\!\!$  3.11\%) & 3.38$\,\cdot\, 10^{-4}$  ($\pm\!\!$  5.40\%) & 1.42$\,\cdot\, 10^{-6}$  ($\pm\!\!$  8.30 \%)\\
1410 & 1588 & 2.92$\,\cdot\, 10^{-3}$  ($\pm\!\!$  8.39\%) & 2.05$\,\cdot\, 10^{-3}$  ($\pm\!\!$  5.86\%) & 7.64$\,\cdot\, 10^{-4}$  ($\pm\!\!$  3.26\%) & 6.51$\,\cdot\, 10^{-5}$  ($\pm\!\!$  5.53\%) & 2.08$\,\cdot\, 10^{-8}$  ($\pm\!\!$ 11.13 \%)\\
1588 & 1784 & 1.14$\,\cdot\, 10^{-3}$  ($\pm\!\!$  8.17\%) & 7.43$\,\cdot\, 10^{-4}$  ($\pm\!\!$  5.43\%) & 2.21$\,\cdot\, 10^{-4}$  ($\pm\!\!$  3.45\%) & 9.80$\,\cdot\, 10^{-6}$  ($\pm\!\!$  5.69\%) & 5.87$\,\cdot\, 10^{-12}$  ($\pm\!\!$ 23.91 \%)\\
1784 & 2000 & 4.25$\,\cdot\, 10^{-4}$  ($\pm\!\!$  7.86\%) & 2.54$\,\cdot\, 10^{-4}$  ($\pm\!\!$  5.02\%) & 5.69$\,\cdot\, 10^{-5}$  ($\pm\!\!$  3.69\%) & 1.02$\,\cdot\, 10^{-6}$  ($\pm\!\!$  5.78\%) & -----\\
2000 & 2238 & 1.51$\,\cdot\, 10^{-4}$  ($\pm\!\!$  7.51\%) & 8.03$\,\cdot\, 10^{-5}$  ($\pm\!\!$  4.65\%) & 1.25$\,\cdot\, 10^{-5}$  ($\pm\!\!$  3.94\%) & 5.62$\,\cdot\, 10^{-8}$  ($\pm\!\!$  5.90\%) & -----\\
2238 & 2500 & 5.00$\,\cdot\, 10^{-5}$  ($\pm\!\!$  7.25\%) & 2.30$\,\cdot\, 10^{-5}$  ($\pm\!\!$  4.27\%) & 2.25$\,\cdot\, 10^{-6}$  ($\pm\!\!$  4.18\%) & 9.11$\,\cdot\, 10^{-10}$  ($\pm\!\!$  6.78\%) & -----\\
2500 & 2787 & 1.52$\,\cdot\, 10^{-5}$  ($\pm\!\!$  6.94\%) & 5.78$\,\cdot\, 10^{-6}$  ($\pm\!\!$  3.93\%) & 3.03$\,\cdot\, 10^{-7}$  ($\pm\!\!$  4.43\%) & 5.78$\,\cdot\, 10^{-13}$  ($\pm\!\!$ 11.27\%) & -----\\
2787 & 3103 & 4.13$\,\cdot\, 10^{-6}$  ($\pm\!\!$  6.64\%) & 1.23$\,\cdot\, 10^{-6}$  ($\pm\!\!$  3.62\%) & 2.69$\,\cdot\, 10^{-8}$  ($\pm\!\!$  4.66\%) & ----- & -----\\
\hline
\end{tabular}
\caption{\label{tab:cms13-ct14-r07-nll}
  Same as Tab.~\ref{tab:cms13-ct14-r04-nll} for the cross sections
  $\sigma_{\rm NLO + NLL}$ with $R=0.7$.
}
\end{center}
\end{table}

\begin{table}[h!]
\renewcommand{\arraystretch}{1.3}
\begin{center}
\begin{tabular}{|r|r|r|r|r|r|r|}
  \hline
  \multicolumn{7}{l}{$\mathbf{\sigma_{\rm NLO}}$ \bf{for LHC at} $\mathbf{\sqrt{S}=13~\mbox{\bf TeV,}}$ $\mathbf{R=0.7}$}\\ 
  \hline
  \hline
\multicolumn{2}{|c|}{$p_T$ [GeV]} &
\multicolumn{5}{|c|}{$d\sigma/dp_T/d|y|$ [pb] \quad ($\pm \Delta \sigma_{\rm th.~scale}~\mbox{in}~\%) $} \\
\multicolumn{1}{|c|}{$p_T^{\min}$} &
\multicolumn{1}{|c|}{$p_T^{\max}$} &
\multicolumn{1}{|c|}{$0 \le |y| < 0.5$} &
\multicolumn{1}{|c|}{$0.5 \le |y| < 1.0$} &
\multicolumn{1}{|c|}{$1.0 \le |y| < 1.5$} &
\multicolumn{1}{|c|}{$1.5 \le |y| < 2.0$} &
\multicolumn{1}{|c|}{$2.0 \le |y| < 2.5$} \\
\hline
  56 &   74 & 1.84$\,\cdot\, 10^{+5}$  ($\pm\!\!$  2.52\%) & 1.77$\,\cdot\, 10^{+5}$  ($\pm\!\!$  4.96\%) & 1.58$\,\cdot\, 10^{+5}$  ($\pm\!\!$  7.67\%) & 1.41$\,\cdot\, 10^{+5}$  ($\pm\!\!$  2.60\%) & 1.15$\,\cdot\, 10^{+5}$  ($\pm\!\!$  4.91 \%)\\
  74 &   97 & 5.05$\,\cdot\, 10^{+4}$  ($\pm\!\!$  4.62\%) & 4.81$\,\cdot\, 10^{+4}$  ($\pm\!\!$  4.13\%) & 4.41$\,\cdot\, 10^{+4}$  ($\pm\!\!$  4.96\%) & 3.68$\,\cdot\, 10^{+4}$  ($\pm\!\!$  3.87\%) & 2.95$\,\cdot\, 10^{+4}$  ($\pm\!\!$  3.42 \%)\\
  97 &  133 & 1.25$\,\cdot\, 10^{+4}$  ($\pm\!\!$  5.74\%) & 1.17$\,\cdot\, 10^{+4}$  ($\pm\!\!$  3.70\%) & 1.04$\,\cdot\, 10^{+4}$  ($\pm\!\!$  4.13\%) & 8.74$\,\cdot\, 10^{+3}$  ($\pm\!\!$  5.03\%) & 6.79$\,\cdot\, 10^{+3}$  ($\pm\!\!$  5.29 \%)\\
 133 &  174 & 2.85$\,\cdot\, 10^{+3}$  ($\pm\!\!$  5.12\%) & 2.66$\,\cdot\, 10^{+3}$  ($\pm\!\!$  4.51\%) & 2.31$\,\cdot\, 10^{+3}$  ($\pm\!\!$  3.14\%) & 1.91$\,\cdot\, 10^{+3}$  ($\pm\!\!$  5.38\%) & 1.43$\,\cdot\, 10^{+3}$  ($\pm\!\!$  5.03 \%)\\
 174 &  220 & 7.70$\,\cdot\, 10^{+2}$  ($\pm\!\!$  5.16\%) & 7.21$\,\cdot\, 10^{+2}$  ($\pm\!\!$  5.68\%) & 6.23$\,\cdot\, 10^{+2}$  ($\pm\!\!$  5.58\%) & 4.93$\,\cdot\, 10^{+2}$  ($\pm\!\!$  5.74\%) & 3.53$\,\cdot\, 10^{+2}$  ($\pm\!\!$  5.93 \%)\\
 220 &  272 & 2.37$\,\cdot\, 10^{+2}$  ($\pm\!\!$  6.07\%) & 2.19$\,\cdot\, 10^{+2}$  ($\pm\!\!$  6.15\%) & 1.86$\,\cdot\, 10^{+2}$  ($\pm\!\!$  6.13\%) & 1.44$\,\cdot\, 10^{+2}$  ($\pm\!\!$  6.40\%) & 9.58$\,\cdot\, 10^{+1}$  ($\pm\!\!$  5.84 \%)\\
 272 &  330 & 7.85$\,\cdot\, 10^{+1}$  ($\pm\!\!$  6.25\%) & 7.20$\,\cdot\, 10^{+1}$  ($\pm\!\!$  6.17\%) & 6.01$\,\cdot\, 10^{+1}$  ($\pm\!\!$  6.28\%) & 4.51$\,\cdot\, 10^{+1}$  ($\pm\!\!$  6.69\%) & 2.78$\,\cdot\, 10^{+1}$  ($\pm\!\!$  6.67 \%)\\
 330 &  395 & 2.78$\,\cdot\, 10^{+1}$  ($\pm\!\!$  6.32\%) & 2.52$\,\cdot\, 10^{+1}$  ($\pm\!\!$  6.34\%) & 2.07$\,\cdot\, 10^{+1}$  ($\pm\!\!$  6.29\%) & 1.49$\,\cdot\, 10^{+1}$  ($\pm\!\!$  6.86\%) & 8.18$\,\cdot\, 10^{+0}$  ($\pm\!\!$  6.35 \%)\\
 395 &  468 & 1.03$\,\cdot\, 10^{+1}$  ($\pm\!\!$  6.52\%) & 9.29$\,\cdot\, 10^{+0}$  ($\pm\!\!$  6.81\%) & 7.42$\,\cdot\, 10^{+0}$  ($\pm\!\!$  6.42\%) & 5.02$\,\cdot\, 10^{+0}$  ($\pm\!\!$  6.91\%) & 2.40$\,\cdot\, 10^{+0}$  ($\pm\!\!$  7.07 \%)\\
 468 &  548 & 3.95$\,\cdot\, 10^{+0}$  ($\pm\!\!$  7.14\%) & 3.53$\,\cdot\, 10^{+0}$  ($\pm\!\!$  6.99\%) & 2.76$\,\cdot\, 10^{+0}$  ($\pm\!\!$  7.37\%) & 1.73$\,\cdot\, 10^{+0}$  ($\pm\!\!$  7.39\%) & 6.82$\,\cdot\, 10^{-1}$  ($\pm\!\!$  6.59 \%)\\
 548 &  638 & 1.56$\,\cdot\, 10^{+0}$  ($\pm\!\!$  7.08\%) & 1.38$\,\cdot\, 10^{+0}$  ($\pm\!\!$  7.02\%) & 1.04$\,\cdot\, 10^{+0}$  ($\pm\!\!$  7.08\%) & 5.94$\,\cdot\, 10^{-1}$  ($\pm\!\!$  7.02\%) & 1.85$\,\cdot\, 10^{-1}$  ($\pm\!\!$  7.37 \%)\\
 638 &  737 & 6.30$\,\cdot\, 10^{-1}$  ($\pm\!\!$  7.37\%) & 5.48$\,\cdot\, 10^{-1}$  ($\pm\!\!$  7.44\%) & 3.96$\,\cdot\, 10^{-1}$  ($\pm\!\!$  7.54\%) & 2.00$\,\cdot\, 10^{-1}$  ($\pm\!\!$  7.40\%) & 4.57$\,\cdot\, 10^{-2}$  ($\pm\!\!$  7.58 \%)\\
 737 &  846 & 2.59$\,\cdot\, 10^{-1}$  ($\pm\!\!$  7.55\%) & 2.22$\,\cdot\, 10^{-1}$  ($\pm\!\!$  7.76\%) & 1.51$\,\cdot\, 10^{-1}$  ($\pm\!\!$  8.05\%) & 6.57$\,\cdot\, 10^{-2}$  ($\pm\!\!$  7.77\%) & 1.00$\,\cdot\, 10^{-2}$  ($\pm\!\!$  6.62 \%)\\
 846 &  967 & 1.07$\,\cdot\, 10^{-1}$  ($\pm\!\!$  8.03\%) & 8.93$\,\cdot\, 10^{-2}$  ($\pm\!\!$  7.88\%) & 5.67$\,\cdot\, 10^{-2}$  ($\pm\!\!$  8.06\%) & 2.04$\,\cdot\, 10^{-2}$  ($\pm\!\!$  7.78\%) & 1.87$\,\cdot\, 10^{-3}$  ($\pm\!\!$  6.80 \%)\\
 967 & 1101 & 4.42$\,\cdot\, 10^{-2}$  ($\pm\!\!$  8.03\%) & 3.60$\,\cdot\, 10^{-2}$  ($\pm\!\!$  8.29\%) & 2.08$\,\cdot\, 10^{-2}$  ($\pm\!\!$  8.40\%) & 5.87$\,\cdot\, 10^{-3}$  ($\pm\!\!$  7.85\%) & 2.73$\,\cdot\, 10^{-4}$  ($\pm\!\!$  6.93 \%)\\
1101 & 1248 & 1.82$\,\cdot\, 10^{-2}$  ($\pm\!\!$  8.46\%) & 1.43$\,\cdot\, 10^{-2}$  ($\pm\!\!$  8.48\%) & 7.37$\,\cdot\, 10^{-3}$  ($\pm\!\!$  8.44\%) & 1.53$\,\cdot\, 10^{-3}$  ($\pm\!\!$  8.10\%) & 2.77$\,\cdot\, 10^{-5}$  ($\pm\!\!$  9.85 \%)\\
1248 & 1410 & 7.42$\,\cdot\, 10^{-3}$  ($\pm\!\!$  8.74\%) & 5.55$\,\cdot\, 10^{-3}$  ($\pm\!\!$  8.76\%) & 2.49$\,\cdot\, 10^{-3}$  ($\pm\!\!$  8.53\%) & 3.52$\,\cdot\, 10^{-4}$  ($\pm\!\!$  7.92\%) & 1.49$\,\cdot\, 10^{-6}$  ($\pm\!\!$ 16.79 \%)\\
1410 & 1588 & 2.96$\,\cdot\, 10^{-3}$  ($\pm\!\!$  9.13\%) & 2.09$\,\cdot\, 10^{-3}$  ($\pm\!\!$  9.06\%) & 7.87$\,\cdot\, 10^{-4}$  ($\pm\!\!$  8.82\%) & 6.78$\,\cdot\, 10^{-5}$  ($\pm\!\!$  7.78\%) & 2.20$\,\cdot\, 10^{-8}$  ($\pm\!\!$ 30.07 \%)\\
1588 & 1784 & 1.15$\,\cdot\, 10^{-3}$  ($\pm\!\!$  9.34\%) & 7.58$\,\cdot\, 10^{-4}$  ($\pm\!\!$  9.53\%) & 2.28$\,\cdot\, 10^{-4}$  ($\pm\!\!$  9.20\%) & 1.03$\,\cdot\, 10^{-5}$  ($\pm\!\!$  7.52\%) & 6.36$\,\cdot\, 10^{-12}$  ($\pm\!\!$ 89.39 \%)\\
1784 & 2000 & 4.29$\,\cdot\, 10^{-4}$  ($\pm\!\!$  9.77\%) & 2.58$\,\cdot\, 10^{-4}$  ($\pm\!\!$  9.63\%) & 5.88$\,\cdot\, 10^{-5}$  ($\pm\!\!$  9.38\%) & 1.08$\,\cdot\, 10^{-6}$  ($\pm\!\!$  8.06\%) & -----\\
2000 & 2238 & 1.52$\,\cdot\, 10^{-4}$  ($\pm\!\!$ 10.32\%) & 8.16$\,\cdot\, 10^{-5}$  ($\pm\!\!$ 10.11\%) & 1.30$\,\cdot\, 10^{-5}$  ($\pm\!\!$  9.56\%) & 5.96$\,\cdot\, 10^{-8}$  ($\pm\!\!$  7.88\%) & -----\\
2238 & 2500 & 4.99$\,\cdot\, 10^{-5}$  ($\pm\!\!$ 10.70\%) & 2.33$\,\cdot\, 10^{-5}$  ($\pm\!\!$ 10.69\%) & 2.33$\,\cdot\, 10^{-6}$  ($\pm\!\!$  9.90\%) & 9.83$\,\cdot\, 10^{-10}$  ($\pm\!\!$ 15.84\%) & -----\\
2500 & 2787 & 1.51$\,\cdot\, 10^{-5}$  ($\pm\!\!$ 11.32\%) & 5.85$\,\cdot\, 10^{-6}$  ($\pm\!\!$ 11.09\%) & 3.15$\,\cdot\, 10^{-7}$  ($\pm\!\!$ 10.25\%) & 6.47$\,\cdot\, 10^{-13}$  ($\pm\!\!$ 26.69\%) & -----\\
2787 & 3103 & 4.06$\,\cdot\, 10^{-6}$  ($\pm\!\!$ 11.93\%) & 1.24$\,\cdot\, 10^{-6}$  ($\pm\!\!$ 11.59\%) & 2.81$\,\cdot\, 10^{-8}$  ($\pm\!\!$ 11.01\%) & ----- & -----\\
\hline
\end{tabular}
\caption{\label{tab:cms13-ct14-r07}
  Same as Tab.~\ref{tab:cms13-ct14-r07-nll} for the cross sections $\sigma_{\rm NLO}$.
}
\end{center}
\end{table}

\twocolumngrid

\bibliographystyle{apsrev}
\bibliography{bibliography}

\end{document}